\documentclass[journal,twoside]{IEEEtran}
\usepackage{cite}
\usepackage{amsmath,amssymb,amsfonts}
\usepackage{algorithmic}
\usepackage{graphicx}
\usepackage{textcomp}
\usepackage{xcolor}
\usepackage{booktabs}
\usepackage{threeparttable}
\usepackage{multirow}
\usepackage{marvosym}
\usepackage{colortbl}
\usepackage{hyperref}
\usepackage{amsmath}
\usepackage{amssymb}
\usepackage{bm}
\usepackage{pifont}
\definecolor{deepgreen}{rgb}{0,0.7,0}

\begin{document}
\title{\vspace{-0.3em}In-Loop Filtering Using Learned Look-Up Tables for Video Coding}	
\author{	
Zhuoyuan Li,
Jiacheng Li, 
Yao Li, 
Jialin Li,
Li Li, \IEEEmembership{Senior Member, IEEE,}
Dong Liu, \IEEEmembership{Senior Member, IEEE,}\\
and Feng Wu, \IEEEmembership{Fellow, IEEE}\vspace{-1.6em}

\thanks{The authors are with the MOE Key Laboratory of Brain-Inspired Intelligent Perception and Cognition, University of Science and Technology of
China, Hefei 230093, China (e-mail: \{zhuoyuanli, jclee, mrliyao, jialin.li\}@mail.ustc.edu.cn, \{lil1, dongeliu, fengwu\}@ustc.edu.cn).	(\textit{Corresponding Author: Dong Liu.})
}
}

\markboth{Under Review}
{Li \MakeLowercase{\textit{et al.}}: In-Loop Filtering using Learned Look-Up Tables for Video Coding}

\maketitle

\begin{abstract}
In-loop filtering (ILF) is a key technology in video coding standards to reduce artifacts and enhance visual quality. Recently, neural network-based ILF schemes have achieved remarkable coding gains, emerging as a powerful candidate for next-generation video coding standards. However, the use of deep neural networks (DNN) brings significant computational and time complexity or high demands for dedicated hardware, making it challenging for general use. To address this limitation, we study a practical ILF solution by adopting look-up tables (LUTs). After training a DNN with a restricted reference range for ILF, all possible inputs are traversed, and the output values of the DNN are cached into LUTs. During the coding process, the filtering process is performed by simply retrieving the filtered pixel through locating the input pixels and interpolating between the cached values, instead of relying on heavy inference computations. In this paper, we propose a universal LUT-based ILF framework, termed LUT-ILF++. First, we introduce the cooperation of multiple kinds of filtering LUTs and propose a series of customized indexing mechanisms to enable better filtering reference perception with limited storage consumption. Second, we propose the cross-component indexing mechanism to enable the filtering of different color components jointly. Third, in order to make our solution practical for coding uses, we propose the LUT compaction scheme to enable the LUT pruning, achieving a lower storage cost of the entire solution.  The proposed framework is implemented in the Versatile Video Coding reference software. Experimental results show that the proposed framework achieves on average 0.82\%/2.97\%/1.63\% and 0.85\%/4.11\%/2.06\% bitrate reduction for common test sequences, under the all-intra and random-access configurations, respectively. Compared to DNN-based solutions, our proposed solution has much lower time complexity and storage cost.
\end{abstract}
\begin{IEEEkeywords}
Deep neural network, in-loop filtering, look-up table, Versatile Video Coding, video coding.
\end{IEEEkeywords}\vspace{-0.5em}

\section{Introduction}
In-loop filtering (ILF) has been widely adopted in advanced video coding standards, such as H.265/HEVC\cite{sullivan2012overview},  H.266/VVC\cite{bross2021overview, karczewicz2021vvc}, AV1 and AV2 \cite{han2021technical, AOM}. To enhance the objective and subjective reconstructed quality of decoded frames, various hand-crafted in-loop filters make a major contribution to these standards and play a key role in the hybrid coding framework, such as deblocking filter (DBF)\cite{list2003adaptive, karczewicz2021vvc}, sample adaptive offset (SAO)\cite{fu2012sample, kuo2022cross, karczewicz2021vvc}, and adaptive loop filtering (ALF)\cite{tsai2013adaptive, meng2021optimized, karczewicz2021vvc}, etc. Recently, deep neural network (DNN) based coding tools (e.g., ILF\cite{dai2017convolutional, dai2018cnn, li2021convolutional, li2023idam, man2025content, jia2019content, wang2021combining, kathariya2023joint, T0088, AB0068, AC0177}, intra/inter prediction\cite{li2021neural, li2024object, feng2024efficient, li2024ustc, feng2025partition}, reference frame generation\cite{jia2023deep}, sampling\cite{li2023designs, li2017convolutional, lin2025low}, etc.) have been rapidly developed for next-generation video coding standards\cite{liu2020deep}. Notably, neural network-based ILF (NNLF) has emerged as a particularly promising tool, achieving significant progress in advanced standardization activities, such as neural network-based video coding (NNVC)\cite{li2023designs} and AV2\cite{AOM}. These DNN-based ILF tools leverage data-driven capabilities to learn effective filters and adaptive filtering strategies, surpassing the hand-crafted filtering techniques. However, their significant computational/time complexity, along with high demands for dedicated hardware, pose challenges for practical applications.

To improve their practicality, a series of optimization solutions have been proposed in previous research to facilitate the integration of these DNN-based models into codecs for practical application and deployment, such as lightweight network designs\cite{AF0181, AF0206, AH0050, AJ0066}, model ensembling\cite{AG0174, huang2021adaptive, AH0195, AJ0054, li2023lightweight}, low-complexity neural operators\cite{AG0155, AH0077, AD0211}, dimensionality reduction\cite{AG0069, AH0207, AI0173, liu2024nn}, computational complexity re-allocation\cite{AD0157, AH0188, zhang2023lightweight,man2025content}, re-parameterization\cite{AH0077}, etc. Particularly, in the standardization activities (such as NNVC\cite{li2023designs}), different target computational complexity (operation point) configurations of NNLF, including Very Low Operation Point (VLOP) 1$\sim$3 \cite{AG0057, AH0051, AI0107}, Low Operation Point (LOP) 1$\sim$5 \cite{AE0281, AF0043, AH0014, AG0069, AI0014}, High Operation Point (HOP) 1$\sim$5 \cite{AD0380, AG0174, AH0014, AH0189, AH0205, AI0014, AI0172}, have been introduced and investigated to provide different trade-offs between compression efficiency and computational resource consumption, which allows for flexible adaptation to diverse deployment scenarios. Although the above schemes can reduce and control the network complexity and further improve model efficiency to some extent, the inherently heavy inference computation burden in codecs remains unavoidable.

\begin{figure}
	\centering
	\includegraphics[width=75mm]{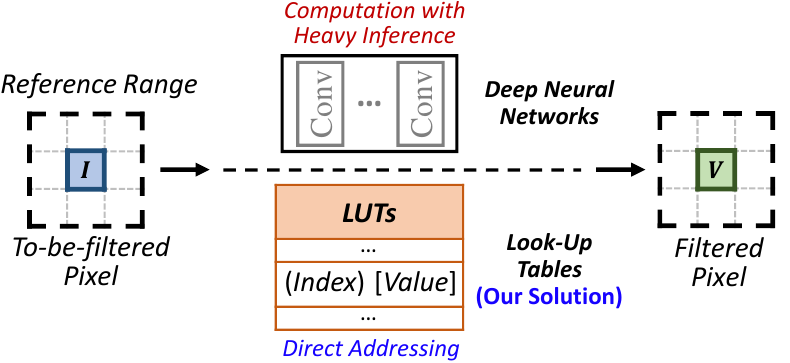}
	\vspace{-0.5em}
	\caption{Illustration of the concept of different deployment manners of DNN-based coding tools in the codec.}
	\label{concept}
	\vspace{-1.5em}
\end{figure}

To address this limitation, inspired by our previous studies in image restoration tasks \cite{li2022mulut, li2024toward, li2024look}, we rethink the efficient practical deployment of these DNN-based coding tools in video coding. In our solution, as shown in Fig.~\ref{concept}, we study a look-up table (LUT)-based approach, and our basic concept is to adopt the look-up operation of LUT to replace the heavy computational process of DNN inference in the coding process. To achieve this goal, the DNN is first trained with a restricted reference range for ILF. Then, the output values of the trained DNN are cached into a LUT by traversing all possible inputs. In the coding process, the filtered pixel is obtained by locating the combination of the to-be-filtered pixel and its reference pixels in the LUT, and interpolating among the cached key values. By introducing a minimal possible storage consumption of LUTs in the codec, we aim to achieve a good trade-off between computational/time complexity and coding performance for practical use, while also being friendly for hardware acceleration with far fewer floating-point operations. 

Building upon this basic concept, in our preliminary work\cite{li2024loop}, we propose a basic \textbf{LUT}-based \textbf{i}n-\textbf{l}oop \textbf{f}iltering framework, termed \textbf{LUT-ILF}. Inspired by the design principle of typical hand-crafted and DNN-based in-loop filters, we combine the LUT attributes and attempt two explorations: (1) \textit{Filtering Reference.} As a crucial design factor of ILF, the utilization and selection of reference pixels that are related to the to-be-filtered pixel play a significant role in capturing local structures and reducing artifacts. To achieve a good filtering goal while avoiding the bursty growth of LUT storage size as the dimension of indexing entries increases in the reference expansion, a series of customized indexing mechanisms across LUTs is proposed to handle this challenge by linearly stacking multiple LUTs for the access of more reference pixels. (2) \textit{Storage Constraint.} As a crucial factor of LUT for practical use, the storage cost is essential to ensure its feasibility. To achieve the controllable storage cost, customized LUT training and cache strategies are proposed to constrain the storage of each LUT with its entry dimension strictly. The potential of the basic framework has been verified across the different scales of complexity in our preliminary work.

In this paper, we further investigate and address the existing bottlenecks and propose a universal framework, termed \textbf{LUT-ILF++}. Specifically, we tackle three bottlenecks: (1) \textit{Filtering Reference Perception.} Due to the constrained cache dimension of index entries of LUT for limited storage cost, it becomes challenging to perceive more effective reference information as the filtering reference range expands, impacting the realization of the optimal filtering goal. (2) \textit{Multi-component Reference Collaboration.} The independent filtering process of each chroma component overlooks the inherent correlations among the different components, impacting the performance of the chroma-component filtering while bringing additional complexity.  (3) \textit{Storage Overhead.} Although the customized LUT training and cache strategies can control the storage cost, it is difficult to optimize storage usage based on the specific demands of the filtering goal, impacting its practicality and limiting its flexibility in deployment. To overcome the above existing bottlenecks, we put forward LUT-ILF++. The contributions of this paper are summarized as follows:

\begin{itemize}
	\item We devise a universal framework, termed LUT-ILF++, for efficient in-loop filtering by activating the multiple functions and cooperative manners of LUTs. Our solution explores a new and more practical way for neural network-based coding tools.
	\item To overcome the bottleneck of filtering reference perception, we introduce the cooperation of multiple filtering LUTs with customized indexing mechanisms, and design a series of cooperative manners to construct the entire filtering process, enabling efficient reference modeling.
	\item To address the cross-component filtering, we propose a cross-component indexing mechanism to enable the collaborative filtering of different  components.
	\item We observe the usage statistics of cached pixel relationships of LUT among different reference ranges. Based on the relationship, we introduce a LUT compaction scheme and propose a LUT pruning strategy with a separable indexing mechanism to enable a lower storage cost of the whole filtering framework in practical use.
\end{itemize}

\vspace{-0.5em}
\section{Preliminary}
\vspace{-0.3em}
In this section, first, we introduce our motivation based on the related works, and retrospect the basic solution of our proposed LUT-based ILF solution in the preliminary work \cite{li2024loop}. Second, we analyze the bottlenecks of the LUT-based solution and propose potential directions for further improvement.

\vspace{-0.8em}
\subsection{Motivation and Basic Solution}
\vspace{-0.1em}
LUT has been extensively adopted as an efficient mapping operator, playing a crucial role in the practical applications of image processing tasks, such as photo enhancement \cite{zeng2020learning} and color manipulation \cite{gems2programming, kim2012new, mantiuk2008display}. A LUT comprises a set of discrete index-value pairs, where the indices serve as inputs and the pre-computed values provide the outputs of a complicated function or a series of imaging computations during inference. Their compact structure can be stored in on-device storage, and supports the low-latency, high-throughput execution by just retrieving pre-computed values from the LUT in memory. Recently, with the rapid development of DNNs and their remarkable performance across various image processing tasks, a series of works \cite{jo2021practical, li2022mulut, li2024toward, li2024look} have tentatively explored the deployment of DNNs for different image processing tasks more practically, leveraging the efficient structure of LUTs and building the mapping relationships to cache DNN models into the LUTs. These works have verified that the LUT-based solution can preserve decent performance while significantly reducing time and computational complexity, and improving deployment efficiency on the device. 

\begin{figure}
	\centering
	\includegraphics[width=85mm]{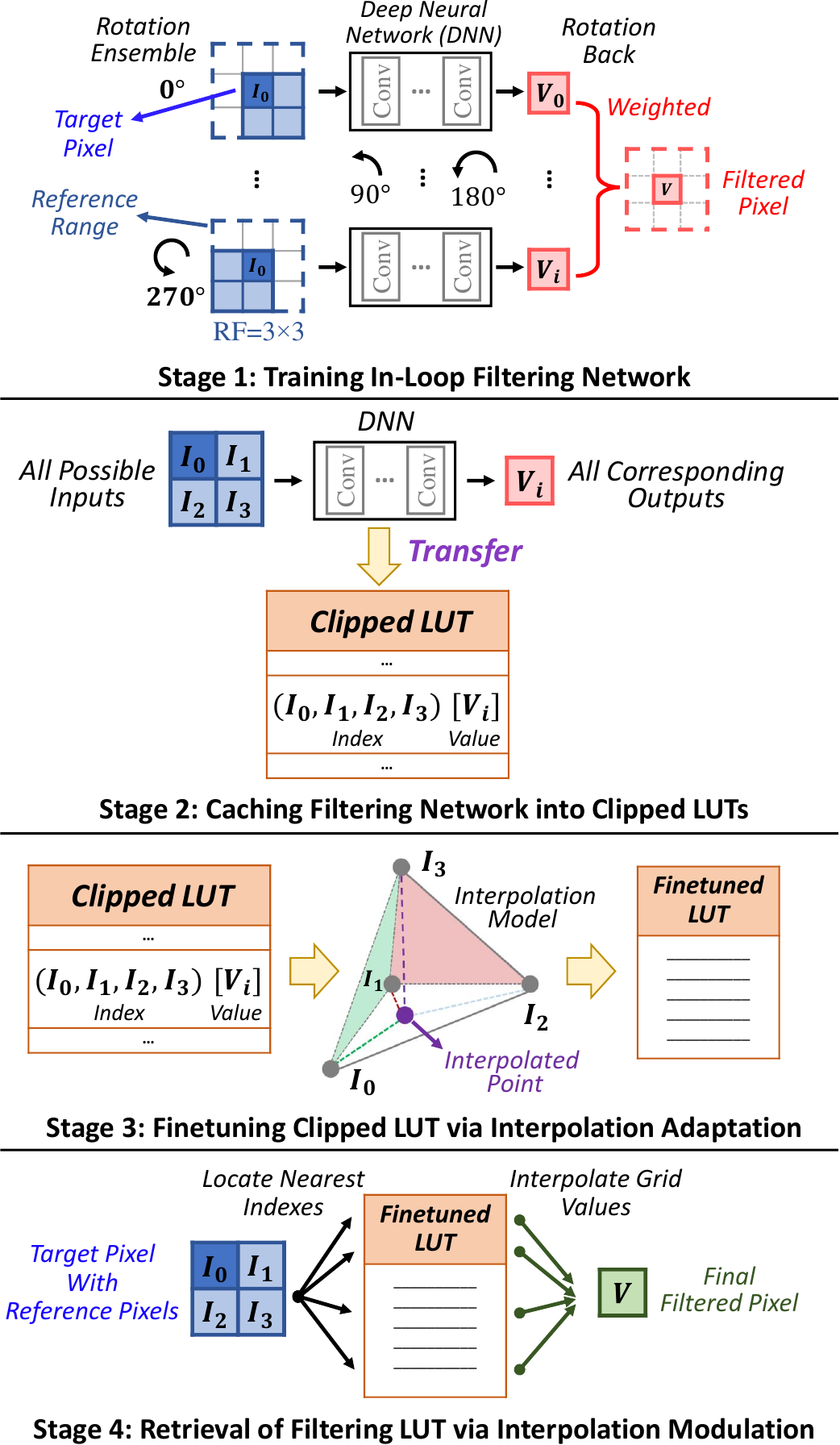}
	\vspace{-0.4em}
	\caption{Overview of the proposed basic LUT-based ILF solution.}
	\label{basic-framework}
	\vspace{-1.7em}
\end{figure}

Motivated by the efficiency of LUT structures in imaging tasks and the significant progress of DNN-based coding tools in advanced standardization activities \cite{dai2017convolutional, dai2018cnn, li2021convolutional, li2023idam, man2025content,jia2019content, wang2021combining, kathariya2023joint, T0088, AB0068, AC0177, li2021neural, jia2023deep,li2023designs, li2017convolutional, lin2025low}, we revisit the practical application issue of these DNN-based coding tools in video coding. In our solution, we study an efficient and practical LUT-based coding tool, and adopt the “\textit{space-trading-for-time-and-computation}” strategy that trades a moderate storage cost of LUTs to replace the huge computational overhead of DNN, thereby reducing the computational and time complexity in practical use. We attempt to apply it to the typical ILF tool by caching the filtering mapping of high-performance NNLF. As shown in Fig.~\ref{basic-framework}, our concept is to adopt the look-up operation (direct addressing) of LUT to replace the heavy computational process of DNN inference of NNLF in the coding process. To achieve the filtering goal, our basic solution comprises four stages to achieve the whole filtering process.

\subsubsection{Training in-loop filtering network}
First, the filtering network is trained with a restricted reference range/receptive field in an end-to-end manner. As a crucial factor of ILF, the effective relationship modeling between the target (to-be-filtered) pixel and its reference pixels plays a significant role in capturing local structures and reducing artifacts. Unlike training a regular filtering DNN to perceive the pixel relationship of an input patch simply, the LUT-based solution requires caching the complex pixel relationships learned by the network into discrete indexing entries of LUT, which directly impacts on the storage consumption. Due to the storage size of LUT growing exponentially as the dimension of indexing entries (i.e., target pixel with reference pixels) increases, we take the 2$\times$2 reference range (4D LUT) as the basic unit, and constrain its relationship up to 4 pixels, which can control the cache entries only depending on a small range of input values. For training, the target pixel ($I_0$) with three surrounding reference pixels (solid line) serves as the input to the network. To enlarge the reference range, the rotation ensemble trick is used to cover the 3$\times$3 reference range (dotted line). The final output value (filtered pixel) is averaged by all outputs of the four rotation inputs ($V_0$$\sim$$V_3$). During training, the filtered and original pixels form a pair, which is supervised by the mean-squared-error loss.

\begin{figure*}
	\centering
	\vspace{-0.5em}
	\includegraphics[width=175mm]{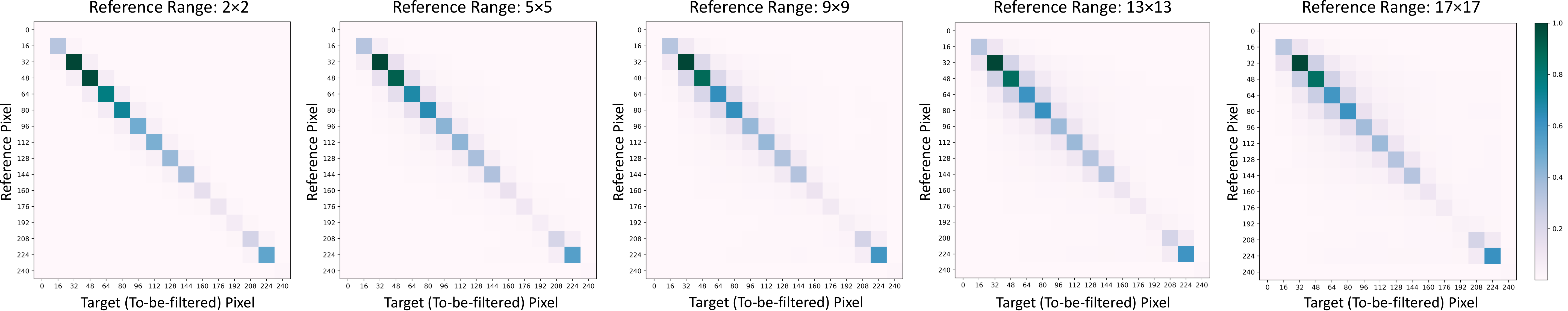}
	\vspace{-0.6em}
	\caption{Pixel value distributions, obtained from pre-filtering frames under All Intra (AI) configuration with QPs of 22, 27, 32, 37, 42, 47 within $2\times2$, $5\times5$, $13\times13$, $17\times17$ reference ranges on the VVC common test sequences~\cite{CTCdocument, liu2021jvet}. The joint distributions between the to-be-filtered pixel and its neighboring pixels (considering the MSB precision) located at 1, 4, 8, 12, and 16 pixels away in the bottom, right, and bottom-right positions are illustrated. The color scale ranges from white (low frequency) to dark blue (high frequency), representing the normalized occurrence frequency.}
	\vspace{-1em}
	\label{fig:pixel_heat}
\end{figure*}

\subsubsection{Caching filtering network into clipped LUTs}
Second, with the filtering network being trained, the filtering LUT (4D) is transferred and cached from the output values of the network via traversing all possible inputs (target pixel with reference pixels, [$0$$\sim$$255$][$0$$\sim$$255$][$0$$\sim$$255$][$0$$\sim$$255$] for $uint8$ case of input), as shown in  Fig.~\ref{basic-framework}. Note that the storage of LUT with a large input/output range will bring heavy storage cost, for example, the full storage size of 4D LUT is $256^4$$\times$1$\times$8 bit $=$ 4 GB, $2^{32}$ bins for possible input value, 1 for 8-bit output value. To avoid the heavy storage cost, the indexing entries of the full LUT are uniformly sampled and stored into the clipped LUT, which only caches the output value of the most significant bits (MSB) of the input pixel value. In our design, the 8-bit input pixel value is uniformly sampled to 4 MSBs, and the 4 MSBs serve as the initial (nearest) index for the indexing of input pixels. For a clipped LUT, the available indexing entries are reduced to [$0$, $16$, $...$, $240$, $255$][$0$, $16$, $...$, $240$, $255$][$0$, $16$, $...$, $240$, $255$][$0$, $16$, $...$, $240$, $255$], and the size is decreased to $17^4$$\times$1$\times$8 bit $=$ 81.56 KB. 

\subsubsection{Finetuning clipped LUT via interpolation adaptation}
Direct sampling of indexing entries can obviously restrict the rapid growth of storage cost, but the non-sampled indexing entries will cause the indexing drift in the retrieval process.
For non-sampled entries, we introduce the interpolation model to estimate the drifting entries, such as the trilinear and 4-simplex model, and the interpolation process is performed to calculate the final filtered pixel by locating the nearest neighbor indices (MSBs) of query indices (pixels) and weighting the cached values of neighbor indices during LUT retrieval. To further compensate for the degradation of non-sampled indexing entries, the finetuning of the clipped LUT is further performed to compensate for the adaptation of the uniform sampling and the interpolation model, facilitating the interpolation of the final retrieved filtered pixel of non-sampled indices from the nearest sampled indices. In finetuning, the cached values of the clipped LUT are activated as the trainable parameters and finetuned by the same setting of filtering network training.

\subsubsection{Retrieving of filtering LUT via interpolation modulation}
During the retrieval of the filtering LUT in the ILF process, the MSBs of the input pixels ($I_0, I_1, I_2, I_3$) are used to locate the nearest indices in the 4D clipped LUT. The corresponding output values, along with the least significant bits (LSBs) of input pixels, are then fed into a linear interpolation model to modulate the final filtered pixel. For the interpolation scheme of clipped LUT in our framework, we follow the same model as \cite{zeng2020learning, jo2021practical}, and use the trilinear/4-$simplex$ interpolation model for 3D/4D LUT in the whole filtering framework.

\vspace{-1em}
\subsection{Bottlenecks}
\vspace{-0.3em}
Based on our preliminary solution \cite{li2024loop}, we re-examine the desirable filtering techniques in emerging coding standards, and we identify three bottlenecks of our proposed LUT-based solution in achieving a practical yet effective alternative. 

\subsubsection{Filtering Reference Perception}
Reference perception modeling has been verified as a crucial factor of filtering performance, such as the diamond filter shape with a 7×7 reference range in ALF\cite{karczewicz2021vvc}, and the flexible neural operators with a wide reference range (receptive field) in DNN-based ILFs\cite{li2023designs}. In the proposed LUT-based ILF solution, the adoption of the “\textit{space-trading-for-time-and-computation}” strategy enables the modeling capability of reference perception to be directly translated into the cached indexing entries of pixel relationships in the LUT, thereby avoiding the online computation of complex filtering functions. However, due to the exhaustive relationship of pixels, the storage size of a single LUT grows exponentially with respect to the increasing input dimension (the number of to-be-filtered and reference pixels). The storage size of a single LUT with the $int8$ precision of input/cache values can be formulated as,
\begin{equation}
	MS = \left(2^{8 - q} + 1\right)^n \times V \times 8 \ bit,
\end{equation}
where $MS$, $q$, $n$, $V$ denote the abbreviation of storage size, the sampling interval of LUT, the cached dimension of the LUT, and the cached value number of each indexing entry in the LUT, respectively. For instance, a fully 4D LUT ($q=0$, $n=4$, $V=1$) requires 64 GB, while extending to a 5D LUT under the same settings would demand over 16 TB; a clipped 4D LUT ($q=4$, $n=4$, $V=1$) requires 81.56 KB, while extending to a 5D LUT would demand over 1 GB. The exponential growth makes it infeasible and impractical to improve the reference perception by increasing the dimension of a single LUT. To address this limitation, our preliminary framework \cite{li2024loop} has proposed that the reference perception can be enlarged to 9×9, 13×13 through the ensemble trick of look-up operations or linearly stacked LUT-driven strategies. However, despite these improvements, the reference perception capability remains significantly limited when compared to that of advanced hand-crafted or DNN-based filtering schemes.

\begin{figure*}
	\centering
	\vspace{-0.3em}
	\includegraphics[width=175mm]{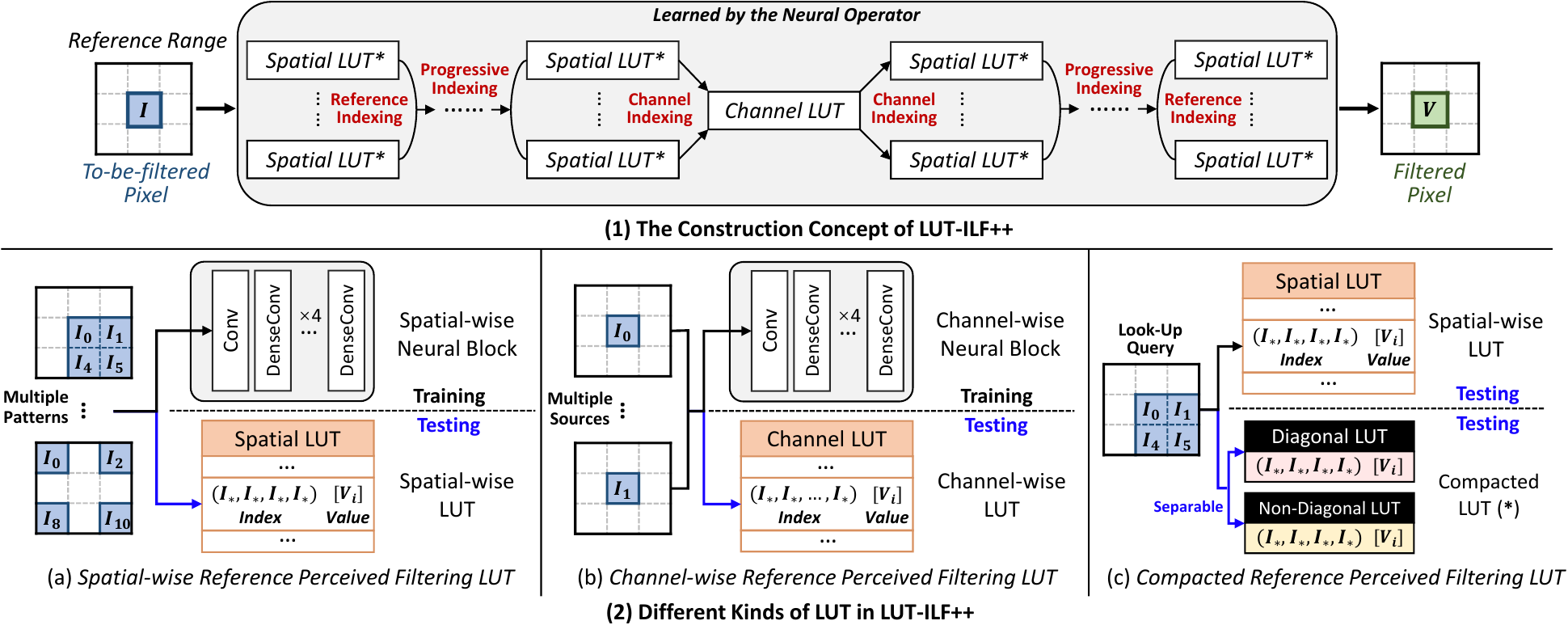}
	\vspace{-0.3em}
	\caption{Overview of LUT-ILF++. (1) The construction concept of LUT-ILF++. The different kinds of LUTs are jointly assembled into a loop filter through coordinated customized indexing mechanisms. The symbol * denotes that the LUT has been pruned via the LUT compaction scheme, and is inferred through the separable indexing mechanism mentioned in Section VI. (2) Different kinds of LUTs in LUT-ILF++. The single LUT is generalized into multiple kinds of elementary reference perceived LUT units for the establishment of a cooperative LUT-based in-loop filtering framework.}
	\label{LUT-ILF++}
	\vspace{-0.5em}
\end{figure*}

\subsubsection{Multi-component Reference Collaboration}
Beyond the direct spatial-wise reference perception of loop filters, various reference perception sources have also been verified to be beneficial for filtering effectiveness, such as the cross-component-guided filtering (e.g., CCALF\cite{karczewicz2021vvc}, CCSAO\cite{kuo2022cross} adopted in VVC), and the spatial-wise multi-reference-range-guided filtering (e.g., rich channel-wise interactions across multiple reference ranges in DNN-based filters\cite{li2023designs}). Due to the inherent attribute of LUT, the direct relationship extension of multi-component reference information within a single LUT will similarly bring the exponential growth of storage consumption, which constrains the collaborative complementary information perception of the whole filtering process.

\subsubsection{Storage Overhead}
In the exhaustive relationship of LUT construction, all possible relationships of pixel values need to be enumerated and stored. Although the design of uniform sampling of the LUT in our basic solution can effectively reduce its storage size by caching only the MSBs of the input, the direct clipping of the LUT ignores the inherent properties of pixel correlations and their actual access relevance of reference perception for the filtering goal, potentially leading to sub-optimal storage compaction. For a simple instance, we observe the retrieved pixel value distributions from the pre-filtering frames under QPs of 22, 27, 32, 37, 42, 47 within different reference ranges on the VVC common test sequences\cite{CTCdocument, liu2021jvet}. In Fig.~\ref{fig:pixel_heat}, we illustrate the relationships between the to-be-filtered pixel and its neighboring (reference) pixels at MSB precision located at different distances in the bottom, right, and bottom-right directions. We can observe a diagonal phenomenon in the occurrence frequency statistics, which reveals two key limitations of the cached LUT. First, a large number of LUT entries are rarely accessed in practice, due to the high similarity between local reference pixels and the to-be-filtered pixel, which leads to concentrated access within a limited subset of LUT indexing entries. Second, as the reference range increases, the retrieved relationships gradually shift toward pixel pairs with larger value differences, indicating that reference perception imposes varying demands on relationship caching across different reference ranges.

\vspace{-0.9em}
\section{The Framework of LUT-ILF++}
\vspace{-0.2em}
To overcome the existing bottlenecks of our preliminary study \cite{li2024loop}, we propose a universal framework, LUT-ILF++, for efficient in-loop filtering by enabling multiple LUTs. In LUT-ILF++, we abandon the utilization of a single-LUT paradigm and introduce the concept of multiple cooperative LUTs, as shown in Fig.~\ref{LUT-ILF++}, which activates the multiple functions and cooperative manners of LUTs. Inspired by the design of hand-crafted and DNN-based loop filters, in the whole filtering process, we treat a single LUT as a basic unit and construct the whole filtering architecture by the cooperation of different kinds of LUTs and the link of customized indexing mechanisms. Specifically, as shown in Fig.~\ref{LUT-ILF++} (1), the framework of LUT-ILF++ is constructed by three kinds of basic LUT units, including spatial-wise, channel-wise, and compacted (*) reference perceived filtering LUTs, which generalizes the single LUT unit to multiple elementary reference perceived LUT units to establish a LUT-based loop filter flexibly. To promote the collaborative capacity of these basic LUT units, a series of cooperative manners and customized indexing mechanisms are introduced to coordinate and manage them, including reference indexing, progressive indexing, channel indexing, and separable indexing (*). It allows each LUT to focus on a specific aspect of reference modeling, achieving effective reference perception with low computation and storage cost. 

During training, as shown in Fig.~\ref{LUT-ILF++}, the different kinds of LUTs are reproduced by their corresponding neural blocks, and the blocks are linked through the customized indexing mechanisms to maintain the structural dependencies required for LUT cooperation. These components are jointly assembled into a unified neural network for end-to-end optimization. With the network training completed, the inputs and outputs of each neural block are cached and compacted (*) into each LUT, and the indexing relationships among neural blocks are retained for LUTs to preserve the cooperative structure learned during training. During inference, all neural computations are replaced by efficient LUT look-up operations. Each LUT independently retrieves the filtering mapping based on its input dimensions, and the indexing relationships ensure the whole filtering pipeline across various LUTs.

In the following sections, we provide a detailed description of three kinds of basic filtering LUT units with their associated cooperative manners and indexing mechanisms, and introduce the whole framework of LUT-ILF++. First, in Section IV, we introduce the spatial and channel-wise filtering LUTs, and explore the cooperation of multiple LUTs through customized indexing mechanisms to establish a better reference perceived filtering paradigm, and put forward the luma filtering framework of LUT-ILF++. Second, in Section V, we introduce the  cross-component cooperation of multiple LUTs to enable the filtering of different chroma components jointly, and put forward the chroma filtering framework in LUT-ILF++. Third, in Section VI, we introduce the LUT compaction scheme with its training strategy, and examine its role in balancing filtering performance and storage efficiency.

\begin{figure*}
	\centering
	\includegraphics[width=175mm]{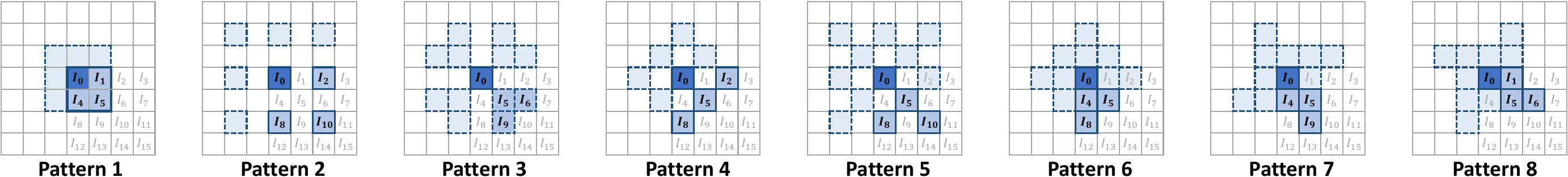}
	\vspace{-0.5em}
	\caption{Illustration of spatial-wise indexing patterns of reference indexing with a large reference range. With the use of pattern 1$\sim$8, LUT-ILF++ can address more reference pixel relationships with 5×5 reference range around $I_0$. The covered reference pixels with the rotation ensemble are marked with dashed boxes.}
	\label{fig:pattern}
\end{figure*}

\begin{figure*}
	\centering
	\vspace{-0.2em}
	\includegraphics[width=179mm]{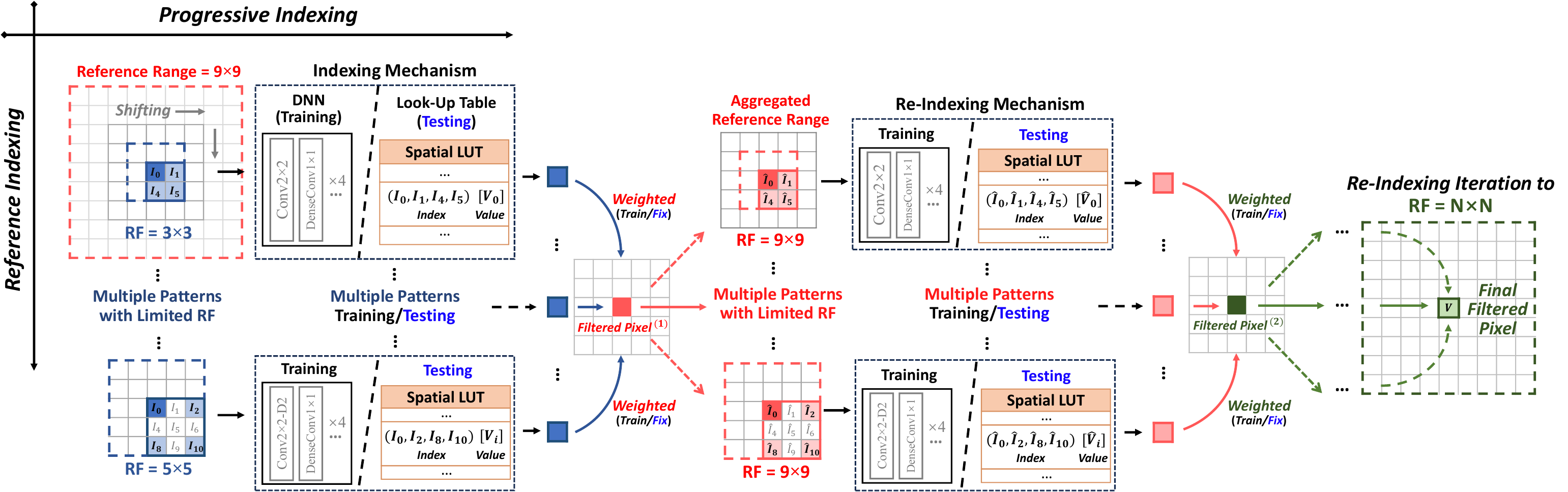}
	\vspace{-0.2em}
	\caption{Illustration of spatial-wise filtering LUTs with reference indexing and progressive indexing mechanisms. The parallel and cascaded spatial-wise neural blocks/LUTs are linked via reference and progressive indexing at the training/testing phase. The covered reference range of each indexing pattern with the rotation ensemble is marked with dashed boxes. The cascaded filtering framework is enabled by iterative re-indexing mechanisms of cascaded spatial-wise LUTs, allowing progressive aggregation of reference information toward an arbitrary target reference range of N×N for a to-be-filter pixel.}
	\label{fig:indexing}
	\vspace{-0.6em}
\end{figure*}

\vspace{-0.3em}
\section{LUT-ILF++ with Cooperation of Multiple LUTs}
In this section, we introduce the cooperative LUT-based filtering scheme that leverages the multiple spatial and channel-wise LUTs to effectively overcome the limitations of filtering reference perception mentioned in Section II.B (1). First, we introduce the design of these basic filtering LUT units with their basic cooperative manners in LUT-ILF++, which enables the LUT-based solution with a large filtering reference range and accurate reference perception in spatial and channel dimensions. Second, we propose the luma filtering framework in LUT-ILF++, and discuss the cooperation of multiple LUTs and the scaling law of the LUT-based ILF framework through a series of spatial and channel-wise modulated explorations.

\vspace{-0.7em}
\subsection{Spatial-wise Filtering LUTs with Reference Indexing}
\vspace{-0.1em}
Local-context-driven filtering has been widely used in the typical loop filters of video coding standards, such as ALF with a diamond filtering template shape and category-specific filters \cite{tsai2013adaptive, meng2021optimized}, and SAO with pre-defined offset classes based on pixel statistics \cite{fu2012sample, kuo2022cross}. In these schemes, different classification strategies or retrievable approaches of surrounding reference pixels are adopted to capture the local reference features and apply various filtering operations. Inspired by these schemes, we propose the first basic cooperative way among the LUT units, spatial-wise LUTs with reference indexing, which introduces the diverse reference indexing to enlarge the reference range of the to-be-filtered pixel by parallelizing more diverse retrievable patterns to address more reference pixels and capture the rich local structures in the spatial dimension.

Based on the standard single pattern with 3$\times$3 reference range of our basic solution (pattern 1 of Fig.~\ref{fig:pattern}\,), we generalize the diverse reference retrievable patterns to support 5$\times$5 reference range. As shown in Fig.~\ref{fig:pattern}, besides the standard pattern 1, the patterns 2$\sim$8 are designed to cover the reference relationships of a 5$\times$5 reference range. For the reference modeling in each pattern, we select representative pixel relationships (filtering template shape) among the to-be-filtered and reference pixels under the constraint of maintaining the invariable LUT dimension. In this way, the total size of cached LUTs grows linearly instead of exponentially, and it can be formulated as,
\begin{equation}
	MS = R \times \left(2^{8 - q} + 1\right)^n \times V \times 8 \ bit,
\end{equation}
where $R$ denotes the number of the used retrievable patterns, $n$ denotes the cached dimension of spatial-wise LUT. In training, as shown in  Fig.~\ref{LUT-ILF++}\,(a) and Fig.~\ref{fig:indexing}, the multiple spatial-wise filtering LUTs with different patterns are reproduced by multiple spatial-wise neural blocks, and these blocks are organized into multiple branches to model the different reference relationships in parallel. Note that the first convolution layer of spatial-wise neural block structure is incorporated with the $reshape$, $unfold$ operations to support the arbitrary indexing patterns with specified coordinates of reference pixels. In inference, as multiple trained spatial-wise blocks are transferred into cached LUTs, the final filtering result is calculated by indexing and weighting their cached filtering results. The filtering process of the whole reference indexing process can be formulated as,
\begin{equation}\fontsize{7.6pt}{6pt}\centering
	\begin{aligned}
		V  = (W_1 \times LUT_{p_1}[I_0][I_1][I_4][I_5]
		+ \text{···} + W_n \times LUT_{p_n}[\text{·}][\text{·}][\text{·}][\text{·}])/n,
	\end{aligned}
\end{equation}
where $V$ denotes the filtered pixel value, $n$ denotes the number of patterns, $LUT[\text{·}]$ denotes the look-up and interpolation operations of LUT retrieval, $P_n$ denotes the pattern ID, $W_n$ denotes the weights of different patterns. For the design of the weighting process, the impact of different reference retrievable patterns on the to-be-filtered pixel is considered. The weights of different patterns are designed as trainable parameters and normalized to the range [0, 1] using the \textit{softmax}() function to reflect the confidence of each reference pattern during training. Once training is completed, these weights are fixed and used by integer operation during inference.

\vspace{-0.5em}
\subsection{Spatial-wise Filtering LUTs with Progressive Indexing}
DNN-based filtering process benefits from the inherent capability of neural networks to progressively modulate the receptive field (RF) through their layer-based architecture \cite{li2023designs}. Compared to hand-crafted schemes, DNN-based schemes can perceive the accurate contextual information with the reference range expansion, enabling the  modeling of complex reference relationships among local structures. Inspired by its efficient RF modulation, we propose the second cooperative way among the LUT units, spatial-wise LUTs with progressive indexing, which introduces the cascaded filtering LUTs to progressively enlarge the reference range of the to-be-filtered pixel and modulate the local structures in the spatial dimension.

As shown in Fig.~\ref{fig:indexing}, building upon the reference indexing pipeline with parallel multiple reference retrievable patterns, we extend the pipeline by adding multiple cascaded LUTs and linking them via the re-indexing mechanisms. The whole cascaded filtering process is supported by iterative look-up operations of cascaded LUTs, enabling progressive aggregation of reference information toward a target N$\times$N reference range for a to-be-filtered pixel. Based on (3), the filtering process of progressive indexing can be formulated as,
\begin{equation}\fontsize{7.6pt}{6pt}
	\begin{aligned}
		V^{(iter)} = (W^{(iter)}_1 \times LUT^{(iter)}_{p_1}[\widehat{I}_0][\widehat{I}_1][\widehat{I}_4][\widehat{I}_5] + 
		W^{(iter)}_2 \times  LUT^{(iter)}_{p_2} \\ [\widehat{I}_0][\widehat{I}_2][\widehat{I}_8][\widehat{I}_{10}]  
		+ \text{···} + W^{(iter)}_n \times LUT^{(iter)}_{p_n}[\widehat{\text{·}}][\widehat{\text{·}}][\widehat{\text{·}}][\widehat{\text{·}}])/n
	\end{aligned}
\end{equation}
where $iter$ denotes the current step number of cascaded iteration, $\widehat{I}$ denotes the output value of previous filtering step that serves as the index of the following filtering LUT. As shown in Fig.~\ref{fig:indexing}, we take the case of $iter=2$ as an example to detail the progressive indexing process. First, in $iter=1$, with the filtering of the to-be-filtered pixel by the reference indexing mechanism, the filtered pixel ensembles the local information of a 5$\times$5 reference range. By shifting the filtering window in a 9$\times$9 reference range, the local information of the 9$\times$9 reference range can be aggregated into a 5$\times$5 aggregated reference range. In $iter=2$, the re-indexing mechanism is used to filter the target pixel based on the aggregated reference pixels to incorporate the larger reference range implicitly. The whole process is similar to cascading multiple convolutional layers in a neural network and achieving information aggregation in the feature domain. In this way, the total size of cached LUTs still grows linearly, which can be formulated as,
\begin{equation}
	MS = P \times R \times \left(2^{8 - q} + 1\right)^n \times V \times 8 \ bit,
\end{equation}
where $P$ denotes the total step number of progressive indexing to the target reference range. In training, as shown in Fig.~\ref{fig:indexing}, each step of progressive indexing is reproduced by the basic multi-branch neural network of reference indexing, and all steps are stacked to form a deep cascaded architecture. Due to the $integer$-$precision$ cache attribute of LUT, the indexing inputs and outputs of LUT are quantized to $integer$ precision, whereas network training uses the $floating$-$point$ gradients. To bridge this gap, we apply the straight-through estimator (STE) strategy\cite{bengio2013estimating} for cascaded architecture training to reproduce the LUT indexing process. During the forward pass, the outputs of the previous progressive indexing step are quantized to $integer$ precision for the follow-up indexing step, while in the backward pass, the gradients are preserved in $floating$-$point$ precision to maintain the end-to-end optimization. In inference, as the deep cascaded architecture is transferred into cached cascaded LUTs, the final filtering result is calculated by step-by-step progressive look-up operations according to Eq. (4).

\begin{figure}
	\centering
	\includegraphics[width=88mm]{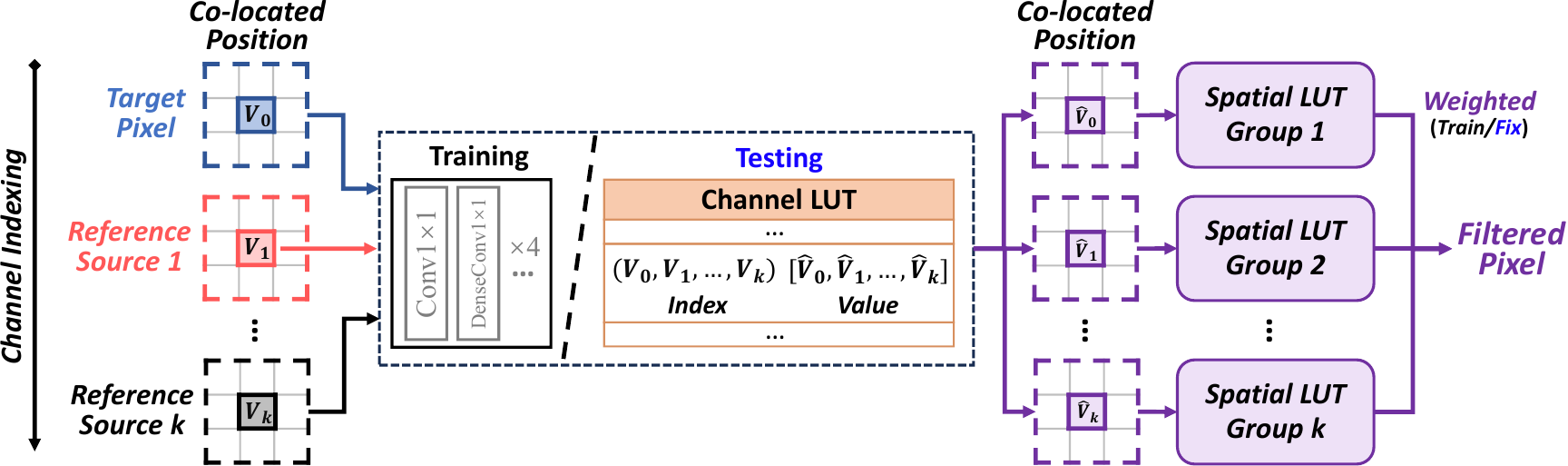}
	\vspace{-1.7em}
	\caption{Illustration of channel-wise LUTs with channel indexing mechanism.}
	\label{fig:channel_indexing}
	\vspace{-1.5em}
\end{figure}

\begin{figure*}
	\centering
	\includegraphics[width=175mm]{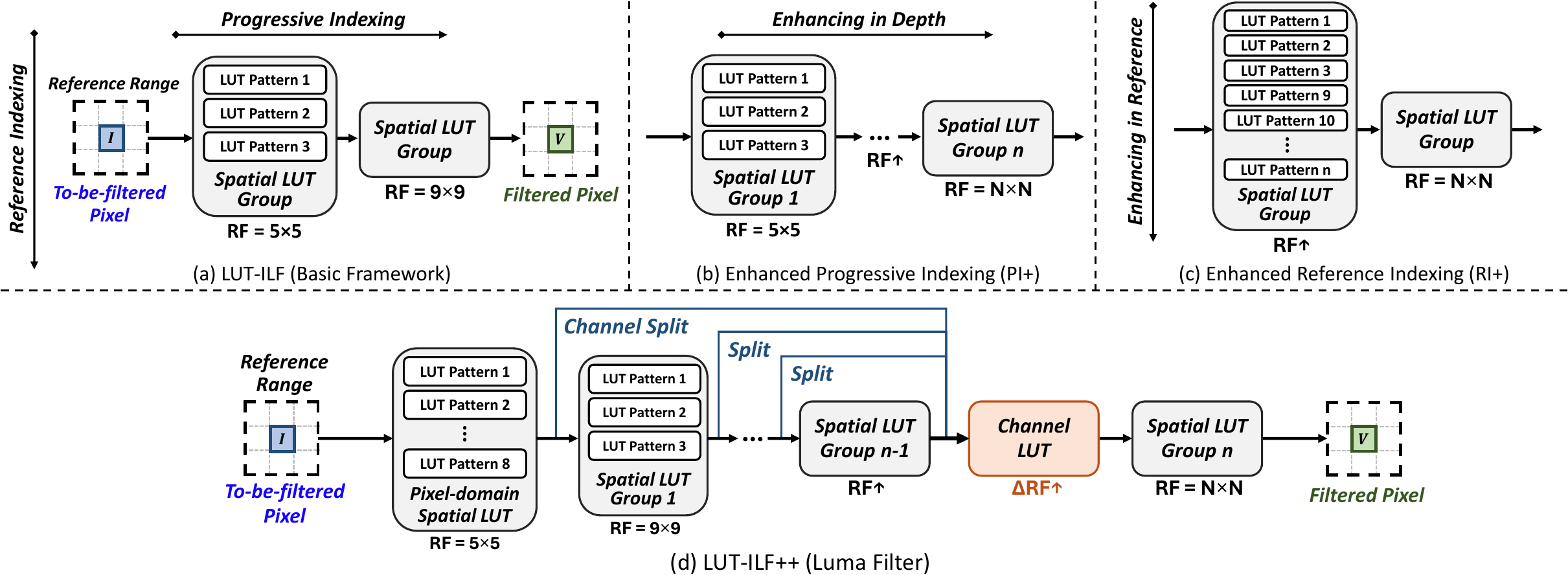}
	\vspace{-0.8em}
	\caption{Overview of the filtering framework of LUT-ILF \cite{li2024loop}, enhanced variants of LUT-ILF, and the luma filtering framework of LUT-ILF++.}
	\label{fig:LUT-ILF++}
	\vspace{-0.8em}
\end{figure*}

\vspace{-1em}
\subsection{Channel-wise Filtering LUTs with Channel Indexing}
\vspace{-0.1em}
Multiple-component-assisted filtering process benefits from its advantage to exploit complementary information of to-be-filtered pixel from different reference sources in coding process, such as cross-component (luma, chroma, etc.) \cite{karczewicz2021vvc, kuo2022cross}, side-information (prediction, residual, etc.), and multi-reference-range guidance \cite{li2023designs}. Compared to the weighted interaction manner of hand-crafted filtering schemes, the flexible high-dimensional channel interaction capability of DNN-based schemes has been verified to provide superior adaptability in modeling the relevance of different reference sources \cite{li2023designs, dong2023temporal}. Inspired by its interaction manner, we propose the third cooperative way among the LUT units, channel-wise LUTs with channel indexing, which introduces the channel-dimensional caching and indexing to ensemble multiple reference sources. 

In contrast to the direct many-to-one mapping relationship of the above spatial-wise LUTs, the correlations among different auxiliary reference sources are more complex than spatial-wise pixel relationships, and the relationship modeling is difficult to represent by a single direct mapping. Therefore, a two-step channel indexing mechanism is introduced to model its complex interactions in the filtering process. As shown in Fig.~\ref{fig:channel_indexing}, inspired by the high-dimensional channel modulation process of DNN models, first, a channel LUT is used to perform many-to-many mappings across the target pixel and different reference sources at co-located spatial positions, achieving channel interaction. The number of input and output dimensions is kept identical to avoid potential bottlenecks caused by insufficient information perception in the channel-wise LUT. Second, each output channel is followed by a spatial-wise LUT group, constructed by basic multi-branch LUTs with reference indexing (patterns 1–3), to refine the individual mixed correlations within each channel. The outputs of all channels are then aggregated through the same weighted mechanism as in reference indexing to produce the final filtered pixel. In this way, the total size of cached LUTs with linear growth can be formulated as,
\begin{equation}\fontsize{9pt}{6pt}
MS = (K \times \left(2^{8 - q} + 1 \right)^K + K \times R \times \left(2^{8 - q} + 1 \right)^n) \times V \times 8 \ bit,
\end{equation}
where $K$ denotes the cached dimension of channel-wise LUT. In training, as shown in Fig.~\ref{fig:channel_indexing}, the $K$-dimensional channel-wise LUT is reproduced by a 1$\times$1 convolution layer with $K$ input and output channels, and the reproduction of spatial-wise LUTs is the same as mentioned in Section IV.A. In inference, the final filtering result is calculated by the cooperation of spatial and channel-wise look-up operations and weighted mechanisms.

\vspace{-0.5em}
\subsection{Cooperation of Multiple LUTs}
In the above subsections, basic spatial and channel-wise reference perceived filtering LUTs with their three basic cooperative ways are proposed, which effectively improve reference perception while ensuring constrained storage consumption growth. In this subsection, based on these basic modules, we discuss the cooperation of multiple LUTs in the LUT-based ILF solution, and put forward the luma filtering framework in LUT-ILF++. The bottleneck of filtering reference perception lies in the exponential exhaustive reference relationship manner of LUTs, which constrains the effective expansion of the reference range required for improved filtering performance, as analyzed in Section II.B (1). Based on the aforementioned series of basic reference perceived modules with linearly constrained storage growth, here we revisit how to perceive a larger reference range in the filtering process, and achieve a good trade-off between filtering gains and complexity with the low storage cost by leveraging cooperative multiple LUTs. 

\begin{table*}
	\renewcommand\arraystretch{1.25}
	\centering
	\scriptsize
	\caption{Comparison Results of LUT-ILF\cite{li2024loop}, Enhanced Variants, and LUT-ILF++ under PSNR metric for Luma Filtering}
	\vspace{-1em}
	\label{tab:cam2}
	\setlength{\tabcolsep}{2.4mm}
	{
\begin{tabular}{c|cl|cc|ccc|cccccc}
	\hline
	\multirow{2}{*}{\textbf{QP}} & \multicolumn{2}{c|}{\textbf{RF = 9 × 9}} & \multicolumn{2}{c|}{\textbf{RF = 13 × 13}}               & \multicolumn{3}{c|}{\textbf{RF = 17 × 17}}                                                                    & \multicolumn{3}{c|}{\textbf{RF = 21 × 21}}                                                                    & \multicolumn{3}{c}{\textbf{RF = 25 × 25}}                                                                    \\ \cline{2-14} 
	& \multicolumn{2}{c|}{\textbf{LUT-ILF\cite{li2024loop}}}       & \multicolumn{1}{c|}{\textbf{RI+}} & \textbf{PI+}           & \multicolumn{1}{c|}{\textbf{RI+}} & \multicolumn{1}{c|}{\textbf{PI+}} & \multicolumn{1}{l|}{\textbf{LUT-ILF++}} & \multicolumn{1}{c|}{\textbf{RI+}} & \multicolumn{1}{c|}{\textbf{PI+}} & \multicolumn{1}{l|}{\textbf{LUT-ILF++}} & \multicolumn{1}{c|}{\textbf{RI+}} & \multicolumn{1}{c|}{\textbf{PI+}} & \multicolumn{1}{l}{\textbf{LUT-ILF++}} \\ \hline
	\textbf{22}                  & \multicolumn{2}{c|}{0.0}          & \multicolumn{1}{c|}{+0.037}  &   +0.040           & \multicolumn{1}{c|}{+0.047}   & \multicolumn{1}{c|}{+0.043}   & \textbf{+0.112}                               & \multicolumn{1}{c|}{+0.045}   & \multicolumn{1}{c|}{+0.050}  & \multicolumn{1}{c|}{\textbf{+0.132} }         & \multicolumn{1}{c|}{+0.033}   & \multicolumn{1}{c|}{+0.057}   & \textbf{+0.117}                              \\ \hline
	\textbf{27}                  & \multicolumn{2}{c|}{0.0}          & \multicolumn{1}{c|}{+0.034}  &   +0.031           & \multicolumn{1}{c|}{+0.042}   & \multicolumn{1}{c|}{+0.036}   & \textbf{+0.093}                              & \multicolumn{1}{c|}{+0.038}   & \multicolumn{1}{c|}{+0.047}  & \multicolumn{1}{c|}{\textbf{+0.098}}         & \multicolumn{1}{c|}{+0.031}   & \multicolumn{1}{c|}{+0.049}   & \textbf{\textbf{+0.087}}                              \\ \hline
	\textbf{32}                  & \multicolumn{2}{c|}{0.0}          & \multicolumn{1}{c|}{+0.058}  &  +0.053            & \multicolumn{1}{c|}{+0.063}   & \multicolumn{1}{c|}{+0.071}   & \textbf{+0.081}                               & \multicolumn{1}{c|}{+0.054}   & \multicolumn{1}{c|}{+0.063}  & \multicolumn{1}{c|}{\textbf{+0.088}}         & \multicolumn{1}{c|}{+0.052}   & \multicolumn{1}{c|}{+0.056}   & \textbf{+0.071}                              \\ \hline
	\textbf{37}                  & \multicolumn{2}{c|}{0.0}          & \multicolumn{1}{c|}{+0.018}  & +0.022             & \multicolumn{1}{c|}{+0.027}   & \multicolumn{1}{c|}{+0.032}   & \textbf{+0.043}                               & \multicolumn{1}{c|}{+0.034}   & \multicolumn{1}{c|}{+0.030}  & \multicolumn{1}{c|}{\textbf{+0.041}}         & \multicolumn{1}{c|}{+0.021}   & \multicolumn{1}{c|}{+0.024}   & \textbf{+0.038}                             \\ \hline
	\textbf{42}                  & \multicolumn{2}{c|}{0.0}          & \multicolumn{1}{c|}{+0.028}  &    +0.033        & \multicolumn{1}{c|}{+0.032}   & \multicolumn{1}{c|}{+0.039}   & \textbf{+0.045}                               & \multicolumn{1}{c|}{+0.037}   & \multicolumn{1}{c|}{+0.040}  & \multicolumn{1}{c|}{\textbf{+0.049}}         & \multicolumn{1}{c|}{+0.030}   & \multicolumn{1}{c|}{+0.038}   & \textbf{+0.043}                             \\ \hline
	\textbf{47}                  & \multicolumn{2}{c|}{0.0}          & \multicolumn{1}{c|}{+0.023}  &   +0.019         & \multicolumn{1}{c|}{+0.023}   & \multicolumn{1}{c|}{+0.027}   & \textbf{+0.034}                               & \multicolumn{1}{c|}{+0.025}   & \multicolumn{1}{c|}{+0.021}  & \multicolumn{1}{c|}{\textbf{+0.029}}         & \multicolumn{1}{c|}{+0.014}   & \multicolumn{1}{c|}{+0.017}   & \textbf{+0.023}                             \\ \hline
\end{tabular}
	}
\label{potential-RF}
\vspace{-1.7em}
\end{table*}

As shown in Fig.~\ref{fig:LUT-ILF++}\, (a), we adopt the basic reference and progressive indexing architectures as the basic framework, which incorporates patterns 1$\sim$3 for each reference indexing (RI) mechanism and a two-step cascaded filtering iteration ($iter$ = 2) for progressive indexing (PI), enabling a 9$\times$9 reference range. The basic framework corresponds to the design of our preliminary framework (LUT-ILF) \cite{li2024loop}. To enlarge the reference range further, following our linear-growth establishment principle, the direct additional expansion of retrievable reference patterns and iteration depth can achieve this goal, as shown in Fig.~\ref{fig:LUT-ILF++}\, (b) and (c). Motivated by these, we conduct a series of modulated explorations to verify the potential of scaling these modules. The explorations are performed on top of VTM-11.0 with all-intra (AI) setting, and the filter is set into the end step of the ILF process without additional rate-distortion optimization.

\begin{figure}
	\centering
	\includegraphics[width=88mm]{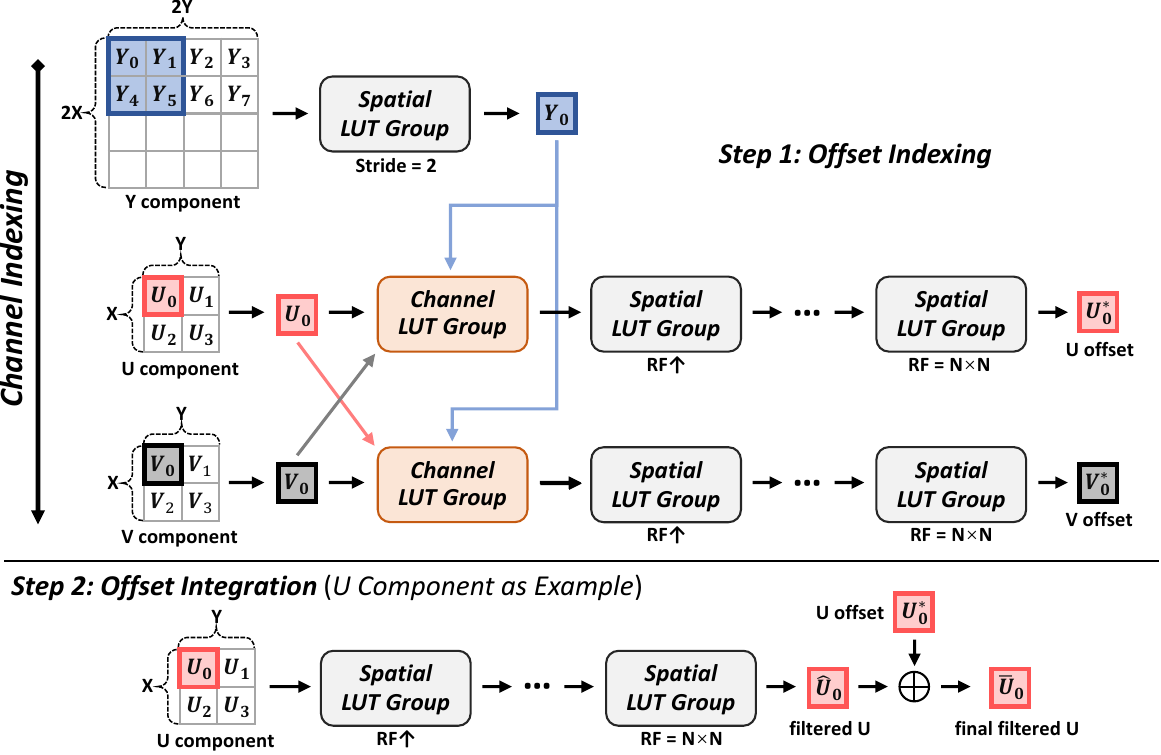}
	\vspace{-1.7em}
	\caption{Overview of the chroma filtering framework of LUT-ILF++.}
	\label{fig:cross_indexing}
	\vspace{-1.5em}
\end{figure}

\begin{figure*}
	\centering
	\includegraphics[width=175mm]{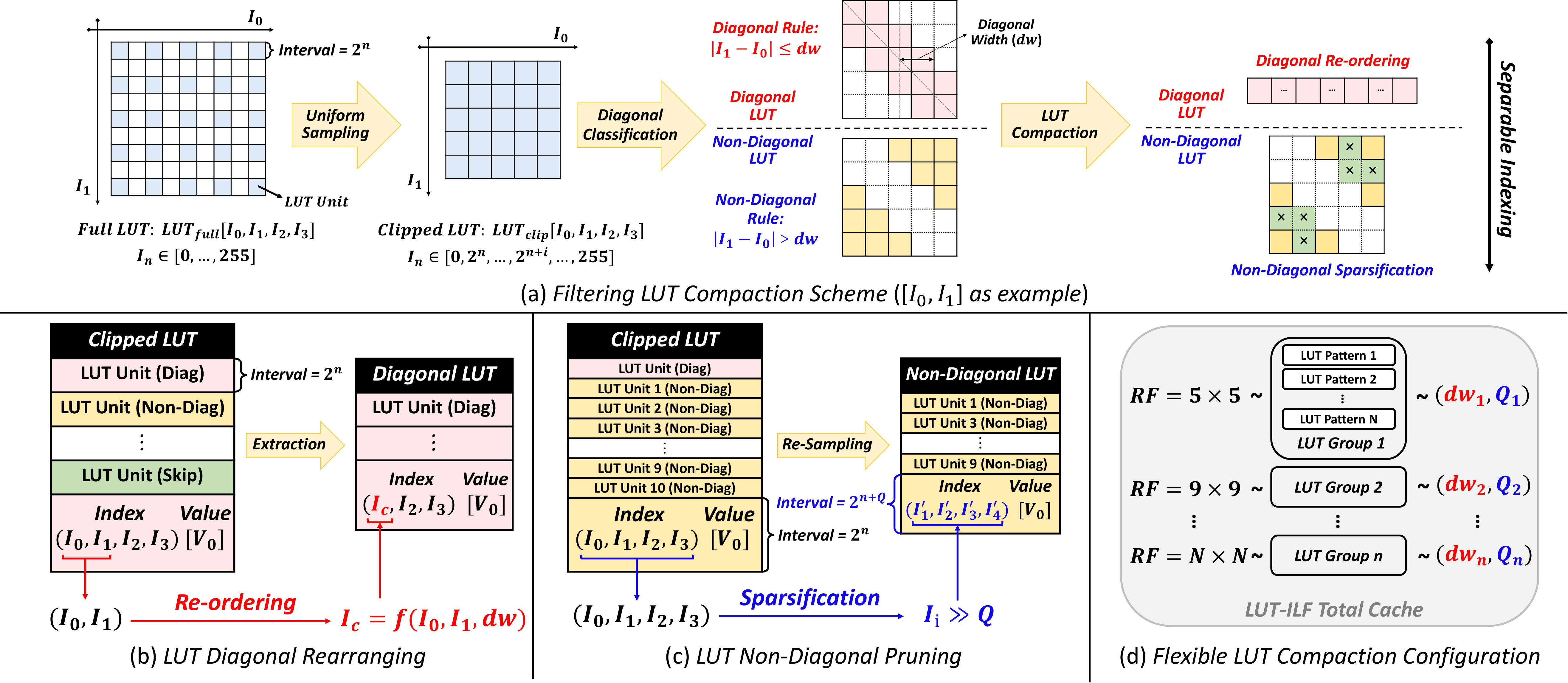}
	\vspace{-0.3em}
	\caption{Overview of LUT compaction scheme in LUT-ILF++.}
	\label{fig:compact-framework}
	\vspace{-0.7em}
\end{figure*} 

As shown in Table~\ref{potential-RF}, the comparison results verify the potential of direct scaling of the basic framework's retrievable patterns (RI+) or cascaded iteration depths (PI+) to a larger target reference range. Although the enhanced variants gain from a larger reference range, the continual expansion of the reference range reaches a bottleneck in accurate reference perception, limiting further noticeable improvements. To ensure the scalability of the cooperative framework, we build upon the basic reference perceived modules and propose a series of scaling strategies that enable its flexible extension. Specifically, as shown in Fig.~\ref{fig:LUT-ILF++}\, (d), based on the scaling architecture of PI+ and RI+ to the target reference range, we further improve the cooperative manner of multiple LUTs from two perspectives.

(1) \textit{Pattern allocation of spatial-wise LUTs with reference indexing.} The number of cached pixel-relationship entries is constrained by the restricted set of retrievable patterns, which makes it inevitable to miss some potentially optimal pixel reference relationships as the reference range increases. To mitigate this issue, more reference retrievable patterns (patterns 1$\sim$8 in Fig.~\ref{fig:pattern}\,) are allocated into the pixel-domain spatial-wise  LUT group to access a broader set of pixel relationships within the proximal reference range, thereby ensuring sufficient perception of the most relevant neighbor reference pixels and maximizing potential optimal dependencies. 

(2) \textit{Channel interaction of spatial-wise LUTs with progressive indexing.} As the reference range expands, the distribution of effective reference relationships becomes increasingly sparse. The growth of ineffective relationships in distant references results in the degradation of useful information density, which hinders the effective learning of reference perception during training. To preserve accurate reference perception as the reference range increases, channel split operation and channel indexing mechanism are used to add channel diversity instead of relying on a purely progressive single-path reference expansion, enabling the additional channel to facilitate the interaction of useful indexing relationships across multiple reference ranges. Meanwhile, the channel interaction also alleviates the non-negligible optimization challenges caused by the integer cache attribute of the LUT-based solution, which helps to avoid the accumulation of errors during the STE strategy-driven training process of the cascaded LUT architecture, thereby improving the convergence stability of neural block reproduction. Note that the channel split operation is performed by doubling the cached value number of each cached indexing entry of the spatial-wise LUT groups. For the neural reproduction of these spatial-wise LUTs, it only requires modifying the output channels of the corresponding convolution layer from 1 to 2. Overall, as shown in Table~\ref{potential-RF}, the proposed scaling strategies in LUT-ILF++ (Luma Filter) improve the scalability of the cooperative filtering framework, enabling effective extension to larger target reference ranges.


\begin{figure*}
	\centering
	\includegraphics[width=177mm]{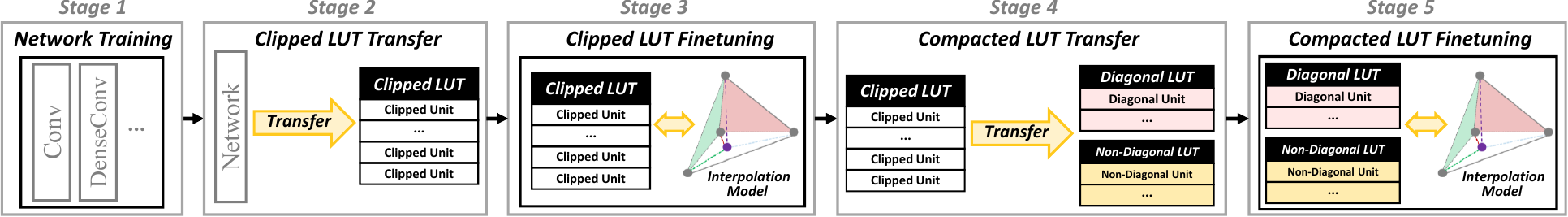}
	\vspace{-0.3em}
	\caption{The training process of compacted LUTs in LUT-ILF++.}
	\vspace{-1.3em}
	\label{fig:LUT-ILF++training}
\end{figure*}

\section{LUT-ILF++ with Cross-Component Cooperation of Multiple LUTs}
In this section, based on the above luma filtering framework of LUT-ILF++, we propose the chroma filtering framework that leverages the interaction ability of channel-wise LUT to effectively overcome the limitation of cross-color-component collaborative filtering mentioned in Section II.B (2). Inspired by the cross-component design of the existing loop filter \cite{karczewicz2021vvc, kuo2022cross}, we propose the cross-component indexing mechanism to exploit the inherent correlations among color components and assist chroma-component filtering, while maintaining low complexity through effective cross-component modeling. 

As shown in Fig.~\ref{fig:cross_indexing}, the whole cross-component cooperative chroma filtering process is designed into two steps, including cross-component offset indexing and integration. First, channel-wise LUTs are used to exploit the cross-component relationships from the proportionally aligned luma and chroma information and modulate the corresponding correction offsets for the filtering of U and V components, respectively. Second, cross-component correction offsets are integrated into the main spatial-wise chroma filtering LUT framework via element-wise addition, yielding the final filtered chroma outputs. In this way, the total cached LUT size for chroma filtering still follows a linearly constrained growth, as it is composed of both spatial and channel-wise LUTs with linear cache design. In training, the neural reproduction of spatial and channel-wise LUTs is the same as mentioned in Section IV.A and Section IV.C. In inference, the final chroma filtering result is generated by the cooperation of spatial and channel-wise look-up operations.

\vspace{-0.6em}
\section{LUT-ILF++ with Cooperation of Compacted LUTs}
In this section, based on the above luma and chroma filtering framework of LUT-ILF++, we propose the LUT compaction scheme that leverages the modular attribute of LUT to overcome the limitation of storage overhead mentioned in Section II.B (3). Based on the bottleneck observation of cached pixel relationships between the to-be-filtered pixel and its reference pixels at MSB precision (Fig.~\ref{fig:pixel_heat}\,), we find that the uniform sampling cache design of LUT indexing entries only intuitively reduces storage cost from the perspective of indexing entry correlation, while ignoring the inherent properties of pixel correlations and their actual access concentration of reference perception for the filtering goal, leading to sub-optimal trade-offs. Inspired by these observations, we introduce the LUT compaction scheme into LUT-ILF++, and propose the LUT pruning strategy with a separable indexing mechanism to achieve accurate storage cost reduction according to the reference perception requirements in practical use. 

\vspace{-0.8em}
\subsection{Compacted Filtering LUTs with Separable Indexing}
As illustrated in Fig.~\ref{fig:pixel_heat}, based on uniform sampling cache design, the diagonal phenomenon of occurrence frequency statistics of accessed pixel value distribution represents the actual access requirements of the LUT-based ILF solution in practice. Based on this phenomenon and the LUT modular attribute, we propose the diagonal compact-oriented LUT pruning strategy, which retains the access-concentrated indexing entries and further eliminates the entry redundancy. As shown in Fig.~\ref{fig:compact-framework}\, (a), we take the compacted process of the first two dimensions (2D, [$I_0$, $I_1$]) of a full LUT (4D, [$I_0$, $I_1$, $I_2$, $I_3$]) as an example to detail the proposed strategy, the detailed calculation process is mentioned in the supplementary material. 

\begin{table*}
	\vspace{-1.3em}
	\renewcommand\arraystretch{1.3}
	\centering
	\fontsize{7.1pt}{9pt}\selectfont
	\caption{BD-Rate and Complexity Results of LUT-ILF++, and Comparison Results with Other In-Loop Filtering Schemes under All Intra (AI) Configuration}
	\label{tab:performance}
	\vspace{-0.8em}
	\setlength{\tabcolsep}{0.7mm}
	\begin{threeparttable}
	{
	{
		\begin{tabular}{cccccccccc}
			\hline
			\multicolumn{10}{c}{\textbf{All Intra (AI) Configuration}}                                                                                                                                                                                                                                                                                                                                                                                                                                                                                                                                                                                                                                                                                                                                                                                                                                                                            \\ \hline
			\multicolumn{1}{c|}{\textbf{Schemes}}                                                                         & \multicolumn{1}{c|}{\textbf{Sequence}} & \multicolumn{1}{c|}{\textbf{\begin{tabular}[c]{@{}c@{}}Y\\ BD-Rate\end{tabular}}} & \multicolumn{1}{c|}{\textbf{\begin{tabular}[c]{@{}c@{}}U\\ BD-Rate\end{tabular}}} & \multicolumn{1}{c|}{\textbf{\begin{tabular}[c]{@{}c@{}}V\\ BD-Rate\end{tabular}}} & \multicolumn{1}{c|}{\textbf{\begin{tabular}[c]{@{}c@{}} Computational \\ Complexity\tnote{3}\end{tabular}}}      & \multicolumn{1}{c|}{\textbf{Storage Cost\tnote{3}}}                                                             & \multicolumn{1}{c|}{\textbf{\begin{tabular}[c]{@{}c@{}} \textbf{Energy Cost\tnote{3}} \\ ($pJ$/pixel)\end{tabular}}}                                                              & \multicolumn{1}{c|}{\textbf{\begin{tabular}[c]{@{}c@{}}Time Complexity\\ (Serial)\end{tabular}}} & \textbf{\begin{tabular}[c]{@{}c@{}}Time Complexity\\ (Parallel)\end{tabular}} \\ \hline
			\multicolumn{1}{c|}{NNVC-VLOP1\tnote{1}\,  (VTM-11.0)}                                                                    & \multicolumn{1}{c|}{Overall}           & \multicolumn{1}{c|}{-3.66\%}                                                      & \multicolumn{1}{c|}{-4.68\%}                                                      & \multicolumn{1}{c|}{-4.54\%}                                                      & \multicolumn{1}{c|}{10.2K Ops/pixel }                                                              & \multicolumn{1}{c|}{
			\begin{tabular}[c]{@{}c@{}}50 KB (int16)\\ 80  KB (float32)\end{tabular}	 }                                                                     & \multicolumn{1}{c|}{\begin{tabular}[c]{@{}c@{}} 3.57K (int16)\\23.46K  (float)\end{tabular}}                                                                          & \multicolumn{1}{c|}{105\%/1951\%}                                                                 & --                                                                   \\ \hline
			\multicolumn{1}{c|}{NNVC-LOP2\tnote{1}\,  (VTM-11.0)}                                                                     & \multicolumn{1}{c|}{Overall}           & \multicolumn{1}{c|}{-4.84\%}                                                      & \multicolumn{1}{c|}{-10.12\%}                                                      & \multicolumn{1}{c|}{-10.31\%}                                                      & \multicolumn{1}{c|}{32.4K Ops/pixel}                                                              & \multicolumn{1}{c|}{\begin{tabular}[c]{@{}c@{}}130 KB (int16)\\ 229 KB (float32)\end{tabular}	}                                                                     & \multicolumn{1}{c|}{\begin{tabular}[c]{@{}c@{}}11.34K (int16)\\74.52K (float)\end{tabular}}                                                                          & \multicolumn{1}{c|}{107\%/3983\%}                                                                 & --                                                                   \\ \hline
			\multicolumn{1}{c|}{NNVC-HOP3\tnote{1}\, (VTM-11.0)}                                                                     & \multicolumn{1}{c|}{Overall}           & \multicolumn{1}{c|}{-9.80\%}                                                      & \multicolumn{1}{c|}{-8.22\%}                                                      & \multicolumn{1}{c|}{-9.79\%}                                                      & \multicolumn{1}{c|}{954.0K Ops/pixel}                                                              & \multicolumn{1}{c|}{
			\begin{tabular}[c]{@{}c@{}}2828 KB (int16)\\ 5595 KB (float32)\end{tabular}	 }                                                                     & \multicolumn{1}{c|}{\begin{tabular}[c]{@{}c@{}}333.90K (int16)\\2194.20K (float)\end{tabular}}                                                                          & \multicolumn{1}{c|}{173\%/55495\%}                                                                 & --                                                                  \\ \hline
			\multicolumn{1}{c|}{LUT-ILF-U\tnote{2}\, (VTM-11.0)}                                                                     & \multicolumn{1}{c|}{Overall}           & \multicolumn{1}{c|}{-0.08\%}                                                      & \multicolumn{1}{c|}{-0.27\%}                                                           & \multicolumn{1}{c|}{-0.11\%}                                                           & \multicolumn{1}{c|}{0.30K Ops/pixel}                                                              & \multicolumn{1}{c|}{492 KB (int8)}                                                                     & \multicolumn{1}{c|}{0.27K}                                                                          & \multicolumn{1}{c|}{103\%/147\%}                                                                 & 102\%/106\%                                                                   \\ \hline
			\multicolumn{1}{c|}{LUT-ILF-V\tnote{2}\, (VTM-11.0)}                                                                     & \multicolumn{1}{c|}{Overall}           & \multicolumn{1}{c|}{-0.32\%}                                                      & \multicolumn{1}{c|}{-0.51\%}                                                           & \multicolumn{1}{c|}{-0.26\%}                                                           & \multicolumn{1}{c|}{0.83K Ops/pixel}                                                              & \multicolumn{1}{c|}{1476 KB (int8)}                                                                     & \multicolumn{1}{c|}{0.75K}                                                                          & \multicolumn{1}{c|}{105\%/198\%}                                                                 & 104\%/111\%                                                                   \\ \hline
			\multicolumn{1}{c|}{LUT-ILF-F\tnote{2}\, (VTM-11.0)}                                                                     & \multicolumn{1}{c|}{Overall}           & \multicolumn{1}{c|}{-0.47\%}                                                      & \multicolumn{1}{c|}{-0.95\%}                                                           & \multicolumn{1}{c|}{-0.64\%}                                                           & \multicolumn{1}{c|}{1.91K Ops/pixel}                                                              & \multicolumn{1}{c|}{3444 KB (int8)}                                                                    & \multicolumn{1}{c|}{1.69K}                                                                        & \multicolumn{1}{c|}{105\%/223\%}                                                                 & 105\%/119\%                                                                   \\ \hline
			\multicolumn{1}{c|}{\multirow{7}{*}{\textbf{\begin{tabular}[c]{@{}c@{}}LUT-ILF++\\ (VTM-11.0)\end{tabular}}}} & \multicolumn{1}{c|}{Class A1}          & \multicolumn{1}{c|}{-0.54\%}                                                      & \multicolumn{1}{c|}{-2.21\%}                                                      & \multicolumn{1}{c|}{-1.80\%}                                                      & \multicolumn{1}{c|}{\multirow{7}{*}{\begin{tabular}[c]{@{}c@{}} 3.59K Ops/pixel (Y)\\0.82K Ops/pixel (U)\\0.82K Ops/pixel (V)\vspace{0.5em}\\ \textbf{5.23K Ops/pixel (Total)}\end{tabular}}} & \multicolumn{1}{c|}{\multirow{7}{*}{\begin{tabular}[c]{@{}c@{}} 330 KB (Y)\\241 KB (U)\\241 KB (V)\vspace{0.5em}\\ \textbf{812 KB (Total)}\end{tabular}}} & \multicolumn{1}{c|}{\multirow{7}{*}{\begin{tabular}[c]{@{}c@{}}3.01K (Y)\\ 0.62K (U)\\ 0.62K (V)\vspace{0.5em}\\ \textbf{4.26K (Total)}\end{tabular}}} & \multicolumn{1}{c|}{110\%/275\%}                                                                 & 107\%/188\%                                                                   \\ \cline{2-5} \cline{9-10} 
			\multicolumn{1}{c|}{}                                                                                         & \multicolumn{1}{c|}{Class A2}          & \multicolumn{1}{c|}{-0.73\%}                                                      & \multicolumn{1}{c|}{-2.59\%}                                                      & \multicolumn{1}{c|}{-1.14\%}                                                      & \multicolumn{1}{c|}{}                                                                                  & \multicolumn{1}{c|}{}                                                                                  & \multicolumn{1}{c|}{}                                                                                  & \multicolumn{1}{c|}{111\%/232\%}                                                                 & 109\%/173\%                                                                   \\ \cline{2-5} \cline{9-10} 
			\multicolumn{1}{c|}{}                                                                                         & \multicolumn{1}{c|}{Class B}           & \multicolumn{1}{c|}{-0.72\%}                                                      & \multicolumn{1}{c|}{-3.11\%}                                                      & \multicolumn{1}{c|}{-1.74\%}                                                      & \multicolumn{1}{c|}{}                                                                                  & \multicolumn{1}{c|}{}                                                                                  & \multicolumn{1}{c|}{}                                                                                  & \multicolumn{1}{c|}{107\%/183\%}                                                                 & 103\%/136\%                                                                   \\ \cline{2-5} \cline{9-10} 
			\multicolumn{1}{c|}{}                                                                                         & \multicolumn{1}{c|}{Class C}           & \multicolumn{1}{c|}{-0.70\%}                                                      & \multicolumn{1}{c|}{-3.90\%}                                                      & \multicolumn{1}{c|}{-1.57\%}                                                      & \multicolumn{1}{c|}{}                                                                                  & \multicolumn{1}{c|}{}                                                                                  & \multicolumn{1}{c|}{}                                                                                  & \multicolumn{1}{c|}{106\%/172\%}                                                                 & 105\%/121\%                                                                   \\ \cline{2-5} \cline{9-10} 
			\multicolumn{1}{c|}{}                                                                                         & \multicolumn{1}{c|}{Class D}           & \multicolumn{1}{c|}{-1.23\%}                                                      & \multicolumn{1}{c|}{-3.40\%}                                                      & \multicolumn{1}{c|}{-2.24\%}                                                      & \multicolumn{1}{c|}{}                                                                                  & \multicolumn{1}{c|}{}                                                                                  & \multicolumn{1}{c|}{}                                                                                  & \multicolumn{1}{c|}{102\%/142\%}                                                                 & 99\%/118\%                                                                   \\ \cline{2-5} \cline{9-10} 
			\multicolumn{1}{c|}{}                                                                                         & \multicolumn{1}{c|}{Class E}           & \multicolumn{1}{c|}{-0.86\%}                                                      & \multicolumn{1}{c|}{-2.03\%}                                                      & \multicolumn{1}{c|}{-1.06\%}                                                      & \multicolumn{1}{c|}{}                                                                                  & \multicolumn{1}{c|}{}                                                                                  & \multicolumn{1}{c|}{}                                                                                  & \multicolumn{1}{c|}{112\%/233\%}                                                                 & 107\%/142\%                                                                   \\ \cline{2-5} \cline{9-10} 
			\multicolumn{1}{c|}{}                                                                                         & \multicolumn{1}{c|}{\textbf{Overall}}  & \multicolumn{1}{c|}{\textbf{-0.82\%}}                                             & \multicolumn{1}{c|}{\textbf{-2.97\%}}                                             & \multicolumn{1}{c|}{\textbf{-1.63\%}}                                             & \multicolumn{1}{c|}{}                                                                                  & \multicolumn{1}{c|}{}                                                                                  & \multicolumn{1}{c|}{}                                                                                  & \multicolumn{1}{c|}{\textbf{108\%/207\%}}                                                        & \textbf{105\%/146\%}                                                          \\ \hline
			
		\end{tabular}
	}
}
\end{threeparttable}
\begin{threeparttable}
\begin{tablenotes}    
\fontsize{6pt}{8pt}\selectfont
\item[1] The results of BD-rate, time complexity, computational complexity, storage cost (int/float model) are cited from \cite{AH0051} (VLOP), \cite{AE0281} (LOP) /\cite{AG0174} (HOP), JVET AHG report \cite{AH0014} and open-sourced repository. 
\item[2] As our preliminary work (LUT-ILF \cite{li2024loop}) does not perform chroma-component filtering, we extend its luma architecture to the chroma and retrain the model for performance and complexity comparison. Due to the integration of chroma-component filtering in LUT-ILF-U/V/F, multi-component joint RDO is adopted for filter decision, and the performance of the Y component of LUT-ILF\cite{li2024loop} is also re-evaluated.
\item[3] The computational complexity, storage cost, and energy cost of both luma and chroma models are jointly evaluated and reported for LUT-ILF++ and the comparative schemes.
\end{tablenotes}
\end{threeparttable}
\vspace{-1.0em}
\end{table*}

Starting from a full LUT transferred from a neural block, uniform sampling is applied to roughly reduce redundancy, resulting in a clipped LUT. Based on the clipped LUT, the clipped indexing entries are further extracted and classified into diagonal and non-diagonal entries by judging the diagonal condition rule: $|I_0 - I_1| \leq dw$, where $dw$ stands for diagonal width. For diagonal indexing entries, as shown in Fig.~\ref{fig:compact-framework}\, (b), the diagonal re-ordering is used to rearrange them as the diagonal LUT by mapping the LUT coordinate ($I_c=f_{mapping}(I_0,I_1,dw)$), and the diagonal LUT is then  stored as a low-dimensional LUT indexed by [$I_c$, $I_2$, $I_3$]. For non-diagonal indexing entries, as shown in Fig.~\ref{fig:compact-framework}\, (c), due to their sparse access concentration in practice, their redundancy is pruned by re-sampling their dimensionality (MSBs) with the allocated sparsification shift ($Q$), and the non-diagonal LUT is then stored with an exponentially reduced storage cost. Note that the flexible LUT compaction configuration is applied to allocate the $dw$, $Q$ values to different spatial-wise LUTs corresponding to different reference ranges, as shown in Fig.~\ref{fig:compact-framework}\, (d), fitting the diagonal distribution phenomenon with varying diagonal width across different ranges observed in Fig.~\ref{fig:pixel_heat}. The LUT compaction scheme can be generalized to multiple dimensions of LUT indexing entries, such as 3D ([$I_0$, $I_1$, $I_2$]), 4D ([$I_0$, $I_1$, $I_2$, $I_3$]). For a clipped LUT, based on Eq. (1), the total size of its compacted version can be formulated as, 
\begin{equation}\fontsize{8.9pt}{9pt}\selectfont
	\begin{aligned}
	LUT_{diag}&: \ MS = D \times (2^{8 - q} + 1)^{n - p} \times V \times 8 \ bit \\
	LUT_{non-diag}&: \ MS = (2^{8 - q - Q} + 1)^{n} \times V \times 8 \ bit
	\end{aligned}
\end{equation}
where $p$ denotes the compacted dimension number of LUT indexing entries, $D$ denotes the number of diagonal entries determined by $dw$ and $p$. 

During inference, the separable indexing mechanism (Fig.~\ref{LUT-ILF++}) is introduced to guide the query of indexing values of the to-be-filtered and reference pixels, where the diagonal condition rule determines which sub-LUT is accessed.

\vspace{-0.7em}
\subsection{Compacted LUTs with Cascaded Training Strategy}
To achieve efficient resource utilization of the whole solution while maintaining coding gains, the cascaded LUT training strategy is proposed and established on the basic LUT training and fine-tuning mentioned in Section II.A (1) and (3), enabling multiple cooperative LUTs to be effectively learned from data. The training step is shown in Fig.~\ref{fig:LUT-ILF++training}, which contains five stages. Due to the potential performance degradation caused by the iterative transfer process of DNNs, clipped and compacted LUTs, the fine-tuning stage with interpolation adaptation is used to bridge these transformations and ensure that the impact on the coding gain remains minimal. In the training stages, after each transformation of the filtering LUT, a fine-tuning stage is cascaded to fine-tune the collaboration between the LUT and the interpolation model, enabling the substitution of the original DNNs for practical use effectively. 

\vspace{-0.2em}
\section{Experiments}
\subsection{Experimental Settings}
In the experiments, the VVC reference software VTM-11.0 is used as the baseline, which is consistent with the anchor version adopted in NNVC\cite{li2023designs}. The codec adopts the configuration of All Intra (AI) and Random Access (RA) according to the VVC Common Test Condition (CTC) \cite{CTCdocument}. The test sequences, including Classes A to E with different resolutions, are tested as specified in \cite{CTCdocument, liu2021jvet}. For each test sequence, quantization parameter (QP) values are set to 22, 27, 32, 37, 42, 47, and Bjontegaard Delta-rate (BD-rate) \cite{2001Calculation} is used as an objective metric to evaluate coding performance. The BVI-DVC\cite{ma2021bvi} and DIV2K\cite{agustsson2017ntire} are used as the training datasets. 

\begin{table*}
	\renewcommand\arraystretch{1.3}
	\centering
	\fontsize{7.1pt}{9pt}\selectfont
	\caption{BD-Rate and Complexity Results of LUT-ILF++, and Comparison Results with Other In-Loop Filtering Schemes under Random Access (RA) Configuration}
	\label{tab:performance2}
	\vspace{-0.9em}
	\begin{threeparttable}
		\setlength{\tabcolsep}{0.7mm}
		{
			{	
				\begin{tabular}{cccccccccc}
					\hline
					\multicolumn{10}{c}{\textbf{Random Access (RA) Configuration}}                                                                                                                                                                                                                                                                                                                                                                                                                                                                                                                                                                                                                                                                                                                                                                                                                                                                            \\ \hline
					\multicolumn{1}{c|}{\textbf{Schemes}}                                                                         & \multicolumn{1}{c|}{\textbf{Sequence}} & \multicolumn{1}{c|}{\textbf{\begin{tabular}[c]{@{}c@{}}Y\\ BD-Rate\end{tabular}}} & \multicolumn{1}{c|}{\textbf{\begin{tabular}[c]{@{}c@{}}U\\ BD-Rate\end{tabular}}} & \multicolumn{1}{c|}{\textbf{\begin{tabular}[c]{@{}c@{}}V\\ BD-Rate\end{tabular}}} & \multicolumn{1}{c|}{\textbf{\begin{tabular}[c]{@{}c@{}}Computational \\ Complexity\tnote{3}\end{tabular}}}      & \multicolumn{1}{c|}{\textbf{\begin{tabular}[c]{@{}c@{}} \textbf{Storage Cost\tnote{3}} \end{tabular}}}                                                             & \multicolumn{1}{c|}{\textbf{\begin{tabular}[c]{@{}c@{}} \textbf{Energy Cost\tnote{3}} \\ ($pJ$/pixel)\end{tabular}}}                                                              & \multicolumn{1}{c|}{\textbf{\begin{tabular}[c]{@{}c@{}}Time Complexity\\ (Serial)\end{tabular}}} & \textbf{\begin{tabular}[c]{@{}c@{}}Time Complexity\\ (Parallel)\end{tabular}} \\ \hline
					\multicolumn{1}{c|}{NNVC-VLOP1\tnote{1}\,  (VTM-11.0)}                                                                    & \multicolumn{1}{c|}{Overall}           & \multicolumn{1}{c|}{-3.69\%}                                                      & \multicolumn{1}{c|}{-4.83\%}                                                      & \multicolumn{1}{c|}{-4.03\%}                                                      & \multicolumn{1}{c|}{10.2K Ops/pixel}                                                              & \multicolumn{1}{c|}{\begin{tabular}[c]{@{}c@{}}50 KB (int16)\\ 80  KB (float)\end{tabular}}                                                                     & \multicolumn{1}{c|}{\begin{tabular}[c]{@{}c@{}} 3.57K (int16)\\23.46K  (float)\end{tabular}}                                                                          & \multicolumn{1}{c|}{107\%/3625\%}                                                                 & --                                                                   \\ \hline
					\multicolumn{1}{c|}{NNVC-LOP2\tnote{1}\,  (VTM-11.0)}                                                                     & \multicolumn{1}{c|}{Overall}           & \multicolumn{1}{c|}{-5.33\%}                                                      & \multicolumn{1}{c|}{-12.26\%}                                                      & \multicolumn{1}{c|}{-11.15\%}                                                      & \multicolumn{1}{c|}{32.4K Ops/pixel}                                                              & \multicolumn{1}{c|}{\begin{tabular}[c]{@{}c@{}}130 KB (int16)\\ 229 KB (float)\end{tabular}}                                                                     & \multicolumn{1}{c|}{\begin{tabular}[c]{@{}c@{}}11.34K (int16)\\74.52K (float)\end{tabular}}                                                                          & \multicolumn{1}{c|}{110\%/7538\%}                                                                 & --                                                                   \\ \hline
					\multicolumn{1}{c|}{NNVC-HOP3\tnote{1}\,  (VTM-11.0)}                                                                     & \multicolumn{1}{c|}{Overall}           & \multicolumn{1}{c|}{-12.39\%}                                                      & \multicolumn{1}{c|}{-13.14\%}                                                      & \multicolumn{1}{c|}{-13.44\%}                                                      & \multicolumn{1}{c|}{954.0K Ops/pixel}                                                              & \multicolumn{1}{c|}{\begin{tabular}[c]{@{}c@{}}2828 KB (int16)\\ 5595 KB (float)\end{tabular}}                                                                     & \multicolumn{1}{c|}{\begin{tabular}[c]{@{}c@{}}333.90K (int16)\\2194.20K (float)\end{tabular}}                                                                          & \multicolumn{1}{c|}{234\%/113729\%}                                                                 & --                                                                   \\ \hline
					\multicolumn{1}{c|}{LUT-ILF-U\tnote{2}\,  (VTM-11.0)}                                                                     & \multicolumn{1}{c|}{Overall}           & \multicolumn{1}{c|}{-0.07\%}                                                      & \multicolumn{1}{c|}{-0.17\%}                                                           & \multicolumn{1}{c|}{-0.08\%}                                                           & \multicolumn{1}{c|}{0.30K Ops/pixel}                                                              & \multicolumn{1}{c|}{492 KB (int8)}                                                                     & \multicolumn{1}{c|}{0.27K}                                                                          & \multicolumn{1}{c|}{102\%/164\%}                                                                 & 103\%/110\%                                                                   \\ \hline
					\multicolumn{1}{c|}{LUT-ILF-V\tnote{2}\,  (VTM-11.0)}                                                                     & \multicolumn{1}{c|}{Overall}           & \multicolumn{1}{c|}{-0.26\%}                                                      & \multicolumn{1}{c|}{-0.47\%}                                                           & \multicolumn{1}{c|}{-0.22\%}                                                           & \multicolumn{1}{c|}{0.83K Ops/pixel}                                                              & \multicolumn{1}{c|}{1476 KB (int8)}                                                                     & \multicolumn{1}{c|}{0.75K}                                                                          & \multicolumn{1}{c|}{105\%/196\%}                                                                 & 104\%/123\%                                                                   \\ \hline
					\multicolumn{1}{c|}{LUT-ILF-F\tnote{2}\,  (VTM-11.0)}                                                                     & \multicolumn{1}{c|}{Overall}           & \multicolumn{1}{c|}{-0.35\%}                                                      & \multicolumn{1}{c|}{-0.65\%}                                                           & \multicolumn{1}{c|}{-0.44\%}                                                           & \multicolumn{1}{c|}{1.91K Ops/pixel}                                                              & \multicolumn{1}{c|}{3444 KB (int8)}                                                                    & \multicolumn{1}{c|}{1.69K}                                                                        & \multicolumn{1}{c|}{106\%/218\%}                                                                 & 106\%/131\%                                                                   \\ \hline
					\multicolumn{1}{c|}{\multirow{7}{*}{\textbf{\begin{tabular}[c]{@{}c@{}}LUT-ILF++\\ (VTM-11.0)\end{tabular}}}} & \multicolumn{1}{c|}{Class A1}          & \multicolumn{1}{c|}{-0.77\%}                                                      & \multicolumn{1}{c|}{-2.91\%}                                                      & \multicolumn{1}{c|}{-1.57\%}                                                      & \multicolumn{1}{c|}{\multirow{7}{*}{\begin{tabular}[c]{@{}c@{}} 3.59K Ops/pixel (Y)\\0.82K Ops/pixel (U)\\0.82K Ops/pixel (V)\vspace{0.5em}\\ \textbf{5.23K Ops/pixel (Total)}\end{tabular}}} & \multicolumn{1}{c|}{\multirow{7}{*}{\begin{tabular}[c]{@{}c@{}} 330 KB (Y)\\241 KB (U)\\241 KB (V)\vspace{0.5em}\\ \textbf{812 KB (Total)}\end{tabular}}} & \multicolumn{1}{c|}{\multirow{7}{*}{\begin{tabular}[c]{@{}c@{}}3.01K (Y)\\ 0.62K (U)\\ 0.62K (V)\vspace{0.5em}\\ \textbf{4.26K (Total)}\end{tabular}}} & \multicolumn{1}{c|}{112\%/244\%}                                                                 & 110\%/184\%                                                                   \\ \cline{2-5} \cline{9-10} 
					\multicolumn{1}{c|}{}                                                                                         & \multicolumn{1}{c|}{Class A2}          & \multicolumn{1}{c|}{-0.86\%}                                                      & \multicolumn{1}{c|}{-4.03\%}                                                      & \multicolumn{1}{c|}{-0.95\%}                                                      & \multicolumn{1}{c|}{}                                                                                  & \multicolumn{1}{c|}{}                                                                                  & \multicolumn{1}{c|}{}                                                                                  & \multicolumn{1}{c|}{109\%/202\%}                                                                 & 108\%/159\%                                                                   \\ \cline{2-5} \cline{9-10} 
					\multicolumn{1}{c|}{}                                                                                         & \multicolumn{1}{c|}{Class B}           & \multicolumn{1}{c|}{-0.75\%}                                                      & \multicolumn{1}{c|}{-4.65\%}                                                      & \multicolumn{1}{c|}{-2.44\%}                                                      & \multicolumn{1}{c|}{}                                                                                  & \multicolumn{1}{c|}{}                                                                                  & \multicolumn{1}{c|}{}                                                                                  & \multicolumn{1}{c|}{110\%/197\%}                                                                 & 108\%/140\%                                                                   \\ \cline{2-5} \cline{9-10} 
					\multicolumn{1}{c|}{}                                                                                         & \multicolumn{1}{c|}{Class C}           & \multicolumn{1}{c|}{-0.54\%}                                                      & \multicolumn{1}{c|}{-5.42\%}                                                      & \multicolumn{1}{c|}{-2.21\%}                                                      & \multicolumn{1}{c|}{}                                                                                  & \multicolumn{1}{c|}{}                                                                                  & \multicolumn{1}{c|}{}                                                                                  & \multicolumn{1}{c|}{107\%/168\%}                                                                 & 106\%/127\%                                                                   \\ \cline{2-5} \cline{9-10} 
					\multicolumn{1}{c|}{}                                                                                         & \multicolumn{1}{c|}{Class D}           & \multicolumn{1}{c|}{-1.18\%}                                                      & \multicolumn{1}{c|}{-4.47\%}                                                      & \multicolumn{1}{c|}{-2.99\%}                                                      & \multicolumn{1}{c|}{}                                                                                  & \multicolumn{1}{c|}{}                                                                                  & \multicolumn{1}{c|}{}                                                                                  & \multicolumn{1}{c|}{105\%/147\%}                                                                 & 103\%/119\%                                                                   \\ \cline{2-5} \cline{9-10} 
					\multicolumn{1}{c|}{}                                                                                         & \multicolumn{1}{c|}{Class E}           & \multicolumn{1}{c|}{-0.95\%}                                                      & \multicolumn{1}{c|}{-2.26\%}                                                      & \multicolumn{1}{c|}{-1.57\%}                                                      & \multicolumn{1}{c|}{}                                                                                  & \multicolumn{1}{c|}{}                                                                                  & \multicolumn{1}{c|}{}                                                                                  & \multicolumn{1}{c|}{111\%/207\%}                                                                 & 107\%/164\%                                                                   \\ \cline{2-5} \cline{9-10} 
					\multicolumn{1}{c|}{}                                                                                         & \multicolumn{1}{c|}{\textbf{Overall}}  & \multicolumn{1}{c|}{\textbf{-0.85\%}}                                             & \multicolumn{1}{c|}{\textbf{-4.11\%}}                                             & \multicolumn{1}{c|}{\textbf{-2.06\%}}                                             & \multicolumn{1}{c|}{}                                                                                  & \multicolumn{1}{c|}{}                                                                                  & \multicolumn{1}{c|}{}                                                                                  & \multicolumn{1}{c|}{\textbf{109\%/194\%}}                                                        & \textbf{107\%/149\%}                                                          \\ \hline
				\end{tabular}
		}}
	\end{threeparttable}
	\begin{threeparttable}
		\begin{tablenotes}    
			\fontsize{6pt}{8pt}\selectfont
			\item Note: The indication of table notes \tnote{1},\tnote{2}, \tnote{3} is the same as in Table~\ref{tab:performance}.
		\end{tablenotes}
	\end{threeparttable}
	\vspace{-1.5em}
\end{table*}

For the complexity evaluated metrics, time complexity, computational complexity, estimated energy cost ($pJ$/pixel \cite{song2021addersr, sze2017efficient, horowitz20141}), and storage cost (KB) are evaluated. For time complexity, the codec time is tested on a CPU model Intel Core i7-11700, and both serial and parallel implementations of LUT-ILF++ are tested. For computational complexity, since in the LUT-based ILF solution, the number of addition and multiplication operations are significantly imbalanced, and in our implementations, multiplication operations are minimized to improve the hardware efficiency and friendliness of LUT-ILF++. Therefore, we report the total number of multiplication and addition operations (Ops) for the comparison of different schemes. In addition, energy cost is used to evaluate the practical complexity of operations according to \cite{song2021addersr, sze2017efficient, horowitz20141}. For a single addition operation, the operation of \textit{int8/int16/int32/float32} corresponds to 0.03/0.05/0.1/0.9 $pJ$. For a single multiplication operation, the operation of \textit{int8/int16/int32/float32} corresponds to 0.2/0.65/3.1/3.7 $pJ$. 

\vspace{-0.7em}
\subsection{The Construction Settings of LUT-ILF++}
\vspace{-0.1em}
\subsubsection{Luma Filter Setting}
As mentioned in Section IV.D, based on the modulated exploration experiments in Table~\ref{potential-RF}, the target reference range 17$\times$17 is selected as the optimal filtering perception and a better  trade-off setting for the luma filter of LUT-ILF++ (Fig.~\ref{fig:LUT-ILF++}\, (d)). The inference architecture is composed of spatial-wise LUT groups, each corresponding to a distinct reference range size (5$\times$5 with Patterns 1$\sim$8, and 9$\times$9, 13$\times$13, 17$\times$17 with Patterns 1$\sim$3), and the channel split and interaction are applied to all LUT groups before the penultimate reference range scale to perform channel interaction. The training architecture is reproduced by standard convolution and dense connected convolution layers.

\subsubsection{Chroma Filter Setting}
As mentioned in Section V, the chroma filtering framework of LUT-ILF++ is adopted for the filtering of each chroma component.  Due to the inherent resolution proportion between luma and chroma components, the target reference range 13$\times$13 is selected as a better trade-off setting. The inference/training architecture is also composed/reproduced in the same manner as the luma. 

\subsubsection{Filter Compaction Setting}
For the filtering frameworks of both luma and chroma filters, the flexible LUT compaction configuration is applied to minimize the storage cost of each spatial-wise LUT of the entire framework according to the diagonal phenomenon observed in Fig.~\ref{fig:pixel_heat}. For luma, based on the uniform sampling cache design of each LUT with 4D dimensions, all dimensions are compacted in a unified manner. Specifically, the $dw$ is set to 2, 2, 3, 3 for each diagonal LUT corresponding to the reference ranges of $5\times5$, $9\times9$, $13\times13$, $17\times17$, respectively, and $Q$ is set to 1 for each non-diagonal LUT. For chroma, the $dw$ and $Q$ are set to 2 and 1.

\subsubsection{Rare-Distortion Optimization of LUT-ILF++}
For the integration of LUT-ILF++ into the filtering process of VVC (DBF, SAO, ALF), we set it at the end of all filtering processes, and the decision flag of LUT-ILF++ is signaled at the coding tree unit (CTU) level to indicate the use of the proposed filter. The flag is determined by the rate-distortion (RD) cost function that $J= SSD +\lambda\times R_{flag}$, where $R_{flag}$ denotes the rate of decision flag in CABAC-based rate estimation, $SSD$ denotes the sum of squared differences (SSD) between the reconstructed and filtering result, and the decision is jointly optimized across the luma and chroma components. 

\subsubsection{Training and Finetuning Strategy}
For the training and finetuning of LUT-ILF++, we adopt the five-step cascaded training strategy mentioned in Section VI.B, with the loss function using MSE. For stages 1 and 3, the learning rate (lr) follows a cosine annealing schedule between 1e-3 and 1e-4. For stage 5, the lr is fixed at 1e-4. For the training iterations, stage 1 adopts 400,000, while stages 3/5 adopt 20,000 each. 

\subsection{Performance}
First, we show the coding performance and other evaluated metrics of LUT-ILF++ integrated into the VVC reference software VTM-11.0. Second, we comprehensively compare LUT-ILF++ with advanced filtering schemes to verify its good trade-offs and practicability. 

\subsubsection{Overall Performance Under Common Test Conditions}
The experimental results are shown in Tables~\ref{tab:performance} and \ref{tab:performance2}. We can see that our proposed LUT-ILF++ can achieve, on average, 0.82\%/0.85\% (Y), 2.97\%/4.11\% (U), 1.63\%/2.06\% (V) BD-rate reduction on VTM-11.0 under AI and RA configurations. Beyond our preliminary framework (LUT-ILF-U, V, F\cite{li2024loop}\,), we observe that the  improved framework, LUT-ILF++, offers a better trade-off point between performance and efficiency, making the LUT-based ILF solution more universal and cost-effective in practice. Specifically, LUT-ILF++ achieves a significant performance gain with only slight increases in computational and time complexity, while enabling lower storage cost and supporting chroma-component filtering.

\subsubsection{Comparisons with Advanced Schemes}
The comparison results with advanced filtering schemes are also shown in Tables~\ref{tab:performance} and \ref{tab:performance2}. We provide a comprehensive comparison with the different complexity configurations of advanced NNLF in NNVC\cite{li2023designs}. For the quantitative comparisons, the computational complexity and decoding time complexity of the LUT-based ILF solution (LUT-ILF-U/V/F, LUT-ILF++) are 33×$\sim$3180× and 18×$\sim$1034× lower than that of NNLF, and the LUT-based schemes also show good performance potential. In addition, compared to the heavy computational process of DNN inference of NNLF, the LUT-based solution achieves superior hardware deployment friendliness due to its storage-and-lookup inference paradigm, which is easily deployable on hardware architectures with fixed-point arithmetic implementations. Although LUT-based schemes generally require necessary storage, in LUT-ILF++, we verify that the modular attributes of LUTs can support flexible storage allocation, which effectively controls the storage cost and further enhances feasibility in practice. 

\vspace{-0.7em}
\subsection{Performance Analysis}
\vspace{-0.1em}
\subsubsection{Ablation Study}
The ablation study is shown in Tables~\ref{tab:ab1}, \ref{tab:ab2}, \ref{tab:ab3}, where a series of variants are introduced to verify the effectiveness of each core module. 

\begin{table}
	\renewcommand\arraystretch{1.2}
	\centering
	\fontsize{6.5pt}{9pt}\selectfont
	\caption{Ablation Study on Luma Filter of LUT-ILF++}
	\label{tab:ab1}
	\vspace{-0.9em}
	\setlength{\tabcolsep}{0.4mm}
	{
		\begin{tabular}{c|c|c|c|c}
			\hline
			\textbf{Varients}                                                  & \textbf{\begin{tabular}[c]{@{}c@{}}Y BD-Rate\\ (AI/RA)\end{tabular}} & \textbf{\begin{tabular}[c]{@{}c@{}}Computational Complexity\\ (Ops/pixel)\end{tabular}} & \textbf{\begin{tabular}[c]{@{}c@{}}Storage Cost\\ (w/ compaction)\end{tabular}} & \textbf{\begin{tabular}[c]{@{}c@{}}Energy Cost\\ ($pJ$/pixel)\end{tabular}} \\ \hline
			\begin{tabular}[c]{@{}c@{}}w/o Pattern\\  Allocation (PA)\end{tabular}  & -0.63\%/-0.55\%                                                          & 2.66K (Y)                                                                                        & 247 KB (Y)                                                                       & 2.22K (Y)                                                               \\ \hline
			\begin{tabular}[c]{@{}c@{}}w/o Channel\\  Interaction (CI)\end{tabular} & -0.29\%/-0.36\%                                                          & 1.96K (Y)                                                                                      &  189 KB (Y)                                                                      & 1.40K (Y)                                                              \\ \hline
			\textbf{LUT-ILF++}                                                   & \textbf{-0.82\%/-0.85\%}                                                 & \textbf{3.59K (Y)}                                                                               & \textbf{330 KB (Y)}                                                              & \textbf{3.01K (Y)}                                                      \\ \hline
		\end{tabular}	
	}
	\vspace{-1em}
\end{table}

For LUT-ILF++ (luma filter), in Table~\ref{tab:cam2}, the impact of reference range and channel interaction (CI) has been validated with PSNR metric, here we further verify the effectiveness of pattern allocation (PA) and CI with BD-rate metric under the fixed target reference range 17$\times$17, as shown in Table~\ref{tab:ab1}. For w/o PA, all spatial LUT groups adopt the same reference indexing patterns (Patterns 1$\sim$3 of Fig.~\ref{fig:pattern}\,). For w/o CI, LUT-ILF++ is degraded to the PI+ (Fig.~\ref{fig:LUT-ILF++}\, (b)) by removing the channel split and indexing mechanism. The comparison results show that both sub-modules lead to significant performance gains with slight increases in computational and storage costs.

For LUT-ILF++ (chroma filter), we verify the effectiveness of offset indexing and integration step under the fixed target reference range 13$\times$13. For w/o offset indexing, the chroma-component filtering adopts the architecture of offset indexing (step 1 of Fig.~\ref{fig:cross_indexing}\,) to directly filter the chroma component instead of the offset reconstruction. For w/o offset integration, the architecture is also degraded to the PI+ (Fig.~\ref{fig:LUT-ILF++}\, (b)). The comparison results show that the offset-driven cross-component indexing mechanism can effectively assist chroma-component filtering with lower complexity.

\begin{table}
	\renewcommand\arraystretch{0.9}
	\centering
	\fontsize{6.2pt}{9pt}\selectfont
	\caption{Ablation Study on Chroma Filter of LUT-ILF++}
	\label{tab:ab2}
	\vspace{-0.9em}
	\setlength{\tabcolsep}{0.5mm}
	{
		\begin{tabular}{c|c|c|c|c|c}
			\hline
			\textbf{Varients}                                                  & \textbf{\begin{tabular}[c]{@{}c@{}}U BD-Rate\\ (AI/RA)\end{tabular}} & \textbf{\begin{tabular}[c]{@{}c@{}}V BD-Rate\\ (AI/RA)\end{tabular}} & \textbf{\begin{tabular}[c]{@{}c@{}}Computational\\ Complexity\\ (Ops/pixel)\end{tabular}} & \textbf{\begin{tabular}[c]{@{}c@{}}Storage Cost\\ (w/ compaction)\end{tabular}} & \textbf{\begin{tabular}[c]{@{}c@{}}Energy  Cost\\ ($pJ$/pixel)\end{tabular}} \\ \hline
			\begin{tabular}[c]{@{}c@{}}w/o \\ Offset \\ Indexing\end{tabular}     & -2.13\%/-2.81\%                                                          & -0.97\%/-1.06\%                                                          & 0.56K (U/V)                                                                                        & 166 KB (U/V)                                                                         & 0.44K                                                                  \\ \hline
			\begin{tabular}[c]{@{}c@{}}w/o \\ Offset \\ Integration\end{tabular} & -0.77\%/-1.23\%                                                          & -0.59\%/-0.67\%                                                          & 0.21K (U/V)                                                                                       & 76 KB (U/V)                                                                         & 0.19K                                                                  \\ \hline
			\textbf{LUT-ILF++}                                                   & \textbf{-2.97\%/-4.11\%}                                                 & \textbf{-1.63\%/-2.06\%}                                                 & \textbf{0.82K (U/V)}                                                                               & \textbf{241  KB (U/V)}                                                                & \textbf{0.62K (U/V)}                                                         \\ \hline
		\end{tabular}
	}
	\vspace{-1.5em}
\end{table}

\begin{table}
	\renewcommand\arraystretch{1.2}
	\centering
	\fontsize{6.5pt}{9pt}\selectfont
	\caption{Ablation Study on Filter Compaction in LUT-ILF++ }
	\label{tab:ab3}
	\vspace{-1.0em}
	\setlength{\tabcolsep}{0.7mm}
	{
		\begin{tabular}{c|c|c|c|c|c}
			\hline
			\textbf{\begin{tabular}[c]{@{}c@{}} Setting\end{tabular}} & \textbf{\begin{tabular}[c]{@{}c@{}}Compaction Dimension\end{tabular}} & \textbf{$dw$}       & \textbf{$Q$} & \textbf{\begin{tabular}[c]{@{}c@{}}Storage Cost\end{tabular}}  & \textbf{\begin{tabular}[c]{@{}c@{}}Y BD-Rate\\ (AI/RA)\end{tabular}} \\ \hline
			DNN                                                        & --                                                                       & --                 & --                                                                                                                                            & --                                                                & -0.87\%/-0.93\%                                                          \\ \hline
			Clipped                                                        & --                                                                       & --                 & --                                                                                                                                            & 2785 KB (Y)                                                                  & -0.83\%/-0.88\%                                                          \\ \hline
			\multirow{7}{*}{\begin{tabular}[c]{@{}c@{}}Compaction\end{tabular}}                                    & 2D ([$I_0$, $I_1$])                                                                      & 2                 & 1                                                                                                                                             & 996 KB (Y)                                                                 & -0.75\%/-0.79\%                                                          \\ \cline{2-6} 
			& 3D ([$I_0$, $I_1$, $I_2$])                                                                     & 2                 & 1                                                                                                                                          & 449 KB (Y)                                                                  & -0.70\%/-0.69\%                                                          \\ \cline{2-6} 
			& 4D ([$I_0$, $I_1$, $I_2$, $I_3$])                                                                       & 2                 & 1                                                                                                                                           & 298 KB (Y)                                                                  & -0.71\%/-0.75\%                                                          \\ \cline{2-6} 
			& 4D ([$I_0$, $I_1$, $I_2$, $I_3$])                                                                      & 2                 & 2                                                                                                                                           & 101 KB (Y)                                                                  & -0.66\%/-0.62\%                                                          \\ \cline{2-6} 
			& 4D ([$I_0$, $I_1$, $I_2$, $I_3$])                                                                      & 3                 & 1                                                                                                                                            & 390 KB (Y)                                                                  & -0.85\%/-0.90\%                                                          \\ \cline{2-6} 
			& 4D ([$I_0$, $I_1$, $I_2$, $I_3$])                                                                      & 3                 & 2                                                                                                                                             & 193 KB (Y)                                                                  & -0.75\%/-0.82\%                                                          \\ \cline{2-6} 
			& 4D ([$I_0$, $I_1$, $I_2$, $I_3$])                                                                      & 4                & 1                                                                                                                                          & 535 KB (Y)                                                                  & -0.89\%/-0.87\%                                                          \\ \hline
			\begin{tabular}[c]{@{}c@{}}\textbf{LUT-ILF++}\end{tabular}                                            & \textbf{4D ([$I_0$, $I_1$, $I_2$, $I_3$]) }                                                             & \textbf{2$\sim$3 (flexible)} & \textbf{1}                                                                                                                         & \textbf{330 KB (Y)}                                                         & \textbf{-0.82\%/-0.85\%}                                                 \\ \hline
		\end{tabular}
	}
	\vspace{2.0em}
	
	\renewcommand\arraystretch{1.2}
	\centering
	\fontsize{6.5pt}{9pt}\selectfont
	\caption{BD-rate Results of LUT-ILF++ at Low-bitrate Points}
	\label{tab:le}
	\vspace{-0.9em}
	\setlength{\tabcolsep}{1.9mm}
	{
		\begin{tabular}{c|c|c|c|c}
			\hline
			\textbf{Schemes}   & \textbf{\begin{tabular}[c]{@{}c@{}}Y BD-Rate\\ (AI/RA)\end{tabular}} & \textbf{\begin{tabular}[c]{@{}c@{}}Computational\\ Complexity\\ (Ops/pixel)\end{tabular}} & \textbf{\begin{tabular}[c]{@{}c@{}}Storage Cost\end{tabular}} & \textbf{\begin{tabular}[c]{@{}c@{}}Energy  Cost\\ ($pJ$/pixel)\end{tabular}} \\ \hline
			LUT-ILF-V          & -0.74\%/-0.39\%                                                          & 0.55K (Y)                                                                                        & 492 KB (Y)                                                                         & 0.49K (Y)                                                                 \\ \hline
			LUT-ILF-F          & -1.03\%/-0.57\%                                                          & 1.27K (Y)                                                                                        & 1148 KB (Y)                                                                         & 1.12K (Y)                                                                 \\ \hline
			\textbf{LUT-ILF++} & \textbf{-1.49\%/-1.21\%}                                                 & \textbf{3.59K (Y)}                                                                               & \textbf{\begin{tabular}[c]{@{}c@{}}330 KB (Y)\end{tabular}}                                                                & \textbf{3.01K (Y)}                                                         \\ \hline
		\end{tabular}
	}
	\vspace{-1.7em}
\end{table}

For the LUT compaction scheme, based on the diagonal cache statistic observation (Fig.~\ref{fig:pixel_heat}\,), we analyze the impact of different compaction elements, including compacted dimension, diagonal width ($dw$), and sparsification shift ($Q$). As shown in Fig.~\ref{tab:ab3}, starting from the full LUT, the variants of regular clipped LUTs and various compacted LUTs with different settings are introduced to validate the effectiveness of the LUT pruning strategy with a separable indexing mechanism in balancing performance and efficiency. The results verify that the proposed pruning strategy effectively aligns with actual cache access requirements to optimize storage cost while preserving high performance. Furthermore, the flexible  compaction configuration achieves a more fine-grained storage optimization tailored to different reference ranges.

\subsubsection{Low-Bitrate Points Exploration}
To further explore the potential of LUT-ILF++, we evaluate it at low bitrate points (QP 27$\sim$47) and compare it with the preliminary framework, as shown in Table~\ref{tab:le}. The results verify its strong potential.

\vspace{-0.2em}
\section{Conclusion}
In this paper, we rethink the practical deployment problem of emerging DNN-based coding tools in video coding.  In our solution, we propose an efficient look-up table-based approach that does not rely on high-performance computing resources, and attempt to apply it to a typical in-loop filtering module. To achieve a favorable trade-off between performance and complexity for practical use, a series of LUT-oriented filtering designs are proposed to realize multiple filtering functionalities, including reference perception, cross-component filtering, etc, enabling a universal LUT-based filtering framework. The experimental results of the proposed LUT-ILF++ demonstrate that it can effectively achieve a good filtering goal while maintaining lower complexity in advanced VVC, providing a new practical way for neural network-based video coding tools. For future work, we plan to explore two directions: (1) we will further explore a unified filtering framework improved by meta-information (partition, motion vector, etc), aiming to integrate more auxiliary cues for improved filtering; (2) we aim to generalize the proposed solution to other coding tools, including fractional-pixel motion estimation\cite{yan2018convolutional}, reference picture resampling\cite{fu2022efficient}, etc.

\bibliographystyle{IEEEtran}
\bibliography{IEEEexample_standard}

\begin{thebibliography}{10}
\providecommand{\url}[1]{#1}
\csname url@samestyle\endcsname
\providecommand{\newblock}{\relax}
\providecommand{\bibinfo}[2]{#2}
\providecommand{\BIBentrySTDinterwordspacing}{\spaceskip=0pt\relax}
\providecommand{\BIBentryALTinterwordstretchfactor}{4}
\providecommand{\BIBentryALTinterwordspacing}{\spaceskip=\fontdimen2\font plus
\BIBentryALTinterwordstretchfactor\fontdimen3\font minus
  \fontdimen4\font\relax}
\providecommand{\BIBforeignlanguage}[2]{{%
\expandafter\ifx\csname l@#1\endcsname\relax
\typeout{** WARNING: IEEEtran.bst: No hyphenation pattern has been}%
\typeout{** loaded for the language `#1'. Using the pattern for}%
\typeout{** the default language instead.}%
\else
\language=\csname l@#1\endcsname
\fi
#2}}
\providecommand{\BIBdecl}{\relax}
\BIBdecl

\bibitem{sullivan2012overview}
G.~J. Sullivan, J.-R. Ohm, W.-J. Han, and T.~Wiegand, ``Overview of the high
  efficiency video coding ({HEVC}) standard,'' \emph{IEEE Transactions on
  Circuits and Systems for Video Technology}, vol.~22, no.~12, pp. 1649--1668,
  2012.

\bibitem{bross2021overview}
B.~Bross, Y.-K. Wang, Y.~Ye, S.~Liu, J.~Chen, G.~J. Sullivan, and J.-R. Ohm,
  ``Overview of the versatile video coding {(VVC)} standard and its
  applications,'' \emph{IEEE Transactions on Circuits and Systems for Video
  Technology}, vol.~31, no.~10, pp. 3736--3764, 2021.

\bibitem{karczewicz2021vvc}
M.~Karczewicz, N.~Hu, J.~Taquet, C.-Y. Chen, K.~Misra, K.~Andersson, P.~Yin,
  T.~Lu, E.~Fran{\c{c}}ois, and J.~Chen, ``{VVC} in-loop filters,'' \emph{IEEE
  Transactions on Circuits and Systems for Video Technology}, vol.~31, no.~10,
  pp. 3907--3925, 2021.

\bibitem{han2021technical}
J.~Han, B.~Li, D.~Mukherjee, C.-H. Chiang, A.~Grange, C.~Chen, H.~Su,
  S.~Parker, S.~Deng, U.~Joshi \emph{et~al.}, ``A technical overview of
  {AV1},'' \emph{Proceedings of the IEEE}, vol. 109, no.~9, pp. 1435--1462,
  2021.

\bibitem{AOM}
AOM, ``{{AOMedia} {Video} 1, 2 ({AV1}, {AV2})},''
  \emph{\url{https://aomedia.org/}}, 2015.

\bibitem{list2003adaptive}
P.~List, A.~Joch, J.~Lainema, G.~Bjontegaard, and M.~Karczewicz, ``Adaptive
  deblocking filter,'' \emph{IEEE Transactions on Circuits and Systems for
  Video Technology}, vol.~13, no.~7, pp. 614--619, 2003.

\bibitem{fu2012sample}
C.-M. Fu, E.~Alshina, A.~Alshin, Y.-W. Huang, C.-Y. Chen, C.-Y. Tsai, C.-W.
  Hsu, S.-M. Lei, J.-H. Park, and W.-J. Han, ``Sample adaptive offset in the
  {HEVC} standard,'' \emph{IEEE Transactions on Circuits and Systems for Video
  technology}, vol.~22, no.~12, pp. 1755--1764, 2012.

\bibitem{kuo2022cross}
C.-W. Kuo, X.~Xiu, Y.-W. Chen, H.-J. Jhu, W.~Chen, N.~Yan, and X.~Wang,
  ``Cross-component sample adaptive offset,'' in \emph{2022 Data Compression
  Conference (DCC)}.\hskip 1em plus 0.5em minus 0.4em\relax IEEE, 2022, pp.
  359--368.

\bibitem{tsai2013adaptive}
C.-Y. Tsai, C.-Y. Chen, T.~Yamakage, I.~S. Chong, Y.-W. Huang, C.-M. Fu,
  T.~Itoh, T.~Watanabe, T.~Chujoh, M.~Karczewicz \emph{et~al.}, ``Adaptive loop
  filtering for video coding,'' \emph{IEEE Journal of Selected Topics in Signal
  Processing}, vol.~7, no.~6, pp. 934--945, 2013.

\bibitem{meng2021optimized}
X.~Meng, J.~Zhang, C.~Jia, X.~Zhang, S.~Wang, and S.~Ma, ``Optimized adaptive
  loop filter in versatile video coding,'' in \emph{2021 Data Compression
  Conference (DCC)}.\hskip 1em plus 0.5em minus 0.4em\relax IEEE, 2021, pp.
  359--359.

\bibitem{dai2017convolutional}
Y.~Dai, D.~Liu, and F.~Wu, ``A convolutional neural network approach for
  post-processing in {HEVC} intra coding,'' in \emph{MultiMedia Modeling: 23rd
  International Conference, MMM 2017, Reykjavik, Iceland, January 4-6, 2017,
  Proceedings, Part I 23}.\hskip 1em plus 0.5em minus 0.4em\relax Springer,
  2017, pp. 28--39.

\bibitem{dai2018cnn}
Y.~Dai, D.~Liu, Z.-J. Zha, and F.~Wu, ``A {CNN}-based in-loop filter with {CU}
  classification for {HEVC},'' in \emph{2018 IEEE Visual Communications and
  Image Processing (VCIP)}.\hskip 1em plus 0.5em minus 0.4em\relax IEEE, 2018,
  pp. 1--4.

\bibitem{li2021convolutional}
Y.~Li, L.~Zhang, and K.~Zhang, ``Convolutional neural network based in-loop
  filter for {VVC} intra coding,'' in \emph{2021 IEEE International Conference
  on Image Processing (ICIP)}.\hskip 1em plus 0.5em minus 0.4em\relax IEEE,
  2021, pp. 2104--2108.

\bibitem{li2023idam}
------, ``{iDAM}: Iteratively trained deep in-loop filter with adaptive model
  selection,'' \emph{ACM Transactions on Multimedia Computing, Communications
  and Applications}, vol.~19, no.~1s, pp. 1--22, 2023.

\bibitem{man2025content}
H.~Man, H.~Wang, R.~Lu, Z.~Wan, X.~Fan, and D.~Zhao, ``Content-aware dynamic
  in-loop filter with adjustable complexity for {VVC} intra coding,''
  \emph{IEEE Transactions on Circuits and Systems for Video Technology}, 2025.

\bibitem{jia2019content}
C.~Jia, S.~Wang, X.~Zhang, S.~Wang, J.~Liu, S.~Pu, and S.~Ma, ``Content-aware
  convolutional neural network for in-loop filtering in high efficiency video
  coding,'' \emph{IEEE Transactions on Image Processing}, vol.~28, no.~7, pp.
  3343--3356, 2019.

\bibitem{wang2021combining}
D.~Wang, S.~Xia, W.~Yang, and J.~Liu, ``Combining progressive rethinking and
  collaborative learning: A deep framework for in-loop filtering,'' \emph{IEEE
  Transactions on Image Processing}, vol.~30, pp. 4198--4211, 2021.

\bibitem{kathariya2023joint}
B.~Kathariya, Z.~Li, and G.~Van~der Auwera, ``Joint pixel and frequency feature
  learning and fusion via channel-wise transformer for high-efficiency learned
  in-loop filter in {VVC},'' \emph{IEEE Transactions on Circuits and Systems
  for Video Technology}, vol.~34, no.~5, pp. 4070--4083, 2023.

\bibitem{T0088}
``Convolutional neural networks-based in-loop filter,'' \emph{JVET-T0088},
  2020.

\bibitem{AB0068}
``{{EE1-1.6}: {RDO} considering deep in-loop filtering},'' \emph{JVET-AB0068},
  2022.

\bibitem{AC0177}
``Deep in-loop filter with additional input information,'' \emph{JVET-AC0177},
  2022.

\bibitem{li2021neural}
Y.~Li, Y.~Yi, D.~Liu, L.~Li, Z.~Li, and H.~Li, ``Neural-network-based
  cross-channel intra prediction,'' \emph{ACM Transactions on Multimedia
  Computing, Communications, and Applications}, vol.~17, no.~3, pp. 1--23,
  2021.

\bibitem{li2024object}
Z.~Li, Z.~Yuan, L.~Li, D.~Liu, X.~Tang, and F.~Wu, ``Object
  segmentation-assisted inter prediction for versatile video coding,''
  \emph{IEEE Transactions on Broadcasting}, 2024.

\bibitem{feng2024efficient}
X.~Feng, L.~Li, D.~Liu, and F.~Wu, ``Efficient partition map prediction via
  token sparsification for fast {VVC} intra coding,'' in \emph{2024 IEEE 26th
  International Workshop on Multimedia Signal Processing (MMSP)}.\hskip 1em
  plus 0.5em minus 0.4em\relax IEEE, 2024, pp. 1--6.

\bibitem{li2024ustc}
Z.~Li, J.~Liao, C.~Tang, H.~Zhang, Y.~Li, Y.~Bian, X.~Sheng, X.~Feng, Y.~Li,
  C.~Gao \emph{et~al.}, ``{USTC-TD}: A test dataset and benchmark for image and
  video coding in 2020s,'' \emph{IEEE Transactions on Multimedia}, 2025.

\bibitem{feng2025partition}
X.~Feng, Z.~Li, L.~Li, D.~Liu, and F.~Wu, ``Partition map-based fast block
  partitioning for {VVC} inter coding,'' \emph{arXiv preprint
  arXiv:2504.18398}, 2025.

\bibitem{jia2023deep}
J.~Jia, Y.~Zhang, H.~Zhu, Z.~Chen, Z.~Liu, X.~Xu, and S.~Liu, ``Deep reference
  frame generation method for {VVC} inter prediction enhancement,'' \emph{IEEE
  Transactions on Circuits and Systems for Video Technology}, vol.~34, no.~5,
  pp. 3111--3124, 2023.

\bibitem{li2023designs}
Y.~Li, J.~Li, C.~Lin, K.~Zhang, L.~Zhang, F.~Galpin, T.~Dumas, H.~Wang,
  M.~Coban, J.~Str{\"o}m \emph{et~al.}, ``Designs and implementations in neural
  network-based video coding,'' \emph{arXiv preprint arXiv:2309.05846}, 2023.

\bibitem{li2017convolutional}
Y.~Li, D.~Liu, H.~Li, L.~Li, F.~Wu, H.~Zhang, and H.~Yang, ``Convolutional
  neural network-based block up-sampling for intra frame coding,'' \emph{IEEE
  Transactions on Circuits and Systems for Video Technology}, vol.~28, no.~9,
  pp. 2316--2330, 2017.

\bibitem{lin2025low}
C.~Lin, Y.~Li, J.~Li, K.~Zhang, and L.~Zhang, ``Low complexity super resolution
  for resampling-based video coding,'' in \emph{2025 IEEE International
  Conference on Acoustics, Speech and Signal Processing (ICASSP)}.\hskip 1em
  plus 0.5em minus 0.4em\relax IEEE, 2025, pp. 1--5.

\bibitem{liu2020deep}
D.~Liu, Y.~Li, J.~Lin, H.~Li, and F.~Wu, ``Deep learning-based video coding: A
  review and a case study,'' \emph{ACM Computing Surveys (CSUR)}, vol.~53,
  no.~1, pp. 1--35, 2020.

\bibitem{AF0181}
``Simplified feature extraction for hop,'' \emph{JVET-AF0181}, 2023.

\bibitem{AF0206}
``{Complexity reduction of NN-based loop-filters},'' \emph{JVET-AF0206}, 2023.

\bibitem{AH0050}
``{EE1-1.1: LOP candidate with Low Complexity TB and new BB structure},''
  \emph{JVET-AH0050}, 2024.

\bibitem{AJ0066}
``{EE1-1.4: Reduced complexity input feature extraction for LOP and VLOP},''
  \emph{JVET-AJ0066}, 2024.

\bibitem{AG0174}
``{EE1-1.1: Report on training with HOP architecture change for EE1-0 (variant
  1)},'' \emph{JVET-AG0174}, 2024.

\bibitem{huang2021adaptive}
Z.~Huang, J.~Sun, X.~Guo, and M.~Shang, ``Adaptive deep reinforcement
  learning-based in-loop filter for {VVC},'' \emph{IEEE Transactions on Image
  Processing}, vol.~30, pp. 5439--5451, 2021.

\bibitem{AH0195}
``{EE1-related: Complexity reduction of NN in-loop filters through early
  cropping},'' \emph{JVET-AH0195}, 2024.

\bibitem{AJ0054}
``{EE1-1.3: Neural network-based in-loop filters using early cropping},''
  \emph{JVET-AJ0054}, 2024.

\bibitem{li2023lightweight}
M.~Li and W.~Ji, ``Lightweight multiattention recursive residual {CNN}-based
  in-loop filter driven by neuron diversity,'' \emph{IEEE Transactions on
  Circuits and Systems for Video Technology}, vol.~33, no.~11, pp. 6996--7008,
  2023.

\bibitem{AG0155}
``{EE1-Related: On Low Complexity Operational Point for In-Loop Filtering},''
  \emph{JVET-AG0155}, 2024.

\bibitem{AH0077}
``{AHG11: A Low-Complexity Neural Network Loop Filter based on Partial
  Convolution and Over-Parameterization},'' \emph{JVET-AH0077}, 2024.

\bibitem{AD0211}
``{EE1-Related: Combination test of EE1-1.3.5 and multi-scale component of
  EE1-1.6},'' \emph{JVET-AD0211}, 2023.

\bibitem{AG0069}
``{AhG11: LOP with inputs transformed},'' \emph{JVET-AG0069}, 2024.

\bibitem{AH0207}
``{EE1 related: LOP input adjustment with trainable input transform},''
  \emph{JVET-AH0207}, 2024.

\bibitem{AI0173}
``{LOP input adjustment with trainable components},'' \emph{JVET-AI0173}, 2024.

\bibitem{liu2024nn}
D.~Liu, J.~Str{\"o}m, M.~Damghanian, and P.~Wennersten, ``{NN}-based in-loop
  filtering with inputs transformed,'' in \emph{IEEE International Conference
  on Image Processing (ICIP)}.\hskip 1em plus 0.5em minus 0.4em\relax IEEE,
  2024, pp. 3737--3743.

\bibitem{AD0157}
``{EE1-related: Neural-network loop filters in EE1-1.1.2 with further
  complexity reduction},'' \emph{JVET-AD0157}, 2023.

\bibitem{AH0188}
``{EE1-2.1: HOP separate models},'' \emph{JVET-AH0188}, 2024.

\bibitem{zhang2023lightweight}
H.~Zhang, C.~Jung, Y.~Liu, and M.~Li, ``Lightweight cnn-based in-loop filter
  for {VVC} intra coding,'' in \emph{IEEE International Conference on Image
  Processing (ICIP)}.\hskip 1em plus 0.5em minus 0.4em\relax IEEE, 2023, pp.
  1635--1639.

\bibitem{AG0057}
``{[AHG11] Study on lower-complexity NNLF},'' \emph{JVET-AG0057}, 2024.

\bibitem{AH0051}
``{EE1-5: Study of the NN architecture at Very Low Operational Point},''
  \emph{JVET-AH0051}, 2024.

\bibitem{AI0107}
``{EE1-1.5: VLOP using LOP3 architecture},'' \emph{JVET-AI0107}, 2024.

\bibitem{AE0281}
``{Unified LOP filter design, training procedure and filter usage},''
  \emph{JVET-AE0281}, 2023.

\bibitem{AF0043}
``{Status of the joint EE1-0 (LOP.2) training},'' \emph{JVET-AF0043}, 2023.

\bibitem{AH0014}
``{NNVC software development (AHG14)},'' \emph{JVET-AH0014}, 2024.

\bibitem{AI0014}
``{NNVC software development (AHG14)},'' \emph{JVET-AI0014}, 2024.

\bibitem{AD0380}
``{BoG report on NN-filter design unification},'' \emph{JVET-AD0380}, 2023.

\bibitem{AH0189}
``{EE1-2.2: Wide activation HOP model},'' \emph{JVET-AH00189}, 2024.

\bibitem{AH0205}
``{EE1-2.3: Integer implementation of HOP In-loop filter with Transformer
  blocks and Attention blocks},'' \emph{JVET-AH0205}, 2024.

\bibitem{AI0172}
``{On the complexity adjustment of HOP4},'' \emph{JVET-AI0172}, 2024.

\bibitem{li2022mulut}
J.~Li, C.~Chen, Z.~Cheng, and Z.~Xiong, ``Mulut: Cooperating multiple look-up
  tables for efficient image super-resolution,'' in \emph{European Conference
  on Computer Vision}.\hskip 1em plus 0.5em minus 0.4em\relax Springer, 2022,
  pp. 238--256.

\bibitem{li2024toward}
------, ``Toward dnn of luts: Learning efficient image restoration with
  multiple look-up tables,'' \emph{IEEE Transactions on Pattern Analysis and
  Machine Intelligence}, 2024.

\bibitem{li2024look}
Y.~Li, J.~Li, and Z.~Xiong, ``Look-up table compression for efficient image
  restoration,'' in \emph{Proceedings of the IEEE/CVF Conference on Computer
  Vision and Pattern Recognition}, 2024, pp. 26\,016--26\,025.

\bibitem{li2024loop}
Z.~Li, J.~Li, Y.~Li, L.~Li, D.~Liu, and F.~Wu, ``In-loop filtering via trained
  look-up tables,'' in \emph{2024 IEEE International Conference on Visual
  Communications and Image Processing (VCIP)}.\hskip 1em plus 0.5em minus
  0.4em\relax IEEE, 2024, pp. 1--5.

\bibitem{zeng2020learning}
H.~Zeng, J.~Cai, L.~Li, Z.~Cao, and L.~Zhang, ``Learning image-adaptive 3d
  lookup tables for high performance photo enhancement in real-time,''
  \emph{IEEE Transactions on Pattern Analysis and Machine Intelligence},
  vol.~44, no.~4, pp. 2058--2073, 2020.

\bibitem{gems2programming}
M.~Pharr and R.~Fernando, ``Programming technics for high performance graphics
  and general purpose computation,'' \emph{Addison-Wesley Professional}, 2005.

\bibitem{kim2012new}
S.~J. Kim, H.~T. Lin, Z.~Lu, S.~S{\"u}sstrunk, S.~Lin, and M.~S. Brown, ``A new
  in-camera imaging model for color computer vision and its application,''
  \emph{IEEE Transactions on Pattern Analysis and Machine Intelligence},
  vol.~34, no.~12, pp. 2289--2302, 2012.

\bibitem{mantiuk2008display}
R.~Mantiuk, S.~Daly, and L.~Kerofsky, ``Display adaptive tone mapping,'' in
  \emph{ACM SIGGRAPH 2008 papers}, 2008, pp. 1--10.

\bibitem{jo2021practical}
Y.~Jo and S.~J. Kim, ``Practical single-image super-resolution using look-up
  table,'' in \emph{Proceedings of the IEEE/CVF Conference on Computer Vision
  and Pattern Recognition}, 2021, pp. 691--700.

\bibitem{CTCdocument}
K.~Suehring and X.~Li, ``{JVET Common Test Conditions and Software Reference
  Configurations},'' \emph{JVET document, JVET-G1010}, 2017.

\bibitem{liu2021jvet}
S.~Liu, A.~Segall, E.~Alshina, and R.-L. Liao, ``{JVET} common test conditions
  and evaluation procedures for neural network-based video coding technology,''
  \emph{JVET-X2016}, 2021.

\bibitem{bengio2013estimating}
Y.~Bengio, N.~L{\'e}onard, and A.~Courville, ``Estimating or propagating
  gradients through stochastic neurons for conditional computation,''
  \emph{arXiv preprint arXiv:1308.3432}, 2013.

\bibitem{dong2023temporal}
C.~Dong, H.~Ma, Z.~Li, L.~Li, and D.~Liu, ``Temporal wavelet transform-based
  low-complexity perceptual quality enhancement of compressed video,''
  \emph{IEEE Transactions on Circuits and Systems for Video Technology},
  vol.~34, no.~5, pp. 4040--4053, 2023.

\bibitem{2001Calculation}
G.~Bjontegaard, ``{Calculation of Average PSNR Differences between
  RD-curves},'' \emph{ITU-T VCEG-M33, April, 2001}, 2001.

\bibitem{ma2021bvi}
D.~Ma, F.~Zhang, and D.~R. Bull, ``{BVI-DVC}: {A} training database for deep
  video compression,'' \emph{IEEE Transactions on Multimedia}, vol.~24, pp.
  3847--3858, 2021.

\bibitem{agustsson2017ntire}
E.~Agustsson and R.~Timofte, ``Ntire 2017 challenge on single image
  super-resolution: Dataset and study,'' in \emph{Proceedings of the IEEE
  Conference on Computer Vision and Pattern Recognition Workshops (CVPRW)},
  2017, pp. 126--135.

\bibitem{song2021addersr}
D.~Song, Y.~Wang, H.~Chen, C.~Xu, C.~Xu, and D.~Tao, ``Addersr: Towards energy
  efficient image super-resolution,'' in \emph{Proceedings of the IEEE/CVF
  Conference on Computer Vision and Pattern Recognition}, 2021, pp.
  15\,648--15\,657.

\bibitem{sze2017efficient}
V.~Sze, Y.-H. Chen, T.-J. Yang, and J.~S. Emer, ``Efficient processing of deep
  neural networks: A tutorial and survey,'' \emph{Proceedings of the IEEE},
  vol. 105, no.~12, pp. 2295--2329, 2017.

\bibitem{horowitz20141}
M.~Horowitz, ``1.1 computing's energy problem (and what we can do about it),''
  in \emph{2014 IEEE International Solid-State Circuits Conference Digest of
  Technical Papers (ISSCC)}.\hskip 1em plus 0.5em minus 0.4em\relax IEEE, 2014,
  pp. 10--14.

\bibitem{yan2018convolutional}
N.~Yan, D.~Liu, H.~Li, B.~Li, L.~Li, and F.~Wu, ``Convolutional neural
  network-based fractional-pixel motion compensation,'' \emph{IEEE Transactions
  on Circuits and Systems for Video Technology}, vol.~29, no.~3, pp. 840--853,
  2018.

\bibitem{fu2022efficient}
T.~Fu, K.~Zhang, L.~Zhang, S.~Wang, and S.~Ma, ``An efficient framework of
  reference picture resampling {(RPR)} for video coding,'' \emph{IEEE
  Transactions on Circuits and Systems for Video Technology}, vol.~32, no.~10,
  pp. 7107--7119, 2022.

\bibitem{gonzalez2009digital}
R.~C. Gonzalez, \emph{Digital image processing}.\hskip 1em plus 0.5em minus
  0.4em\relax Pearson education india, 2009.

\bibitem{szeliski2022computer}
R.~Szeliski, \emph{Computer vision: algorithms and applications}.\hskip 1em
  plus 0.5em minus 0.4em\relax Springer Nature, 2022.

\bibitem{kasson1995performing}
J.~M. Kasson, S.~I. Nin, W.~Plouffe, and J.~L. Hafner, ``Performing color space
  conversions with three-dimensional linear interpolation,'' \emph{Journal of
  Electronic Imaging (JEI)}, vol.~4, no.~3, pp. 226--250, 1995.

\bibitem{liu2023reconstructed}
G.~Liu, Y.~Ding, M.~Li, M.~Sun, X.~Wen, and B.~Wang, ``Reconstructed
  convolution module based look-up tables for efficient image
  super-resolution,'' in \emph{Proceedings of the IEEE/CVF International
  Conference on Computer Vision}, 2023, pp. 12\,217--12\,226.

\bibitem{yin2024online}
G.~Yin, Z.~Qu, X.~Jiang, S.~Jiang, Z.~Han, N.~Zheng, H.~Yang, X.~Liu, Y.~Yang,
  D.~Li \emph{et~al.}, ``{Online Streaming Video Super-Resolution With
  Convolutional Look-Up Table},'' \emph{IEEE Transactions on Image Processing},
  vol.~33, pp. 2305--2317, 2024.

\end{thebibliography}

\clearpage


\twocolumn[
\begin{center}
	\Huge \vspace{0.5em}\textit{Supplementary Material for} \\In-Loop Filtering Using Learned Look-Up Tables for Video Coding
	\vspace{10mm}
\end{center}
]

\appendices
\section{Supplementary Methodology}
In this section, we provide supplementary descriptions and implementation details for some modules of LUT-ILF++ that are only briefly introduced in the main text, including LUT interpolation, LUT compaction implementation. By elaborating on the specific procedures and internal workflows, we aim to improve the clarity and reproducibility of LUT-ILF++. 

\subsection{LUT-ILF++ with Diverse Interpolation}
Direct uniform sampling of indexing entries of the full filtering LUT is applied to restrict the rapid growth of storage cost in LUT-ILF++, as mentioned in Section II.A (3) and Section VI of the main text. Due to the non-sampled indexing entries will cause the indexing drift in the LUT retrieve process, the interpolation model is introduced to estimate the drifting entries, and the interpolation process is performed to calculate the final filtered pixel by locating the nearest neighbor indexes (most significant bits) of query indexes (to-be-filtered pixel with reference pixels) and weighting the cached values of neighbor indexes during LUT retrieval. In LUT-ILF++, due to the various types of multi-dimensional LUTs constructed to support diverse filtering operations and goals, such as the reference/progressive indexing with 4D spatial-wise LUTs and cross-component offset indexing with 3D channel-wise LUTs, various types of interpolation model are introduced to support their interpolation process with different dimension numbers, respectively. Specifically, a tri-linear interpolation model is adopted for 3D LUTs, and a 4-$simplex$ interpolation model is used for 4D LUTs, corresponding to the dimensionality of the indexing space in each case. Here we detail the usage of these interpolation models in LUT-ILF++. 

\subsubsection{Trilinear-based LUT Indexing Entry Interpolation}
\begin{figure}
	\centering
	\includegraphics[width=69mm]{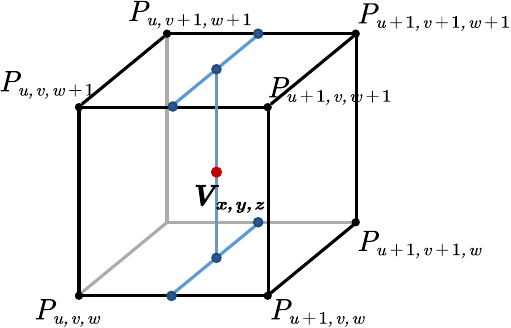}
	\vspace{-0.3em}
	\caption{Illustration of the trilinear interpolation model of a 3D LUT.}
	\label{fig:trilinear}
	\vspace{-1.5em}
\end{figure}

The core solution of the linear interpolation scheme is to ensemble known sample values to perform linear weighted interpolations based on the combination of linear distances in each dimension, such as bilinear, bicubic\cite{gonzalez2009digital}, and trilinear model\cite{szeliski2022computer}. For 3D LUTs, the trilinear interpolation model \cite{zeng2020learning} is adopted to estimate the non-sampled indexing entries, as shown in Fig.~\ref{fig:trilinear}\,.

\begin{figure*}
	\centering
	\vspace{-1em}
	\includegraphics[width=160mm]{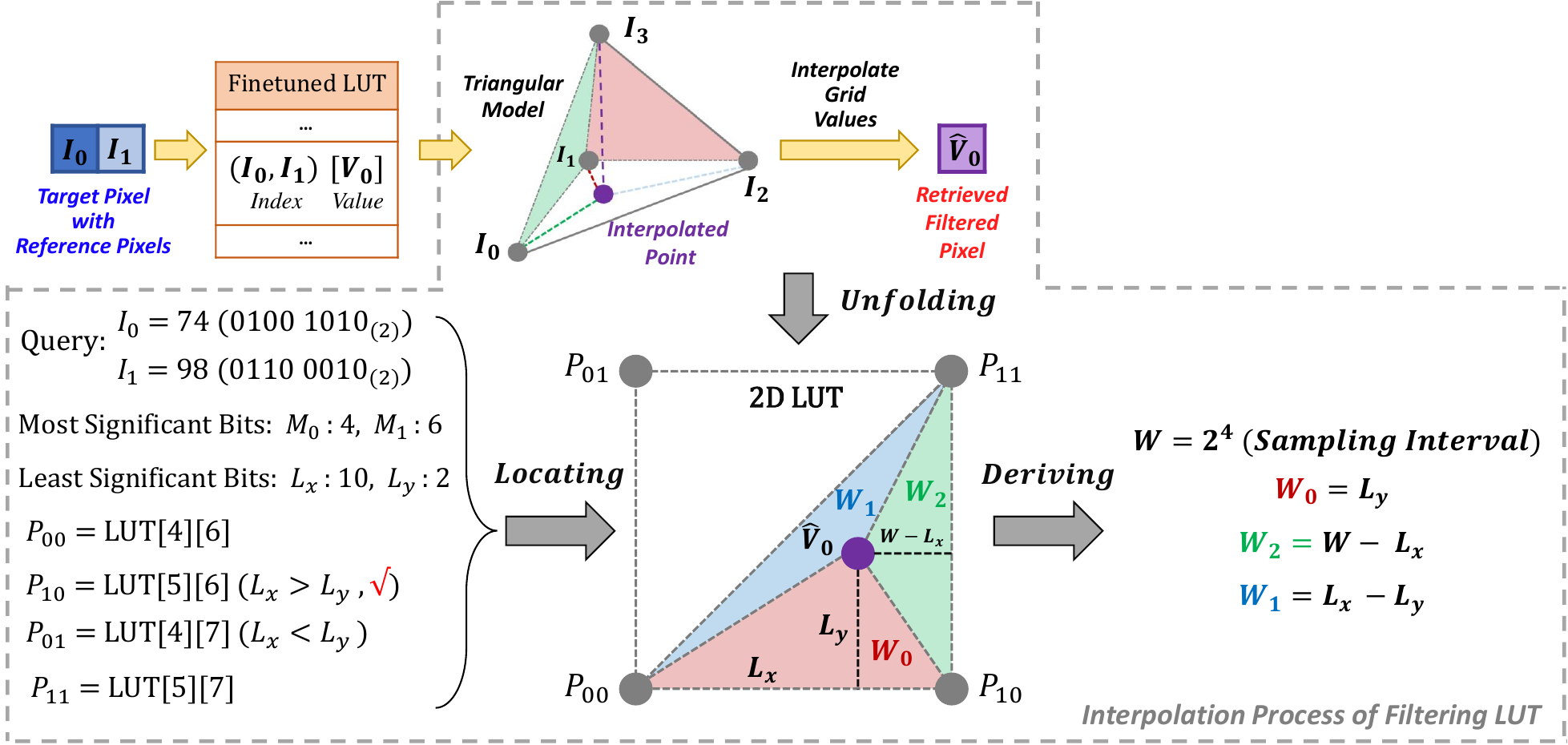}
	\vspace{-0.3em}
	\caption{Illustration of the example of the triangular interpolation process of a 2D LUT. Based on this interpolation principle applied in 2D space, the simplex interpolation process for a 4D LUT can be directly extended to the 4D space.}
	\label{fig:2D-triangular}
	\vspace{-1.2em}
\end{figure*}

To perform trilinear interpolation in the 3D LUT space, first, a local cube is defined by the eight nearest neighboring indices (grid points) ranging from $[u,v,w]$ to $[u+1,v+1,w+1]$. 
During the interpolation process, the most significant bits (MSBs) of the query index (to-be-filtered pixel and its two reference pixel values) are used to locate the local cube (nearest neighboring indices) in the 3D LUT space. Specifically, given the target-reference pixel triplet $\left(x,y,z\right)$, the corresponding index coordinates can be computed as $u = x \gg L$, $v = y \gg L$, and $w = z \gg L$, where $L$ is the length of the least significant bits (LSBs). These indices serve as the origin of the interpolation cube, from which the eight surrounding entries are retrieved for weighted averaging. Second, the interpolation weights along each axis are computed based on the fractional offset between the query pixel pair index and its corresponding nearest neighboring indices. These offsets are implicitly encoded in the least significant bits (LSBs), and indicate the relative position of the query index within the interpolation cube. Specifically, $d_x$, $d_y$, and $d_z$ represent the normalized distances from the nearest neighboring indices to the query index along the $x$, $y$, and $z$ dimensions, respectively. These values are then used to derive the trilinear interpolation weights. Using the cached index values of the eight nearest neighboring indices within the local cube, the final filtered (interpolated) value of the target pixel is computed as a weighted sum:\vspace{-0.3em}
\begin{equation}
	\footnotesize
	\begin{aligned}
		V_{x,y,z}&=
		\left( 1\!-\!d_x \right) \left( 1\!-\!d_y \right) \left( 1\!-\!d_z \right) P_{u,v,w}
		\!+\!d_x\left( 1\!-\!d_y \right) \left( 1\!-\!d_z \right) P_{u+1,v,w}
		\\
		&+\!\left( 1\!-\!d_x \right) d_y\left( 1\!-\!d_z \right) P_{u,v+1,w}\!+\!d_xd_y\left( 1\!-\!d_z \right) P_{u+1,v+1,w}
		\\
		&+\!\left( 1\!-\!d_x \right) \left( 1\!-\!d_y \right) d_zP_{u,v,w+1}\!+\!d_x\left( 1\!-\!d_y \right) d_zP_{u+1,v,w+1}
		\\
		&+\!\left( 1\!-\!d_x \right) d_yd_zP_{u,v+1,w+1}+d_xd_yd_zP_{u+1,v+1,w+1},
	\end{aligned}
\end{equation}
where $V_{x,y,z}$ denotes the final interpolated (filtered) value of query index $[x, y, z]$, and $P_{u,v,w}$ denotes the cached index value of the $[u, v, w]$ in the filtering LUT. The interpolation weights $d_x = \frac{x - (u \ll L)}{1 \ll L}$, $d_y = \frac{y - (v \ll L)}{1 \ll L}$, and $d_z = \frac{z - (w \ll L)}{1 \ll L}$ are the normalized distances between the query index and the nearest neighboring indices along the $x$, $y$, and $z$ dimensions, respectively, where $L$ denotes the number of LSBs.

\subsubsection{4-$simplex$-based LUT Indexing Entry Interpolation}
The formulation of the above linear interpolation scheme can be directly extended to high-dimensional spaces; however, it may face challenges in terms of computational complexity and adaptability to irregularly distributed data in high-dimensional spaces. To achieve this goal with lower computational complexity, we utilize the well-established simplex interpolation model \cite{kasson1995performing} for a high-dimensional LUT interpolation process, which follows the same model and interpolation scheme used in LUT-based image processing tasks \cite{jo2021practical, li2022mulut, liu2023reconstructed, yin2024online, li2024toward}. The core concept is to perform interpolation within a simplex. For an interpolation process in an $N$-dimensional space, $N+1$ known query indices are identified from the nearest neighboring  indices to construct a simplex that encloses the target index (to-be-filtered pixel and its three reference pixel values). The barycentric coordinates of the target index within the simplex are then calculated to determine the interpolation weights relative to these $N+1$ vertices.

To detail the 4-$simplex$ interpolation process of 4D LUT in LUT-ILF++, we serve the 2D LUT with the triangular interpolation process as a simple equivalent case, as shown in Fig.~\ref{fig:2D-triangular}\,, from which the 4D interpolation process naturally extends by following the same interpolation principle. In the 2D interpolation process, first, for the query indexing of the target pixel and its one adjacent reference pixel values, four nearest neighboring query indices are retrieved: $P_{00}$, $P_{01}$, $P_{10}$, and $P_{11}$. Second, among them, three vertices that form a triangle enclosing the target index are selected for interpolation. As the example shown in Fig.~\ref{fig:2D-triangular}\,, the interpolation falls within the bottom right triangle in the case where the LSBs of the reference pixel value are smaller than those of the target to-be-filtered pixel, i.e., $L_{x} > L_{y}$. According to the barycentric coordinates of the target index, the interpolation weights can be easily derived as: $w_{0}=L_{y}$, $w_{1}=L_{x}-L_{y}$, $w_{2}=W-L_{x}$, and $W$ denotes the sampling interval. Finally, the interpolation value can be derived as the weighted sum as follows: 
\begin{equation}
	\widehat{V_0}=\left( w_0\times P_{11}+w_1\times P_{10}+w_2\times P_{00} \right).
\end{equation}

\begin{table}
	\centering
	\caption{Summary Rules of 4-$simplex$ Interpolation Model\\ (followed by \cite{jo2021practical, li2022mulut, liu2023reconstructed, yin2024online, li2024toward}.)}
	\vspace{-0.5em}
	\footnotesize
	\setlength{\tabcolsep}{1pt}
	\begin{tabular}{@{}lccccccccc@{}}
		\toprule
		Condition & $w_0$ & $w_1$ & $w_2$ & $w_3$ & $w_4$ & $O_1$ & $O_2$ & $O_3$ \\ \midrule
		$L_x \!\!>\!\! L_y \!\!>\!\! L_z \!\!>\!\! L_t$ & $W \!\!-\!\! L_x$ & $L_x \!\!-\!\! L_y$ & $L_y \!\!-\!\! L_z$ & $L_z \!\!-\!\! L_t$ & $L_t$ & $P_{1000}$ & $P_{1100}$ & $P_{1110}$ \\
		$L_x \!\!>\!\! L_y \!\!>\!\! L_t \!\!>\!\! L_z$ & $W \!\!-\!\! L_x$ & $L_x \!\!-\!\! L_y$ & $L_y \!\!-\!\! L_t$ & $L_t \!\!-\!\! L_z$ & $L_z$ & $P_{1000}$ & $P_{1100}$ & $P_{1101}$ \\
		$L_x \!\!>\!\! L_t \!\!>\!\! L_y \!\!>\!\! L_z$ & $W \!\!-\!\! L_x$ & $L_x \!\!-\!\! L_t$ & $L_t \!\!-\!\! L_y$ & $L_y \!\!-\!\! L_z$ & $L_z$ & $P_{1000}$ & $P_{1001}$ & $P_{1101}$ \\
		$L_t \!\!>\!\! L_x \!\!>\!\! L_y \!\!>\!\! L_z$ & $W \!\!-\!\! L_t$ & $L_t \!\!-\!\! L_x$ & $L_x \!\!-\!\! L_y$ & $L_y \!\!-\!\! L_z$ & $L_z$ & $P_{0001}$ & $P_{1001}$ & $P_{1101}$ \\
		$L_x \!\!>\!\! L_z \!\!>\!\! L_y \!\!>\!\! L_t$ & $W \!\!-\!\! L_x$ & $L_x \!\!-\!\! L_z$ & $L_z \!\!-\!\! L_y$ & $L_y \!\!-\!\! L_t$ & $L_t$ & $P_{1000}$ & $P_{1010}$ & $P_{1110}$ \\
		$L_x \!\!>\!\! L_z \!\!>\!\! L_t \!\!>\!\! L_y$ & $W \!\!-\!\! L_x$ & $L_x \!\!-\!\! L_z$ & $L_z \!\!-\!\! L_t$ & $L_t \!\!-\!\! L_y$ & $L_y$ & $P_{1000}$ & $P_{1010}$ & $P_{1101}$ \\
		$L_t \!\!>\!\! L_z \!\!>\!\! L_x \!\!>\!\! L_y$ & $W \!\!-\!\! L_t$ & $L_t \!\!-\!\! L_z$ & $L_z \!\!-\!\! L_x$ & $L_x \!\!-\!\! L_y$ & $L_y$ & $P_{0100}$ & $P_{1100}$ & $P_{1110}$ \\
		$L_t \!\!>\!\! L_y \!\!>\!\! L_x \!\!>\!\! L_z$ & $W \!\!-\!\! L_t$ & $L_t \!\!-\!\! L_y$ & $L_y \!\!-\!\! L_x$ & $L_x \!\!-\!\! L_z$ & $L_z$ & $P_{0100}$ & $P_{1100}$ & $P_{1110}$ \\
		$L_y \!\!>\!\! L_z \!\!>\!\! L_x \!\!>\!\! L_t$ & $W \!\!-\!\! L_y$ & $L_y \!\!-\!\! L_z$ & $L_z \!\!-\!\! L_x$ & $L_x \!\!-\!\! L_t$ & $L_t$ & $P_{0010}$ & $P_{1001}$ & $P_{1101}$ \\
		$L_y \!\!>\!\! L_z \!\!>\!\! L_t \!\!>\!\! L_x$ & $W \!\!-\!\! L_y$ & $L_y \!\!-\!\! L_z$ & $L_z \!\!-\!\! L_t$ & $L_t \!\!-\!\! L_x$ & $L_x$ & $P_{0010}$ & $P_{1010}$ & $P_{1101}$ \\
		$L_y \!\!>\!\! L_x \!\!>\!\! L_z \!\!>\!\! L_t$ & $W \!\!-\!\! L_y$ & $L_y \!\!-\!\! L_x$ & $L_x \!\!-\!\! L_z$ & $L_z \!\!-\!\! L_t$ & $L_t$ & $P_{0100}$ & $P_{1001}$ & $P_{1101}$ \\
		$L_y \!\!>\!\! L_x \!\!>\!\! L_t \!\!>\!\! L_z$ & $W \!\!-\!\! L_y$ & $L_y \!\!-\!\! L_x$ & $L_x \!\!-\!\! L_t$ & $L_t \!\!-\!\! L_z$ & $L_z$ & $P_{0100}$ & $P_{1010}$ & $P_{1101}$ \\
		$L_z \!\!>\!\! L_y \!\!>\!\! L_x \!\!>\!\! L_t$ & $W \!\!-\!\! L_z$ & $L_z \!\!-\!\! L_y$ & $L_y \!\!-\!\! L_x$ & $L_x \!\!-\!\! L_t$ & $L_t$ & $P_{0010}$ & $P_{1001}$ & $P_{1101}$ \\
		$L_z \!\!>\!\! L_x \!\!>\!\! L_y \!\!>\!\! L_t$ & $W \!\!-\!\! L_z$ & $L_z \!\!-\!\! L_x$ & $L_x \!\!-\!\! L_y$ & $L_y \!\!-\!\! L_t$ & $L_t$ & $P_{0010}$ & $P_{1001}$ & $P_{1101}$ \\
		$L_z \!\!>\!\! L\textbf{}_x \!\!>\!\! L_t \!\!>\!\! L_y$ & $W \!\!-\!\! L_z$ & $L_z \!\!-\!\! L_x$ & $L_x \!\!-\!\! L_t$ & $L_t \!\!-\!\! L_y$ & $L_y$ & $P_{0010}$ & $P_{1010}$ & $P_{1101}$ \\
		$L_z \!\!>\!\! L_t \!\!>\!\! L_x \!\!>\!\! L_y$ & $W \!\!-\!\! L_z$ & $L_z \!\!-\!\! L_t$ & $L_t \!\!-\!\! L_x$ & $L_x \!\!-\!\! L_y$ & $L_y$ & $P_{0010}$ & $P_{1010}$ & $P_{1101}$ \\
		$L_z \!\!>\!\! L_t \!\!>\!\! L_y \!\!>\!\! L_x$ & $W \!\!-\!\! L_z$ & $L_z \!\!-\!\! L_t$ & $L_t \!\!-\!\! L_y$ & $L_y \!\!-\!\! L_x$ & $L_x$ & $P_{0010}$ & $P_{1010}$ & $P_{1101}$ \\ 
		else & $W \!\!-\!\! L_t$ & $L_t \!\!-\!\! L_z$ & $L_z \!\!-\!\! L_y$ & $L_y \!\!-\!\! L_x$ & $L_x$ & $P_{0001}$ & $P_{0011}$ & $P_{0111}$ \\ 
		\bottomrule
	\end{tabular}
	\label{tab:simplex_condition}
	\vspace{-1.8em}
\end{table}

Based on the above 2D interpolation case, the 4-$simplex$ interpolation process of 4D LUT can be naturally extended in a similar manner, with the establishment of interpolation simplex selection conditions. The rule of vertices selection can be characterized by the relative magnitudes of the LSB values of query index in the 4D dimension case, $L_{x}$, $L_{y}$, $L_{z}$, and $L_{t}$ denote the LSB values for each dimension. Let $\pi$ denote a permutation of these four LSB values, and $\pi\left(i\right)$ represents the $i$-th largest LSB value. Then the binary index code of the $k$-th selected node, $I\left(O_{k}\right)$, can be formulated as:
\begin{equation}
	I\left( O_k \right) =\sum_{i=1}^k{2^{3-\mathrm{index}\left( \pi \left( i \right) \right)}},  k=1,2,3,
\end{equation}
where $\mathrm{index}\left( \pi \left( i \right) \right) $ denotes the index of LSB value $\pi\left(i\right)$ in list $\left[ L_x,L_y,L_z,L_t \right] $, with indices starting from zero.
The weights can be expressed in the following form with respect to 
$\pi$:
\begin{equation}
	w_{k}=\begin{cases}
		W-\pi \left( k \right)  &   k=1 \\
		\pi \left( k \right) -\pi \left( k+1 \right)    &   2\leqslant k\leqslant 3 \\
		\pi \left( 4 \right)    &   k=4
	\end{cases},
\end{equation}
where $W$ denotes the samping interval.
For example, in the case $L_x > L_y > L_t > L_z$, the binary index array can be derive as $\left[0b1000,0b1000+0b0100,0b1000+0b0100+0b0001\right]$.
This corresponds to selecting the vertices $P_{1000}$, $P_{1100}$, and $P_{1101}$. 
Another two vertices $P_{0000}$, and $P_{1111}$ are fixed for all cases. The corresponding weights can be expressed as $W - L_x$, $L_x - L_y$, $L_y - L_t$, $L_t - L_z$, and $L_z$. The summary rules of simplex selection and the interpolation weights of 4-$simplex$ interpolation model are shown in Table~\ref{tab:simplex_condition}\,.

\subsection{The Compaction Implementation of LUT-ILF++}
In this subsection, we detail the LUT pruning strategy of the LUT compaction scheme in LUT-ILF++, including the LUT diagonal re-ordering and non-diagonal pruning. Here we also take the first two dimensions (2D, [$I_0$, $I_1$]) as an example to detail them, as shown in Fig.10 of the main text.

\subsubsection{LUT Diagonal Rearranging}
Based on the judgment of the diagonal condition rule ($|I_0 - I_1| \leq diagonal\ width\ (dw)$) in the clipped LUT, the clipped indexing entries are retrieved and classified into diagonal and non-diagonal indexing entries. For diagonal indexing entries, the diagonal re-ordering is used to rearrange them as the diagonal LUT by mapping the LUT coordinate ($I_c=f_{mapping}(I_0,I_1,dw)$), and the diagonal LUT is then stored as a low-dimensional LUT indexed by [$I_c$, $I_2$, $I_3$]. In the mapping process, the total number of diagonal indexing entries ($D$) can be calculated as, 
\begin{equation}
	D = (2\times dw + 1)\times L - dw \times (dw + 1),
\end{equation}
among the compacted dimensions in 2D space case, and the first two dimensions of 4D LUT are compacted into one dimension by mapping the index [$I_0$, $I_1$] to compacted index $I_c$. The indexing relationship conversion between the index [$I_0$, $I_1$] and compacted index $I_c$ can be formulated as, 
\begin{equation}
	I_c = f_{mapping}(I_0,I_1,dw) = I_1 \times (2\times dw + 1) + r - 1,
\end{equation}
where $L$ indicates the size of each dimension, $r$ indicates the relative distance between $I_0$ and $I_1$, and it can be calculated as,
\begin{equation}
	r = I_0 - I_1 + dw, \ (0 \leq r \leq 2\times dw).
\end{equation} 
Through the above mapping manner, the low-dimensional 3D diagonal LUT is compacted from the 4D clipped LUT, and the final index $[I_c, I_2, I_3]$ is used to retrieve the low-dimensional diagonal LUT. In general use, the LUT re-ordering process can be easily generalized to multiple dimensions of LUT indexing entries.

\subsubsection{LUT Non-Diagonal Pruning}
For non-diagonal indexing entries, inspired by the observation of their sparse access concentration in practice (Fig.3 of the main text), their redundancy is pruned by re-sampling their dimensionality (MSBs) with the allocated sparsification shift ($Q$), achieving exponentially reduced storage cost. Starting from the 4D clipped LUT with storage dimension $(2^{8-q} + 1) \times (2^{8-q} + 1) \times (2^{8-q} + 1) \times (2^{8-q} + 1)$, as shown in Fig.10 and mentioned in Section II.B (1) of the main text, the whole clipped LUT is further directly re-sampled and sparsified its dimensionality (MSBs) to obtain the non-diagonal LUT with storage dimension $(2^{8-q-Q} + 1) \times (2^{8-q-Q} + 1) \times (2^{8-q-Q} + 1) \times (2^{8-q-Q} + 1)$. Note that some cached indexing entries of diagonal LUT are overlapped and cached into non-diagonal LUT to predict the values that do not follow the diagonal condition but are close to the diagonal boundary.

\begin{table*}
	\renewcommand\arraystretch{1.15}
	\centering
	\fontsize{7.2pt}{9pt}\selectfont
	\vspace{-2.5em}
	\caption{BD-rate Results of Our Proposed LUT-ILF++ Compared to VTM-11.0 on CTC Test Sequences with Regular-bitrate Points\\ (QP 22$\sim$42), and Comparison Results with the Other In-Loop Filtering Schemes \cite{li2024loop} under All Intra (AI) Configuration}
	\label{tab:DBR1}
	\vspace{-0.9em}
	\setlength{\tabcolsep}{1.5mm}
	{
		\begin{tabular}{cclcccclcccclcccc}
			\hline
			\multicolumn{17}{c}{\textbf{All Intra (AI) Configuration (\%)}}                                                                                                                                                                                                                                                                                                                                                                                     \\ \hline
			\multirow{2}{*}{\textbf{Class}}                                                         & \textbf{Sequence}         &                                        & \multicolumn{4}{c}{\textbf{LUT-ILF-V~\cite{li2024loop}}}                           &                      & \multicolumn{4}{c}{\textbf{LUT-ILF-F~\cite{li2024loop}}}                         &                      & \multicolumn{4}{c}{\textbf{LUT-ILF++}}                           \\ \cline{2-2} \cline{4-7} \cline{9-12} \cline{14-17} 
			& Name                      &                                        & Y                & U                & V                & Ratio            & \multicolumn{1}{c}{} & Y                & U                & V                & Ratio            &                      & Y                & U       & V                & Ratio            \\ \cline{1-2} \cline{4-7} \cline{9-12} \cline{14-17} 
			\multirow{4}{*}{\textbf{\begin{tabular}[c]{@{}c@{}}ClassA1\\ (3840x2160)\end{tabular}}} & \textit{Tango2}           &                                        & -0.21\%          & -0.16\%          & -0.17\%          & 23.13\%          &                      & -0.33\%          & -0.44\%          & -0.53\%          & 39.15\%          &                      & -0.67\%          & -5.10\%          & -2.77\%          & 41.57\%          \\
			& \textit{FoodMarket4}      &                                        & -0.13\%          & -0.41\%          & -0.03\%          & 21.33\%          &                      & -0.26\%          & -0.24\%          & -0.41\%          & 33.67\%          &                      & -0.49\%          & -1.01\%          & -0.68\%          & 30.74\%          \\
			& \textit{Campfire}         &                                        & -0.24\%          & -0.45\%          & -0.18\%          & 27.60\%          &                      & -0.32\%          & -1.03\%          & -0.72\%          & 40.33\%          &                      & -0.47\%          & -0.52\%          & -1.95\%          & 47.93\%          \\
			& \textit{\textbf{Average}} &                                        & \textbf{-0.19\%} & \textbf{-0.34\%} & \textbf{-0.13\%} & \textbf{24.02\%} &                      & \textbf{-0.30\%} & \textbf{-0.57\%} & \textbf{-0.55\%} & \textbf{37.72\%} &                      & \textbf{-0.54\%} & \textbf{-2.21\%} & \textbf{-1.80\%} & \textbf{40.08\%} \\ \cline{1-2} \cline{4-7} \cline{9-12} \cline{14-17} 
			\multirow{4}{*}{\textbf{\begin{tabular}[c]{@{}c@{}}ClassA2\\ (3840x2160)\end{tabular}}} & \textit{CatRobot1}        &                                        & -0.41\%          & -0.61\%          & -0.53\%          & 33.44\%          &                      & -0.60\%          & -0.88\%          & -0.42\%          & 43.16\%          &                      & -0.91\%          & -4.80\%          & -1.60\%          & 51.07\%          \\
			& \textit{DaylightRoad2}    &                                        & -0.44\%          & -0.34\%          & -0.13\%          & 32.13\%          &                      & -0.57\%          & -0.82\%          & -0.49\%          & 46.56\%          &                      & -0.51\%          & -2.78\%          & -1.10\%          & 45.70\%          \\
			& \textit{ParkRunning3}     &                                        & -0.21\%          & -0.28\%          & -0.15\%          & 29.73\%          &                      & -0.42\%          & -0.46\%          & -0.34\%          & 42.23\%          &                      & -0.77\%          & -0.19\%          & -0.72\%          & 49.69\%          \\
			& \textit{\textbf{Average}} &                                        & \textbf{-0.36\%} & \textbf{-0.41\%} & \textbf{-0.27\%} & \textbf{31.77\%} &                      & \textbf{-0.53\%} & \textbf{-0.73\%} & \textbf{-0.42\%} & \textbf{43.98\%} &                      & \textbf{-0.73\%} & \textbf{-2.59\%} & \textbf{-1.14\%} & \textbf{48.82\%} \\ \cline{1-2} \cline{4-7} \cline{9-12} \cline{14-17} 
			\multirow{6}{*}{\textbf{\begin{tabular}[c]{@{}c@{}}ClassB\\ (1920x1080)\end{tabular}}}  & \textit{MarketPlace}      &                                        & -0.25\%          & -0.11\%          & -0.03\%          & 31.31\%          &                      & -0.54\%          & -1.28\%          & -0.98\%          & 42.18\%          &                      & -0.79\%          & -3.54\%          & -1.95\%          & 49.96\%          \\
			& \textit{RitualDance}      &                                        & -0.38\%          & -0.51\%          & -0.32\%          & 37.32\%          &                      & -0.52\%          & -1.21\%          & -1.18\%          & 43.35\%          &                      & -0.88\%          & -2.82\%          & -2.47\%          & 49.78\%          \\
			& \textit{Cactus}           &                                        & -0.21\%          & -0.38\%          & -0.12\%          & 24.16\%          &                      & -0.31\%          & -0.81\%          & -1.01\%          & 32.27\%          &                      & -0.70\%          & -3.74\%          & -1.38\%          & 41.46\%          \\
			& \textit{BasketballDrive}  &                                        & 0.04\%          & -0.13\%          & 0.04\%           & 10.27\%          &                      & -0.02\%          & -0.59\%          & -0.22\%          & 18.94\%          &                      & -0.71\%          & -2.56\%          & -1.08\%          & 27.41\%          \\
			& \textit{BQTerrace}        &                                        & -0.33\%          & -0.47\%          & -0.42\%          & 31.84\%          &                      & -0.27\%          & -1.19\%          & -0.76\%          & 33.31\%          &                      & -0.51\%          & -2.90\%          & -1.84\%          & 38.27\%          \\
			& \textit{\textbf{Average}} &                                        & \textbf{-0.22\%} & \textbf{-0.32\%} & \textbf{-0.17\%} & \textbf{26.97\%} &                      & \textbf{-0.33\%} & \textbf{-1.02\%} & \textbf{-0.83\%} & \textbf{34.01\%} &                      & \textbf{-0.72\%} & \textbf{-3.11\%} & \textbf{-1.74\%} & \textbf{41.38\%} \\ \cline{1-2} \cline{4-7} \cline{9-12} \cline{14-17} 
			\multirow{5}{*}{\textbf{\begin{tabular}[c]{@{}c@{}}ClassC\\ (832x480)\end{tabular}}}    & \textit{BasketballDrill}  &                                        & -0.30\%          & -0.56\%          & -0.21\%          & 34.31\%          &                      & -0.37\%          & -0.91\%          & -0.33\%          & 41.12\%          &                      & -0.81\%          & -2.78\%          & -1.83\%          & 50.43\%          \\
			& \textit{BQMall}           & \multicolumn{1}{c}{\textit{}}          & -0.31\%          & -0.63\%          & -0.27\%          & 39.26\%          &                      & -0.59\%          & -1.57\%          & -0.27\%          & 51.35\%          &                      & -0.92\%          & -4.56\%          & -1.74\%          & 63.35\%          \\
			& \textit{PartyScene}       & \multicolumn{1}{c}{\textit{}}          & -0.14\%          & -0.41\%          & -0.17\%            & 35.31\%          &                      & -0.18\%          & -1.41\%          & -0.40\%          & 44.67\%          &                      & -0.67\%          & -4.35\%          & -0.99\%          & 64.90\%          \\
			& \textit{RaceHorsesC}      & \multicolumn{1}{c}{\textit{}}          & -0.29\%          & -0.52\%          & 0.06\%           & 37.17\%          &                      & -0.33\%          & -0.91\%          & -0.36\%          & 42.57\%          &                      & -0.41\%          & -3.90\%          & -1.73\%          & 41.73\%          \\
			& \textit{\textbf{Average}} & \multicolumn{1}{c}{\textit{\textbf{}}} & \textbf{-0.26\%} & \textbf{-0.53\%} & \textbf{-0.14\%} & \textbf{36.51\%} &                      & \textbf{-0.37\%} & \textbf{-1.20\%} & \textbf{-0.34\%} & \textbf{44.93\%} &                      & \textbf{-0.70\%} & \textbf{-3.90\%} & \textbf{-1.57\%} & \textbf{55.10\%} \\ \cline{1-2} \cline{4-7} \cline{9-12} \cline{14-17} 
			\multirow{5}{*}{\textbf{\begin{tabular}[c]{@{}c@{}}ClassD\\ (416x240)\end{tabular}}}    & \textit{BasketballPass}   &                                        & -0.46\%          & -1.12\%          & -0.42\%          & 42.93\%          &                      & -0.66\%          & -1.77\%          & -1.02\%          & 53.38\%          &                      & -1.27\%          & -3.28\%          & -3.57\%          & 68.08\%          \\
			& \textit{BQSquare}         &                                        & -0.49\%          & -0.85\%          & -0.62\%          & 41.08\%          &                      & -0.52\%          & -1.12\%          & -0.52\%          & 53.22\%          &                      & -1.34\%          & -1.63\%          & -1.80\%          & 69.58\%          \\
			& \textit{BlowingBubbles}   &                                        & -0.30\%          & -0.33\%          & -0.22\%          & 48.53\%          &                      & -0.41\%          & -1.05\%          & -0.87\%          & 51.77\%          &                      & -0.83\%          & -3.45\%          & -0.63\%          & 66.92\%          \\
			& \textit{RaceHorses}       &                                        & -0.71\%          & -1.02\%          & -0.96\%          & 55.28\%          &                      & -0.97\%          & -1.19\%          & -0.79\%          & 61.63\%          &                      & -1.49\%          & -5.24\%          & -2.94\%          & 74.99\%          \\
			& \textit{\textbf{Average}} &                                        & \textbf{-0.49\%} & \textbf{-0.83\%} & \textbf{-0.56\%} & \textbf{46.96\%} &                      & \textbf{-0.64\%} & \textbf{-1.28\%} & \textbf{-0.80\%} & \textbf{55.01\%} &                      & \textbf{-1.23\%} & \textbf{-3.40\%} & \textbf{-2.24\%} & \textbf{69.89\%} \\ \cline{1-2} \cline{4-7} \cline{9-12} \cline{14-17} 
			\multirow{4}{*}{\textbf{\begin{tabular}[c]{@{}c@{}}ClassE\\ (1280x720)\end{tabular}}}   & \textit{FourPeople}       &                                        & -0.44\%          & -0.92\%          & -0.66\%          & 33.74\%          &                      & -0.59\%          & -1.03\%          & -0.72\%          & 46.61\%          &                      & -1.00\%          & -1.89\%          & -0.80\%          & 47.41\%          \\
			& \textit{Johnny}           & \multicolumn{1}{c}{\textit{}}          & -0.29\%          & -0.20\%          & -0.16\%          & 16.64\%          &                      & -0.54\%          & -0.83\%          & -1.07\%          & 23.88\%          &                      & -0.77\%          & -2.14\%          & -1.47\%          & 36.27\%          \\
			& \textit{KristenAndSara}   & \multicolumn{1}{c}{\textit{}}          & -0.42\%          & -0.77\%          & -0.35\%          & 26.63\%          &                      & -0.61\%          & -0.88\%          & -0.89\%          & 37.65\%          &                      & -0.81\%          & -2.06\%          & -0.90\%          & 41.17\%          \\
			& \textit{\textbf{Average}} & \multicolumn{1}{c}{\textit{\textbf{}}} & \textbf{-0.38\%} & \textbf{-0.63\%} & \textbf{-0.39\%} & \textbf{25.67\%} &                      & \textbf{-0.58\%} & \textbf{-0.91\%} & \textbf{-0.89\%} & \textbf{36.04\%} &                      & \textbf{-0.86\%} & \textbf{-2.03\%} & \textbf{-1.06\%} & \textbf{41.62\%} \\ \hline
			\multicolumn{2}{c}{\textit{\textbf{Overall}}}                                                                       & \multicolumn{1}{c}{}                   & \textbf{-0.32\%} & \textbf{-0.51\%} & \textbf{-0.26\%} & \textbf{31.98\%} & \multicolumn{1}{c}{} & \textbf{-0.47\%} & \textbf{-0.95\%} & \textbf{-0.64\%} & \textbf{41.95\%} & \multicolumn{1}{c}{} & \textbf{-0.82\%} & \textbf{-2.97\%} & \textbf{-1.63\%} & \textbf{49.50\%} \\ \hline
		\end{tabular}
	}
	
	\vspace{2.5em}
	\renewcommand\arraystretch{1.05}
	\centering
	\fontsize{7.2pt}{9pt}\selectfont
	\vspace{-0.5em}
	\caption{BD-rate Results of Our Proposed LUT-ILF++ Compared to VTM-11.0 on CTC Test Sequences with Low-bitrate Points\\ (QP 27$\sim$47), and Comparison Results with the Other In-Loop Filtering Schemes \cite{li2024loop} under All Intra (AI) Configuration}
	\label{tab:DBR2}
	\vspace{-0.8em}
	\setlength{\tabcolsep}{3.1mm}
	{
		\begin{tabular}{cclcclcclcccc}
			\hline
			\multicolumn{13}{c}{\textbf{All Intra (AI) Configuration (\%)}}                                                                                                                                                                                                                                                                                                                                 \\ \hline
			\multirow{2}{*}{\textbf{Class}}                                                         & \textbf{Sequence}         &                                        & \multicolumn{2}{c}{\textbf{LUT-ILF-V \cite{li2024loop}}} &                      & \multicolumn{2}{c}{\textbf{LUT-ILF-F \cite{li2024loop}}} &                      & \multicolumn{4}{c}{\textbf{LUT-ILF++}}                  \\ \cline{2-2} \cline{4-5} \cline{7-8} \cline{10-13} 
			& Name                      &                                        & Y                      & Ratio                  & \multicolumn{1}{c}{} & Y                      & Ratio                  &                      & Y                & U       & V                & Ratio            \\ \cline{1-2} \cline{4-5} \cline{7-8} \cline{10-13} 
			\multirow{4}{*}{\textbf{\begin{tabular}[c]{@{}c@{}}ClassA1\\ (3840x2160)\end{tabular}}} & \textit{Tango2}           &                                        & -0.61\%                & 47.05\%                &                      & -0.95\%                & 65.72\%                &                      & -1.32\%          & -6.67\%          & -2.26\%          & 63.16\%          \\
			& \textit{FoodMarket4}      &                                        & -0.21\%                & 40.35\%                &                      & -0.55\%                & 53.97\%                &                      & -0.84\%          & -1.42\%          & -0.80\%          & 48.46\%          \\
			& \textit{Campfire}         &                                        & -0.64\%                & 44.45\%                &                      & -0.84\%                & 63.07\%                &                      & -1.38\%          & -0.79\%          & -2.97\%          & 59.61\%          \\
			& \textit{\textbf{Average}} &                                        & \textbf{-0.49\%}       & \textbf{43.95\%}       &                      & \textbf{-0.78\%}       & \textbf{60.92\%}       &                      & \textbf{-1.18\%} & \textbf{-2.96\%} & \textbf{-2.01\%} & \textbf{57.07\%} \\ \cline{1-2} \cline{4-5} \cline{7-8} \cline{10-13} 
			\multirow{4}{*}{\textbf{\begin{tabular}[c]{@{}c@{}}ClassA2\\ (3840x2160)\end{tabular}}} & \textit{CatRobot1}        &                                        & -0.98\%                & 52.26\%                &                      & -1.40\%                & 67.08\%                &                      & -1.63\%          & -6.94\%          & -2.78\%          & 69.73\%          \\
			& \textit{DaylightRoad2}    &                                        & -0.79\%                & 60.50\%                &                      & -1.12\%                & 64.96\%                &                      & -1.27\%          & -3.69\%          & -1.39\%          & 64.92\%          \\
			& \textit{ParkRunning3}     &                                        & -0.96\%                & 50.75\%                &                      & -1.20\%                & 65.41\%                &                      & -1.47\%          & -0.63\%          & -0.20\%          & 68.88\%          \\
			& \textit{\textbf{Average}} &                                        & \textbf{-0.91\%}       & \textbf{54.50\%}       &                      & \textbf{-1.24\%}       & \textbf{65.82\%}       &                      & \textbf{-1.45\%} & \textbf{-3.75\%} & \textbf{-1.45\%} & \textbf{67.85\%} \\ \cline{1-2} \cline{4-5} \cline{7-8} \cline{10-13} 
			\multirow{6}{*}{\textbf{\begin{tabular}[c]{@{}c@{}}ClassB\\ (1920x1080)\end{tabular}}}  & \textit{MarketPlace}      &                                        & -0.68\%                & 51.74\%                &                      & -1.08\%                & 67.37\%                &                      & -1.32\%          & -6.00\%          & -1.70\%          & 66.92\%          \\
			& \textit{RitualDance}      &                                        & -0.54\%                & 49.10\%                &                      & -1.02\%                & 65.60\%                &                      & -1.38\%          & -4.19\%          & -3.93\%          & 70.45\%          \\
			& \textit{Cactus}           &                                        & -0.43\%                & 41.77\%                &                      & -0.73\%                & 55.19\%                &                      & -1.32\%          & -7.39\%          & -2.17\%          & 57.31\%          \\
			& \textit{BasketballDrive}  &                                        & -0.10\%                & 25.65\%                &                      & -0.24\%                & 38.51\%                &                      & -1.37\%          & -3.28\%          & -1.64\%          & 37.85\%          \\
			& \textit{BQTerrace}        &                                        & -0.66\%                & 44.98\%                &                      & -0.76\%                & 50.63\%                &                      & -1.03\%          & -4.76\%          & -2.68\%          & 52.98\%          \\
			& \textit{\textbf{Average}} &                                        & \textbf{-0.48\%}       & \textbf{42.65\%}       &                      & \textbf{-0.76\%}       & \textbf{55.46\%}       &                      & \textbf{-1.28\%} & \textbf{-5.12\%} & \textbf{-2.42\%} & \textbf{57.11\%} \\ \cline{1-2} \cline{4-5} \cline{7-8} \cline{10-13} 
			\multirow{5}{*}{\textbf{\begin{tabular}[c]{@{}c@{}}ClassC\\ (832x480)\end{tabular}}}    & \textit{BasketballDrill}  &                                        & -0.77\%                & 57.23\%                &                      & -1.00\%                & 62.40\%                &                      & -1.50\%          & -3.92\%          & -2.36\%          & 70.33\%          \\
			& \textit{BQMall}           & \multicolumn{1}{c}{\textit{}}          & -0.70\%                & 60.61\%                &                      & -1.02\%                & 73.47\%                &                      & -1.55\%          & -7.64\%          & -2.91\%          & 83.14\%          \\
			& \textit{PartyScene}       & \multicolumn{1}{c}{\textit{}}          & -0.53\%                & 59.64\%                &                      & -0.71\%                & 68.09\%                &                      & -1.23\%          & -8.54\%          & -2.08\%          & 84.67\%          \\
			& \textit{RaceHorsesC}      & \multicolumn{1}{c}{\textit{}}          & -0.55\%                & 47.73\%                &                      & -0.64\%                & 69.88\%                &                      & -0.83\%          & -6.55\%          & -3.30\%          & 61.62\%          \\
			& \textit{\textbf{Average}} & \multicolumn{1}{c}{\textit{\textbf{}}} & \textbf{-0.64\%}       & \textbf{56.30\%}       &                      & \textbf{-0.84\%}       & \textbf{68.46\%}       &                      & \textbf{-1.28\%} & \textbf{-6.66\%} & \textbf{-2.66\%} & \textbf{74.94\%} \\ \cline{1-2} \cline{4-5} \cline{7-8} \cline{10-13} 
			\multirow{5}{*}{\textbf{\begin{tabular}[c]{@{}c@{}}ClassD\\ (416x240)\end{tabular}}}    & \textit{BasketballPass}   &                                        & -0.97\%                & 64.91\%                &                      & -1.37\%                & 78.58\%                &                      & -2.09\%          & -5.30\%          & -6.03\%          & 88.08\%          \\
			& \textit{BQSquare}         &                                        & -1.11\%                & 65.08\%                &                      & -1.41\%                & 79.08\%                &                      & -2.35\%          & -2.07\%          & -4.42\%          & 89.58\%          \\
			& \textit{BlowingBubbles}   &                                        & -0.69\%                & 67.50\%                &                      & -0.93\%                & 72.91\%                &                      & -1.47\%          & -6.00\%          & -1.45\%          & 86.92\%          \\
			& \textit{RaceHorses}       &                                        & -1.53\%                & 79.25\%                &                      & -1.87\%                & 84.91\%                &                      & -2.43\%          & -7.69\%          & -4.39\%          & 94.83\%          \\
			& \textit{\textbf{Average}} &                                        & \textbf{-1.08\%}       & \textbf{69.18\%}       &                      & \textbf{-1.40\%}       & \textbf{78.87\%}       &                      & \textbf{-2.09\%} & \textbf{-5.26\%} & \textbf{-4.07\%} & \textbf{89.85\%} \\ \cline{1-2} \cline{4-5} \cline{7-8} \cline{10-13} 
			\multirow{4}{*}{\textbf{\begin{tabular}[c]{@{}c@{}}ClassE\\ (1280x720)\end{tabular}}}   & \textit{FourPeople}       &                                        & -1.04\%                & 56.50\%                &                      & -1.47\%                & 71.10\%                &                      & -1.78\%          & -3.25\%          & -1.99\%          & 67.28\%          \\
			& \textit{Johnny}           & \multicolumn{1}{c}{\textit{}}          & -0.73\%                & 33.14\%                &                      & -1.12\%                & 46.24\%                &                      & -1.42\%          & -3.09\%          & -2.77\%          & 55.11\%          \\
			& \textit{KristenAndSara}   & \multicolumn{1}{c}{\textit{}}          & -1.08\%                & 45.24\%                &                      & -1.36\%                & 59.49\%                &                      & -1.61\%          & -3.31\%          & -1.84\%          & 61.03\%          \\
			& \textit{\textbf{Average}} & \multicolumn{1}{c}{\textit{\textbf{}}} & \textbf{-0.95\%}       & \textbf{44.96\%}       &                      & \textbf{-1.32\%}       & \textbf{58.94\%}       &                      & \textbf{-1.60\%} & \textbf{-3.22\%} & \textbf{-2.20\%} & \textbf{61.15\%} \\ \hline
			\multicolumn{2}{c}{\textit{\textbf{Overall}}}                                                                       & \multicolumn{1}{c}{}                   & \textbf{-0.74\%}       & \textbf{51.92\%}       & \multicolumn{1}{c}{} & \textbf{-1.03\%}       & \textbf{64.74\%}       & \multicolumn{1}{c}{} & \textbf{-1.49\%} & \textbf{-4.69\%} & \textbf{-2.55\%} & \textbf{67.99\%} \\ \hline
		\end{tabular}
	}
	\vspace{0.3em}
\end{table*}

\begin{table*}
	\renewcommand\arraystretch{1.2}
	\centering
	\fontsize{7.2pt}{9pt}\selectfont
	\vspace{-3.2em}
	\caption{BD-rate Results of Our Proposed LUT-ILF++ Compared to VTM-11.0 on CTC Test Sequences with Regular-bitrate Points \\(QP 22$\sim$42), and Comparison Results with the Other In-Loop Filtering Schemes \cite{li2024loop} under Random Access (RA) Configuration}
	\label{tab:DBR3}
	\vspace{-0.9em}
	\setlength{\tabcolsep}{1.5mm}
	{
		\begin{tabular}{cclcccclccclccc}
			\hline
			\multicolumn{14}{c}{\textbf{Random Access (RA) Configuration (\%)}} \\ \hline
			\multirow{2}{*}{\textbf{Class}} & \textbf{Sequence} & & \multicolumn{3}{c}{\textbf{LUT-ILF-V~\cite{li2024loop}}} & & \multicolumn{3}{c}{\textbf{LUT-ILF-F~\cite{li2024loop}}} & & \multicolumn{3}{c}{\textbf{LUT-ILF++}} \\ 
			\cline{2-2} \cline{4-6} \cline{8-10} \cline{12-14}
			& Name & & Y & U & V & & \multicolumn{1}{c}{Y} & U & V & & \multicolumn{1}{c}{Y} & U & V \\ 
			\cline{1-2} \cline{4-6} \cline{8-10} \cline{12-14}
			\multirow{4}{*}{\textbf{\begin{tabular}[c]{@{}c@{}}ClassA1\\ (3840x2160)\end{tabular}}} 
			& \textit{Tango2} & & -0.27\% & -0.26\% & -0.13\% & & -0.39\% & -0.36\% & -0.31\% & & -0.61\% & -6.82\% & -1.55\% \\
			& \textit{FoodMarket4} & & -0.13\% & -0.92\% & -0.16\% & & -0.10\% & -0.80\% & -0.43\% & & -0.58\% & -1.23\% & -0.42\% \\
			& \textit{Campfire} & & -0.20\% & -0.16\% & -0.05\% & & -0.14\% & -0.07\% & -0.25\% & & -1.13\% & -0.69\% & -2.75\% \\
			& \textit{\textbf{Average}} & & \textbf{-0.20\%} & \textbf{-0.45\%} & \textbf{-0.12\%} & & \textbf{-0.21\%} & \textbf{-0.41\%} & \textbf{-0.33\%} & & \textbf{-0.77\%} & \textbf{-2.91\%} & \textbf{-1.57\%} \\ 
			\cline{1-2} \cline{4-6} \cline{8-10} \cline{12-14}
			\multirow{4}{*}{\textbf{\begin{tabular}[c]{@{}c@{}}ClassA2\\ (3840x2160)\end{tabular}}} 
			& \textit{CatRobot1} & & -0.30\% & -0.71\% & -0.51\% & & -0.49\% & -0.74\% & -0.28\% & & -0.90\% & -6.12\% & -2.34\% \\
			& \textit{DaylightRoad2} & & -0.25\% & -0.44\% & -0.31\% & & -0.43\% & -0.71\% & -0.30\% & & -0.79\% & -5.24\% & -0.40\% \\
			& \textit{ParkRunning3} & & -0.21\% & -0.38\% & 0.07\% & & -0.13\% & -0.30\% & -0.12\% & & -0.90\% & -0.72\% & -0.10\% \\
			& \textit{\textbf{Average}} & & \textbf{-0.25\%} & \textbf{-0.51\%} & \textbf{-0.25\%} & & \textbf{-0.35\%} & \textbf{-0.58\%} & \textbf{-0.23\%} & & \textbf{-0.86\%} & \textbf{-4.03\%} & \textbf{-0.95\%} \\ 
			\cline{1-2} \cline{4-6} \cline{8-10} \cline{12-14}
			\multirow{6}{*}{\textbf{\begin{tabular}[c]{@{}c@{}}ClassB\\ (1920x1080)\end{tabular}}} 
			& \textit{MarketPlace} & & -0.30\% & -0.48\% & -0.14\% & & -0.66\% & -0.65\% & -0.79\% & & -0.99\% & -5.23\% & -2.72\% \\
			& \textit{RitualDance} & & -0.23\% & -0.38\% & -0.04\% & & -0.20\% & -0.39\% & -0.74\% & & -0.92\% & -4.83\% & -3.26\% \\
			& \textit{Cactus} & & -0.25\% & -0.42\% & -0.07\% & & -0.37\% & -0.57\% & -0.88\% & & -0.86\% & -6.90\% & -2.48\% \\
			& \textit{BasketballDrive} & & -0.10\% & -0.25\% & -0.11\% & & -0.19\% & -0.04\% & -0.55\% & & -0.45\% & -2.48\% & -1.65\% \\
			& \textit{BQTerrace} & & -0.19\% & -0.57\% & -0.30\% & & -0.15\% & -0.01\% & -0.51\% & & -0.53\% & -3.82\% & -2.06\% \\
			& \textit{\textbf{Average}} & & \textbf{-0.21\%} & \textbf{-0.42\%} & \textbf{-0.14\%} & & \textbf{-0.31\%} & \textbf{-0.33\%} & \textbf{-0.70\%} & & \textbf{-0.75\%} & \textbf{-4.65\%} & \textbf{-2.44\%} \\ 
			\cline{1-2} \cline{4-6} \cline{8-10} \cline{12-14}
			\multirow{5}{*}{\textbf{\begin{tabular}[c]{@{}c@{}}ClassC\\ (832x480)\end{tabular}}} 
			& \textit{BasketballDrill} & & -0.23\% & -0.66\% & -0.19\% & & -0.25\% & -0.75\% & -0.11\% & & -0.45\% & -2.61\% & -1.71\% \\
			& \textit{BQMall} & & -0.13\% & -0.73\% & -0.04\% & & -0.27\% & -1.41\% & -0.35\% & & -0.75\% & -5.82\% & -2.06\% \\
			& \textit{PartyScene} & & -0.09\% & -0.51\% & -0.66\% & & -0.17\% & -0.76\% & -0.70\% & & -0.54\% & -6.96\% & -1.68\% \\
			& \textit{RaceHorsesC} & & -0.13\% & -0.59\% & -0.14\% & & -0.09\% & -1.24\% & 0.09\% & & -0.43\% & -6.30\% & -3.38\% \\
			& \textit{\textbf{Average}} & & \textbf{-0.14\%} & \textbf{-0.62\%} & \textbf{-0.26\%} & & \textbf{-0.19\%} & \textbf{-1.04\%} & \textbf{-0.26\%} & & \textbf{-0.54\%} & \textbf{-5.42\%} & \textbf{-2.21\%} \\ 
			\cline{1-2} \cline{4-6} \cline{8-10} \cline{12-14}
			\multirow{5}{*}{\textbf{\begin{tabular}[c]{@{}c@{}}ClassD\\ (416x240)\end{tabular}}} 
			& \textit{BasketballPass} & & -0.38\% & -0.24\% & -0.38\% & & -0.41\% & -0.96\% & -0.30\% & & -1.04\% & -1.47\% & -4.25\% \\
			& \textit{BQSquare} & & -0.48\% & -0.95\% & -0.51\% & & -0.53\% & -1.01\% & -0.34\% & & -1.34\% & -1.89\% & -2.94\% \\
			& \textit{BlowingBubbles} & & -0.22\% & -0.66\% & -0.65\% & & -0.38\% & -0.89\% & -0.57\% & & -0.77\% & -6.33\% & -0.44\% \\
			& \textit{RaceHorses} & & -0.30\% & -0.73\% & -0.16\% & & -0.55\% & -1.12\% & -0.52\% & & -1.59\% & -8.19\% & -4.33\% \\
			& \textit{\textbf{Average}} & & \textbf{-0.35\%} & \textbf{-0.65\%} & \textbf{-0.41\%} & & \textbf{-0.46\%} & \textbf{-0.99\%} & \textbf{-0.44\%} & & \textbf{-1.18\%} & \textbf{-4.47\%} & \textbf{-2.99\%} \\ 
			\cline{1-2} \cline{4-6} \cline{8-10} \cline{12-14}
			\multirow{4}{*}{\textbf{\begin{tabular}[c]{@{}c@{}}ClassE\\ (1280x720)\end{tabular}}} 
			& \textit{FourPeople} & & -0.51\% & -0.46\% & -0.64\% & & -0.52\% & -0.97\% & -0.90\% & & -1.23\% & -2.02\% & -1.10\% \\
			& \textit{Johnny} & & -0.35\% & 0.38\% & -0.19\% & & -0.61\% & -0.10\% & -0.56\% & & -0.77\% & -1.97\% & -1.98\% \\
			& \textit{KristenAndSara} & & -0.46\% & -0.35\% & 0.53\% & & -0.73\% & -0.43\% & -0.41\% & & -0.84\% & -2.79\% & -1.62\% \\
			& \textit{\textbf{Average}} & & \textbf{-0.44\%} & \textbf{-0.14\%} & \textbf{-0.10\%} & & \textbf{-0.62\%} & \textbf{-0.50\%} & \textbf{-0.61\%} & & \textbf{-0.95\%} & \textbf{-2.26\%} & \textbf{-1.57\%} \\ \hline
			\multicolumn{2}{c}{\textit{\textbf{Overall}}} & & \textbf{-0.26\%} & \textbf{-0.47\%} & \textbf{-0.22\%} & & \textbf{-0.35\%} & \textbf{-0.65\%} & \textbf{-0.44\%} & & \textbf{-0.85\%} & \textbf{-4.11\%} & \textbf{-2.06\%} \\ \hline
		\end{tabular}
	}
	
	\vspace{2em}
	\renewcommand\arraystretch{1.1}
	\centering
	\fontsize{7.2pt}{9pt}\selectfont
	\vspace{-0.5em}
	\caption{BD-rate Results of Our Proposed LUT-ILF++ Compared to VTM-11.0 on CTC Test Sequences with Low-bitrate Points\\(QP 27$\sim$47), and Comparison Results with the Other In-Loop Filtering Schemes \cite{li2024loop} under Random Access (RA) Configuration}
	\label{tab:DBR4}
	\vspace{-0.8em}
	\setlength{\tabcolsep}{2.5mm}
	{
		\begin{tabular}{cccccccccc}
			\hline
			\multicolumn{10}{c}{\textbf{Random Access (RA) Configuration (\%)}} \\ \hline
			\multirow{2}{*}{\textbf{Class}} & \textbf{Sequence} & & \multicolumn{1}{c}{\textbf{LUT-ILF-V~\cite{li2024loop}}} & & \multicolumn{1}{c}{\textbf{LUT-ILF-F~\cite{li2024loop}}} && \multicolumn{3}{c}{\textbf{LUT-ILF++}} \\ 
			\cline{2-2} \cline{4-4} \cline{6-6} \cline{8-10}
			& Name & & \multicolumn{1}{c}{Y} & & \multicolumn{1}{c}{Y} && \multicolumn{1}{c}{Y} & U & V \\ 
			\cline{1-2} \cline{4-4} \cline{6-6}  \cline{8-10}
			\multirow{4}{*}{\textbf{\begin{tabular}[c]{@{}c@{}}ClassA1\\ (3840x2160)\end{tabular}}} 
			& \textit{Tango2}       & & -0.52\% & & -0.75\% && -0.91\% & -7.32\% & -1.14\% \\
			& \textit{FoodMarket4}  & & -0.24\% & & -0.36\% && -0.70\% & -1.17\% & -0.58\% \\
			& \textit{Campfire}     & & -0.21\% & & -0.24\% && -1.40\% & -1.06\% & -2.24\% \\
			& \textit{\textbf{Average}} & & \textbf{-0.33\%} & & \textbf{-0.45\%} && \textbf{-1.00\%} & \textbf{-3.19\%} & \textbf{-1.32\%} \\ 
			\cline{1-2} \cline{4-4} \cline{6-6}  \cline{8-10}
			\multirow{4}{*}{\textbf{\begin{tabular}[c]{@{}c@{}}ClassA2\\ (3840x2160)\end{tabular}}} 
			& \textit{CatRobot}       & & -0.63\% & & -0.56\% && -1.17\% & -6.54\% & -3.00\% \\
			& \textit{DaylightRoad2}  & & -0.28\% & & -0.94\% && -1.06\% & -4.67\% & -0.13\% \\
			& \textit{ParkRunning3}   & & -0.23\% & & -0.40\% && -1.35\% & -1.03\% & -0.20\% \\
			& \textit{\textbf{Average}} & & \textbf{-0.38\%} && \textbf{-0.64\%} && \textbf{-1.19\%} & \textbf{-4.08\%} & \textbf{-1.11\%} \\ 
			\cline{1-2} \cline{4-4} \cline{6-6}  \cline{8-10}
			\multirow{6}{*}{\textbf{\begin{tabular}[c]{@{}c@{}}ClassB\\ (1920x1080)\end{tabular}}} 
			& \textit{MarketPlace}      & & -0.41\% && -0.73\% && -1.08\% & -4.86\% & -2.93\% \\
			& \textit{RitualDance}      & & -0.26\% && -0.40\% && -1.05\% & -4.76\% & -2.90\% \\
			& \textit{Cactus}           & & -0.24\% && -0.44\% && -0.94\% & -7.55\% & -2.36\% \\
			& \textit{BasketballDrive}  & & -0.49\% && -0.30\% && -0.82\% & -2.28\% & -1.39\% \\
			& \textit{BQTerrace}        & & -0.39\% && -0.33\% && -1.02\% & -4.59\% & -2.11\% \\
			& \textit{\textbf{Average}} & & \textbf{-0.36\%} && \textbf{-0.44\%} && \textbf{-0.98\%} & \textbf{-4.81\%} & \textbf{-2.34\%} \\ 
			\cline{1-2} \cline{4-4} \cline{6-6}  \cline{8-10}
			\multirow{5}{*}{\textbf{\begin{tabular}[c]{@{}c@{}}ClassC\\ (832x480)\end{tabular}}} 
			& \textit{BasketballDrill} & & -0.42\% && -0.66\% && -0.76\% & -2.28\% & -1.80\% \\
			& \textit{BQMall}          & & -0.23\% && -0.63\% && -1.06\% & -7.08\% & -2.85\% \\
			& \textit{PartyScene}      & & -0.11\% && -0.24\% && -0.88\% & -9.73\% & -2.25\% \\
			& \textit{RaceHorses}      & & -0.38\% && -0.39\% && -0.99\% & -7.20\% & -2.76\% \\
			& \textit{\textbf{Average}} & & \textbf{-0.28\%} && \textbf{-0.47\%} && \textbf{-0.92\%} & \textbf{-6.57\%} & \textbf{-2.42\%} \\ 
			\cline{1-2} \cline{4-4} \cline{6-6}  \cline{8-10}
			\multirow{5}{*}{\textbf{\begin{tabular}[c]{@{}c@{}}ClassD\\ (416x240)\end{tabular}}} 
			& \textit{BasketballPass} & & -0.52\% && -0.74\% && -1.40\% & -1.98\% & -4.42\% \\
			& \textit{BQSquare}       & & -0.36\% && -0.62\% && -2.18\% & -1.86\% & -4.82\% \\
			& \textit{BlowingBubbles} & & -0.42\% && -0.47\% && -1.24\% & -7.49\% & -0.86\% \\
			& \textit{RaceHorses}     & & -0.47\% && -0.57\% && -1.93\% & -8.89\% & -3.68\% \\
			& \textit{\textbf{Average}} & & \textbf{-0.44\%} && \textbf{-0.59\%} && \textbf{-1.69\%} & \textbf{-5.06\%} & \textbf{-3.45\%} \\ 
			\cline{1-2} \cline{4-4} \cline{6-6}  \cline{8-10}
			\multirow{4}{*}{\textbf{\begin{tabular}[c]{@{}c@{}}ClassE\\ (1280x720)\end{tabular}}} 
			& \textit{FourPeople}      & & -0.75\% && -0.76\% && -1.69\% & -2.22\% & -1.41\% \\
			& \textit{Johnny}          & & -0.36\% && -0.84\% && -1.16\% & -2.30\% & -2.93\% \\
			& \textit{KristenAndSara}  & & -0.53\% && -0.81\% && -1.45\% & -3.10\% & -1.89\% \\
			& \textit{\textbf{Average}} & & \textbf{-0.55\%} && \textbf{-0.80\%} && \textbf{-1.43\%} & \textbf{-2.54\%} & \textbf{-2.08\%} \\ \hline
			\multicolumn{2}{c}{\textit{\textbf{Overall}}} & & \textbf{-0.39\%} && \textbf{-0.57\%} && \textbf{-1.21\%} & \textbf{-4.54\%} & \textbf{-2.21\%} \\ \hline
		\end{tabular}
	}
	\vspace{0.3em}
\end{table*}

\begin{figure*}
	\centering
	\vspace{-1.6em}
	\includegraphics[width=165mm]{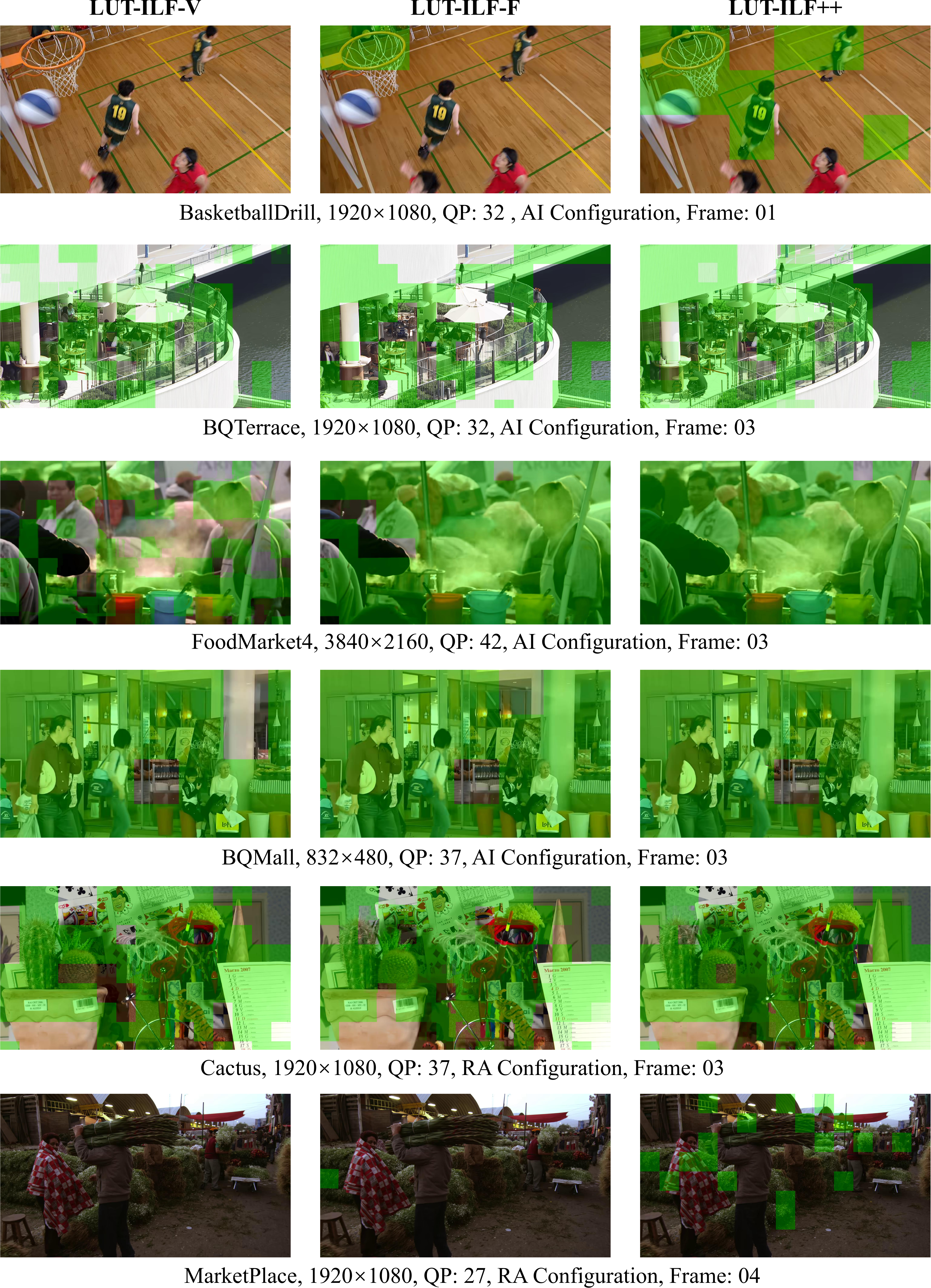}
	\vspace{-0.3em}
	\caption{The filter selection comparison results of different LUT-based ILF schemes \cite{li2024loop} (LUT-ILF-V, LUT-ILF-F) and the proposed LUT-ILF++ on several test sequences with rich textures and complex scenes under AI or RA configuration are presented. For each sequence, we select the most significant comparisons from the first five coding frames to demonstrate the advantages of our proposed LUT-ILF++. The green blocks indicate the regions filtered by the corresponding schemes, and a larger area covered in green indicates a higher usage ratio of the corresponding filter.}
	\label{fig:partition}
	\vspace{-1.2em}
\end{figure*}

\begin{figure*}
	\centering
	\vspace{-1.6em}
	\includegraphics[width=160mm]{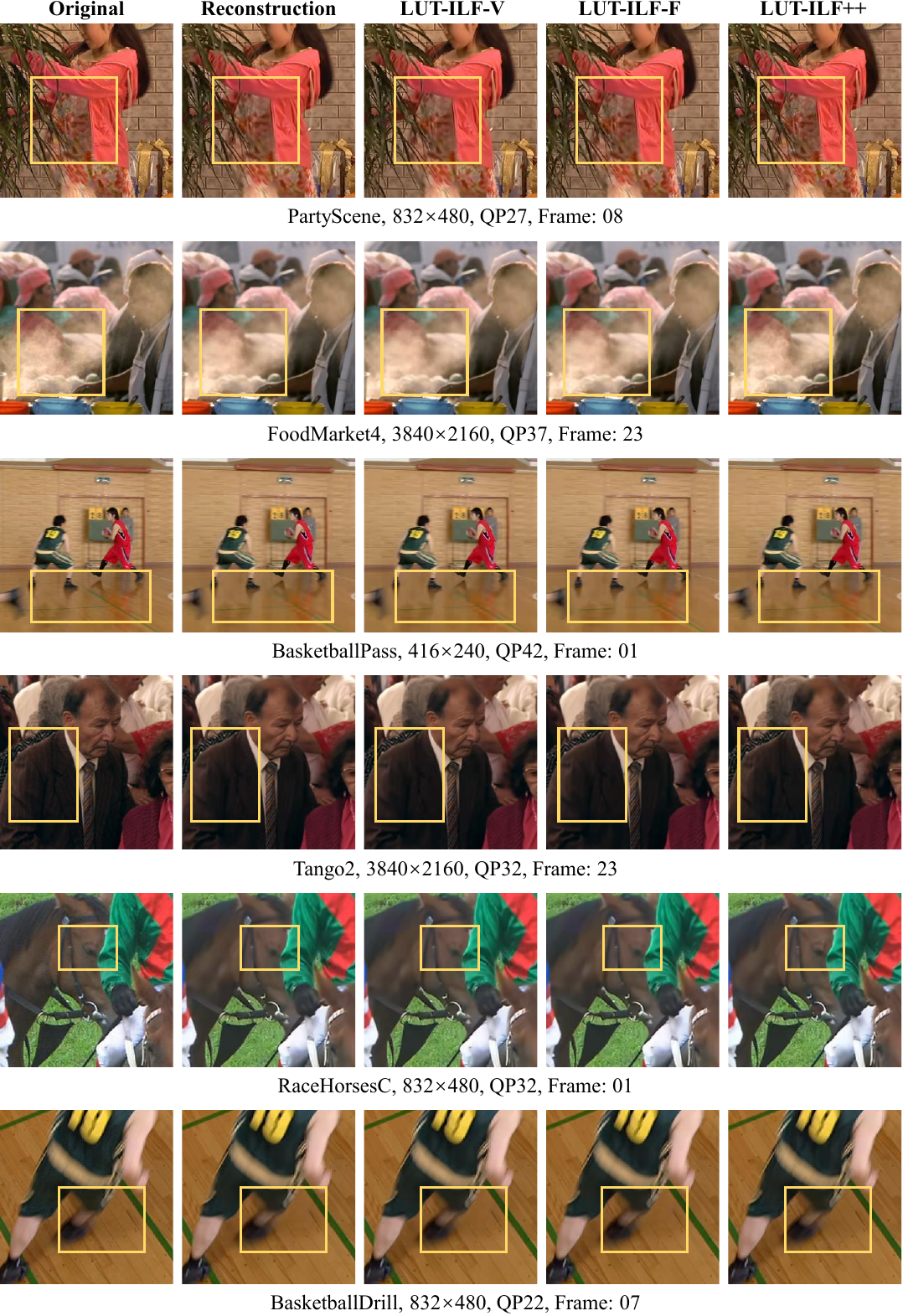}
	\vspace{-0.6em}
	\caption{The subjective comparison results of original frames, reconstructed frames, the filtered results of different LUT-based ILF schemes \cite{li2024loop} (LUT-ILF-V, LUT-ILF-F), and the filtered results of the proposed LUT-ILF++ under AI or RA configuration are presented. For each sequence, we select the most significant comparisons to demonstrate the advantages of our proposed LUT-ILF++. The yellow blocks indicate the regions filtered by the corresponding filtering schemes.}
	\label{fig:subjective}
	\vspace{-1.2em}
\end{figure*}

\section{Supplementary Experiments and Analyses}
In this section, first, we supplement the detailed experimental results of \textit{LUT-ILF++}, and detailed comparison results of each sequence with different LUT-based ILF schemes\cite{li2024loop}, including the detailed BD-rate results of each sequence on regular and low bitrate points, the filter usage ratio and selection results, and subjective exhibition. Second, we supplement some detailed experimental analyses, including the deep analyses of LUT finetuning of different LUT-based ILF schemes\cite{li2024loop} and LUT-ILF++, the detailed calculation manner of computational complexity and energy cost, and the deployment discussion of LUT-based ILF solution integrated into the codec.

\begin{table*}
	\renewcommand\arraystretch{1.8}
	\centering
	\fontsize{7.7pt}{9pt}\selectfont
	\vspace{-3em}
	\caption{The Computational Complexity Results and Specific Operation Num of \textbf{Luma} Component of Our Basic Filtering Framework (\textit{LUT-ILF-U/V/F}) and Our Proposed Improved Luma Filtering Framework (LUT-ILF++)  on the Pixel (per pixel) and Frame level\\ (a 1920 × 1080 HD frame)}
	\label{tab:cc}
	\vspace{-0.8em}
	\begin{threeparttable}
		\setlength{\tabcolsep}{0.5mm}
		{
			\begin{tabular}{ccccccc}
				\hline
				\multicolumn{1}{c|}{\textbf{Operation}}                                                                                   & \multicolumn{1}{c|}{\textbf{Level}}                       & \multicolumn{1}{c|}{\begin{tabular}[c]{@{}c@{}}\textit{\textbf{\textit{Operation Num of}}} \\ \textit{\textbf{LUT-ILF-U}}\end{tabular}} & \multicolumn{1}{c|}{\begin{tabular}[c]{@{}c@{}}\textbf{\textit{Operation Num of}} \\ \textit{\textbf{LUT-ILF-V}}\end{tabular}} & \multicolumn{1}{c|}{\begin{tabular}[c]{@{}c@{}}\textbf{\textit{Operation Num of}} \\ \textit{\textbf{LUT-ILF-F}}\end{tabular}} & \multicolumn{1}{c|}{\textit{\begin{tabular}[c]{@{}c@{}}\textbf{Operation Num of} \\ \textbf{LUT-ILF++ (w/o Compaction)}\end{tabular}}} & \textit{\begin{tabular}[c]{@{}c@{}}\textbf{Operation Num of} \\ \textbf{LUT-ILF++ (w/ Compaction)}\end{tabular}} \\ \hline
				\multicolumn{1}{c|}{int8 Add}                                                                                             & \multicolumn{1}{c|}{\multirow{6}{*}{Pixel-wise}}          & \multicolumn{1}{c|}{70}                          & \multicolumn{1}{c|}{206} & \multicolumn{1}{c|}{478} & \multicolumn{1}{c|}{1001} &  1553                      \\ \cline{1-1}\cline{3-7}
				\multicolumn{1}{c|}{int8 Multiply}                                                                                        & \multicolumn{1}{c|}{}                                     & \multicolumn{1}{c|}{4}                           & \multicolumn{1}{c|}{4} & \multicolumn{1}{c|}{4}   & \multicolumn{1}{c|}{14}   & 14                          \\ \cline{1-1}\cline{3-7}
				\multicolumn{1}{c|}{int32 Add}                                                                                            & \multicolumn{1}{c|}{}                                     & \multicolumn{1}{c|}{68}                          & \multicolumn{1}{c|}{190}  & \multicolumn{1}{c|}{446}  & \multicolumn{1}{c|}{1107
				}  & 1107                          \\ \cline{1-1}\cline{3-7}
				\multicolumn{1}{c|}{int32 Multiply}                                                                                       & \multicolumn{1}{c|}{}                                     & \multicolumn{1}{c|}{55}                          & \multicolumn{1}{c|}{152}   & \multicolumn{1}{c|}{344}  & \multicolumn{1}{c|}{919
				}     & 919                       \\ \cline{1-1}\cline{3-7}
				\multicolumn{1}{c|}{Total Add}                                                                                            & \multicolumn{1}{c|}{}                                     & \multicolumn{1}{c|}{138}                         & \multicolumn{1}{c|}{396}  & \multicolumn{1}{c|}{924}  & \multicolumn{1}{c|}{2108
				}  & 2660                             \\ \cline{1-1}\cline{3-7}
				\multicolumn{1}{c|}{Total Multiply}                                                                                       & \multicolumn{1}{c|}{}                                     & \multicolumn{1}{c|}{59}                          & \multicolumn{1}{c|}{156}  & \multicolumn{1}{c|}{348}   & \multicolumn{1}{c|}{933
				}   & 933                            \\ \hline
				\multicolumn{1}{c|}{int8 Add}                                                                                             & \multicolumn{1}{c|}{\multirow{4}{*}{Frame-wise}}          & \multicolumn{1}{c|}{145,152,000}                 & \multicolumn{1}{c|}{427,161,600}     & \multicolumn{1}{c|}{991,180,800}  & \multicolumn{1}{c|}{2,075,673,600} & 3,220,300,800
				\\ \cline{1-1}\cline{3-7}
				\multicolumn{1}{c|}{int8 Multiply}                                                                                        & \multicolumn{1}{c|}{}                                     & \multicolumn{1}{c|}{8,294,400} & \multicolumn{1}{c|}{8,294,400} & \multicolumn{1}{c|}{8,294,400} & \multicolumn{1}{c|}{29,030,400
				}                   & 29,030,400
				\\ \cline{1-1}\cline{3-7}
				\multicolumn{1}{c|}{int32 Add}                                                                                            & \multicolumn{1}{c|}{}                                     & \multicolumn{1}{c|}{141,004,800}                 & \multicolumn{1}{c|}{393,984,000} & \multicolumn{1}{c|}{924,825,600} & \multicolumn{1}{c|}{2,295,475,200
					
				} & 2,295,475,200
				\\ \cline{1-1}\cline{3-7}
				\multicolumn{1}{c|}{int32 Multiply}                                                                                       & \multicolumn{1}{c|}{}                                     & \multicolumn{1}{c|}{114,048,000}                 & \multicolumn{1}{c|}{315,187,200}  & \multicolumn{1}{c|}{713,318,400
				} & \multicolumn{1}{c|}{1,905,638,400
				} & 1,905,638,400
				\\ \hline
				\multicolumn{1}{c|}{\textbf{Total Add}}                                                                                   & \multicolumn{1}{c|}{\multirow{2}{*}{\textbf{Frame-wise}}} & \multicolumn{1}{c|}{\textbf{286,156,800}}        & \multicolumn{1}{c|}{\textbf{821,145,600}}   & \multicolumn{1}{c|}{\textbf{1,916,006,400}}  & \multicolumn{1}{c|}{\textbf{4,371,148,800
				}} & \textbf{5,515,776,000}
				\\ \cline{1-1} \cline{3-7} 
				\multicolumn{1}{c|}{\textbf{Total Multiply}}                                                                              & \multicolumn{1}{c|}{}                                     & \multicolumn{1}{c|}{\textbf{122,342,400}}        & \multicolumn{1}{c|}{\textbf{323,481,600}} & \multicolumn{1}{c|}{\textbf{721,612,800}}  & \multicolumn{1}{c|}{\textbf{1,934,668,800
				}}  & \textbf{1,934,668,800
				}      \\ \hline
				\multicolumn{1}{c|}{\textbf{\begin{tabular}[c]{@{}c@{}} Computational\\ Complexity \,(Ops/pixel)\end{tabular}}} & \multicolumn{1}{c|}{\textbf{Pixel-wise}}                  & \multicolumn{1}{c|}{\textbf{0.20}}               & \multicolumn{1}{c|}{\textbf{0.55}} & \multicolumn{1}{c|}{\textbf{1.27}} & \multicolumn{1}{c|}{\textbf{3.04}}  & \textbf{3.59}              \\ \hline
				\multicolumn{1}{c|}{\textbf{Energy Cost\tnote{1}  \,($pJ$/pixel)}}                                                                            & \multicolumn{1}{c|}{\textbf{Pixel-wise}}                  & \multicolumn{1}{c|}{\textbf{180.2}}              & \multicolumn{1}{c|}{\textbf{497.2}} 
				& \multicolumn{1}{c|}{\textbf{1126.1}}  	& \multicolumn{1}{c|}{\textbf{2992.43
				}} & \textbf{3008.99}           \\ \hline
			\end{tabular}
		}
	\end{threeparttable}
	\begin{threeparttable}
		\begin{tablenotes}    
			\fontsize{7pt}{8pt}\selectfont
			\item[1] The energy cost is calculated according to \cite{song2021addersr, sze2017efficient, horowitz20141} mentioned in Section VII.A of main text. 
		\end{tablenotes}
	\end{threeparttable}
	
	\vspace{4em}
	\renewcommand\arraystretch{1.6}
	\centering
	\fontsize{7.7pt}{9pt}\selectfont
	\vspace{-1em}
	\caption{The Computational Complexity Results and Specific Operation Num of \textbf{Luma} Component of Our Adopted 3D/4D Channel LUT  and  Each \textbf{Chroma} Component of Our Proposed Improved Chroma Filtering Framework (LUT-ILF++) \\ on the Pixel (per pixel) and Frame level (a 1920 × 1080 HD frame)}
	\label{tab:cc2}
	\vspace{-0.8em}
	\begin{threeparttable}
		\setlength{\tabcolsep}{1.9mm}
		{
			\begin{tabular}{ccccccccc}
				\hline
				\multicolumn{1}{c|}{\textbf{Operation}}                                                                                   & \multicolumn{1}{c|}{\textbf{Level}}                       & \multicolumn{1}{c|}{\begin{tabular}[c]{@{}c@{}}\textbf{\textit{Operation Num of}} \\ \textit{\textbf{Trilinear-based}} \\ \textit{\textbf{3D Channel LUT}}\end{tabular}} & \multicolumn{1}{c|}{\begin{tabular}[c]{@{}c@{}}\textbf{\textit{Operation Num of}} \\ \textit{\textbf{4-$simplex$-based}} \\ \textit{\textbf{4D Channel LUT}}\end{tabular}} & \multicolumn{1}{c|}{\textit{\begin{tabular}[c]{@{}c@{}}\textbf{Operation Num of} \\ \textbf{LUT-ILF++ (w/o Compaction)} \\ \textbf{on Each Chroma Component}\end{tabular}}} & \textit{\begin{tabular}[c]{@{}c@{}}\textbf{Operation Num of} \\ \textbf{LUT-ILF++ (w/ Compaction)}\\ \textbf{on Each Chroma Component}\end{tabular}} \\ \hline
				\multicolumn{1}{c|}{int8 Add}                                                                                             & \multicolumn{1}{c|}{\multirow{6}{*}{Pixel-wise}}          & \multicolumn{1}{c|}{37}                          & \multicolumn{1}{c|}{21.5
				}  & \multicolumn{1}{c|}{964
					
				} &  1612
				\\ \cline{1-1}\cline{3-6}
				\multicolumn{1}{c|}{int8 Multiply}                                                                                        & \multicolumn{1}{c|}{}                                     & \multicolumn{1}{c|}{2}                           & \multicolumn{1}{c|}{2
				}   & \multicolumn{1}{c|}{20
					
				}   & 20
				\\ \cline{1-1}\cline{3-6}
				\multicolumn{1}{c|}{int32 Add}                                                                                            & \multicolumn{1}{c|}{}                                     & \multicolumn{1}{c|}{27}                          & \multicolumn{1}{c|}{28
				}   & \multicolumn{1}{c|}{882
					
				}  & 882
				
				\\ \cline{1-1}\cline{3-6}
				\multicolumn{1}{c|}{int32 Multiply}                                                                                       & \multicolumn{1}{c|}{}                                     & \multicolumn{1}{c|}{77}                          & \multicolumn{1}{c|}{27
				}   & \multicolumn{1}{c|}{761
					
				}     & 761
				
				\\ \cline{1-1}\cline{3-6}
				\multicolumn{1}{c|}{Total Add}                                                                                            & \multicolumn{1}{c|}{}                                     & \multicolumn{1}{c|}{64}                         & \multicolumn{1}{c|}{49.5
				}  & \multicolumn{1}{c|}{1846
					
				}  & 2494
				\\ \cline{1-1}\cline{3-6}
				\multicolumn{1}{c|}{Total Multiply}                                                                                       & \multicolumn{1}{c|}{}                                     & \multicolumn{1}{c|}{79}                          & \multicolumn{1}{c|}{29
				}   & \multicolumn{1}{c|}{781
					
				}   & 781
				
				\\ \hline
				\multicolumn{1}{c|}{int8 Add}                                                                                             & \multicolumn{1}{c|}{\multirow{4}{*}{Frame-wise}}          & \multicolumn{1}{c|}{76,723,200
				}                 & \multicolumn{1}{c|}{44,582,400
				}      & \multicolumn{1}{c|}{499,737,600
				} & 835,660,800

				\\ \cline{1-1}\cline{3-6}
				\multicolumn{1}{c|}{int8 Multiply}                                                                                        & \multicolumn{1}{c|}{}                                     & \multicolumn{1}{c|}{4,147,200
				}  & \multicolumn{1}{c|}{4,147,200
				} & \multicolumn{1}{c|}{10,368,000
				}                   & 10,368,000

				\\ \cline{1-1}\cline{3-6}
				\multicolumn{1}{c|}{int32 Add}                                                                                            & \multicolumn{1}{c|}{}                                     & \multicolumn{1}{c|}{55,987,200
				}                 & \multicolumn{1}{c|}{58,060,800
				} & \multicolumn{1}{c|}{457,228,800

				} & 457,228,800

				\\ \cline{1-1}\cline{3-6}
				\multicolumn{1}{c|}{int32 Multiply}                                                                                       & \multicolumn{1}{c|}{}                                     & \multicolumn{1}{c|}{159,667,200
				}                 & \multicolumn{1}{c|}{55,987,200
				}  & \multicolumn{1}{c|}{394,502,400
					
				} & 394,502,400

				\\ \hline
				\multicolumn{1}{c|}{\textbf{Total Add}}                                                                                   & \multicolumn{1}{c|}{\multirow{2}{*}{\textbf{Frame-wise}}} & \multicolumn{1}{c|}{\textbf{132,710,400
				}}        & \multicolumn{1}{c|}{\textbf{102,643,200
				}}     & \multicolumn{1}{c|}{\textbf{956,966,400}} & \textbf{1,292,889,600
				}
				\\ \cline{1-1} \cline{3-6} 
				\multicolumn{1}{c|}{\textbf{Total Multiply}}                                                                              & \multicolumn{1}{c|}{}                                     & \multicolumn{1}{c|}{\textbf{163,814,400
				}}        & \multicolumn{1}{c|}{\textbf{60,134,400
				}}  & \multicolumn{1}{c|}{\textbf{404,870,400
				}}  & \textbf{404,870,400
				}      \\ \hline
				\multicolumn{1}{c|}{\textbf{\begin{tabular}[c]{@{}c@{}} Computational\\ Complexity \,(Ops/pixel)\end{tabular}}} & \multicolumn{1}{c|}{\textbf{Pixel-wise}}                  & \multicolumn{1}{c|}{\textbf{0.14}}               & \multicolumn{1}{c|}{\textbf{0.08}} & \multicolumn{1}{c|}{\textbf{0.66}}  & \textbf{0.82}              \\ \hline
				\multicolumn{1}{c|}{\textbf{Energy Cost\tnote{1}  \,($pJ$/pixel)}}                                                                            & \multicolumn{1}{c|}{\textbf{Pixel-wise}}                  & \multicolumn{1}{c|}{\textbf{242.91
				}}              & \multicolumn{1}{c|}{\textbf{87.55}} 
				& \multicolumn{1}{c|}{\textbf{620.06
				}} & \textbf{624.92}           \\ \hline
			\end{tabular}
		}
	\end{threeparttable}
	\begin{threeparttable}
		\begin{tablenotes}    
			\fontsize{7pt}{8pt}\selectfont
			\item[1] The energy cost is calculated according to \cite{song2021addersr, sze2017efficient, horowitz20141} mentioned in Section VII.A of main text. 
		\end{tablenotes}
	\end{threeparttable}
	\vspace{-1.5em}
\end{table*}

\begin{table}
	\renewcommand\arraystretch{1.1}
	\centering
	\fontsize{6.5pt}{9pt}\selectfont
	\caption{The Ablation Study of LUT Fintuning of Different LUT-based ILF Models under All Intra (AI) Configuration}
	\label{tab:Fintuning}
	\vspace{-0.7em}
	\setlength{\tabcolsep}{0.8mm}
	{
		\begin{tabular}{c|c|c|c|c|c}
			\hline
			\textbf{Schemes} & \textbf{\begin{tabular}[c]{@{}c@{}}One-Step \\ Finetuning\end{tabular}}  & \textbf{\begin{tabular}[c]{@{}c@{}}Two-Step \\ Finetuning\end{tabular}}  & \textbf{Y BD-rate (\%)} & \textbf{\begin{tabular}[c]{@{}c@{}}Training \\ Iteration \end{tabular}} & \textbf{\begin{tabular}[c]{@{}c@{}}Training \\ Time (days) \end{tabular}} \\ \hline
			LUT-ILF-U (DNN)      &  --   &     --                                & -0.13\%                                                                                        & 400000                                                                        & 3                                                               \\ \hline
			LUT-ILF-U (LUT)      &  \ding{55}   &       --                            &     0.19\%                                                                & --                                                                      &--                                                                   \\ \hline
			LUT-ILF-U (LUT)     &  \ding{51}   &       --                            & -0.08\%                                                                   & 20000                                                                        & 0.5                                                                  \\ \hline
			LUT-ILF-V (DNN)      &  --   &     --                                & -0.36\%                                                                                        & 400000                                                                        & 17                                                               \\ \hline
			LUT-ILF-V (LUT)      &  \ding{55}   &       --                            &     -0.07\%                                                                & --                                                                      &--                                                                   \\ \hline
			LUT-ILF-V (LUT)     &  \ding{51}   &       --                            & -0.32\%                                                                   & 20000                                                                        & 4                                                                  \\ \hline
			LUT-ILF-F (DNN)      &  --   &      --                            & -0.51\%                                                                & 400000                                                                      & 27                                                               \\ \hline
			LUT-ILF-F (LUT)      &  \ding{55}  & --                     & -0.15\%                                                                        & --                                                                       &     --                                                       \\ \hline
			LUT-ILF-F (LUT)      &  \ding{51}  & --                     & -0.47\%                                                                                    & 20000                                                                       & 4                                                                \\ \hline
			\textbf{LUT-ILF++} (DNN)     &  -- &  --                          & -0.87\%                                                               & 400000                                                                        & 32                                                                 \\ \hline
			\textbf{LUT-ILF++} (LUT) & \ding{55}  & \ding{55}                                            & -0.39\%                                                                           & --                                                               & --                                                       \\ \hline
			\textbf{LUT-ILF++} (LUT)     &  \ding{51} &  \ding{55}                     & -0.77\%                                                                  & 20000                                                                        & 4                                                                  \\ \hline
			\textbf{LUT-ILF++} (LUT)     &  \ding{51} &  \ding{51}                         & \textbf{-0.82\%}                                                                             & \textbf{20000}                                                                       & \textbf{7}                                                               \\ \hline
		\end{tabular}
	}
	\vspace{-1.3em}
\end{table}

\subsection{Detailed Performance of LUT-ILF++ under Common Test Condition on Different Coding Configurations}
\subsubsection{BD-rate Performance of LUT-ILF++ on Regular Bitrate Points (QP 22$\sim$42)}
Supplementing Section VII.C (1) of the main text, the R-D performance and usage ratio of the entire LUT-ILF++ filtering framework on the VVC common test sequences is illustrated in Table~\ref{tab:DBR1} and \ref{tab:DBR3}\,. Y, U, and V represent the R-D performance gain of the three channels of YUV, respectively. We can see that LUT-ILF++ can achieve, on average, 0.82\%/2.97\%/1.63\% and 0.85\%/4.11\%/2.06\%, and achieve up to 1.49\% and 1.59\% BD-rate reduction (Y component) with high usage ratio and friendly computational complexity (Table II/III of the main text) on VTM-11.0 for all sequences under AI and RA configurations. The experimental results show that the proposed framework performs better for sequences with rich texture and complex scenes, such as $CatRobot$, $RitualDance$, $BQMall$, and $RaceHorses$. The subjective selection and visual results of different LUT-based ILF schemes~\cite{li2024loop} are also shown and compared in Fig.~\ref{fig:partition}\, and \ref{fig:subjective}\,, which also represents that the LUT-ILF++ can better handle the complex texture regions.

\subsubsection{BD-rate Performance of LUT-ILF++ on Low Bitrate Points (QP 27$\sim$47)}
To explore the potential and robustness of the proposed framework, we also test it on low-bitrate points (QP 27$\sim$47), as shown in Table~\ref{tab:DBR2} and \ref{tab:DBR4}\,. Compared to the results of regular QP points (Table~\ref{tab:DBR1} and \ref{tab:DBR3}\,), it demonstrates the powerful potential and comprehensive effectiveness of the proposed framework on a wide range of QP points. Specifically, the proposed LUT-ILF++ can achieve, on average, 1.49\%/4.69\%/2.55\% and 1.21\%/4.54\%/2.21\%, and achieve up to 2.43\%/8.54\%/6.03\% and 2.18\%/9.73\%/4.82\% BD-rate reduction on VTM-11.0 for all sequences under the AI and RA configurations.

\subsubsection{Usage Ratio}
To verify the efficiency of LUT-ILF++, we also evaluate its usage ratio and compare with other LUT-based ILF schemes\cite{li2024loop}, as shown in Table~\ref{tab:DBR1}\, and \ref{tab:DBR2}\,, which is calculated by,
$Ratio = N_{test}/N_{total}$, where $N_{test}$ indicates the number of coding tree units (CTUs) selected and filtered by the corresponding filter, and $N_{total}$ indicates the total number of CTUs. Through the comparison of usage ratio, we can observe that the proposed framework achieves a significantly higher selection proportion, averaging 49.50\%/67.99\% on regular/low bitrate points under AI configuration, and achieving up to 74.99\%/94.83\%, respectively. The results demonstrate its remarkable filtering capability and effectiveness. (Note that the usage ratio of different schemes is not shown and compared under RA configuration, because some schemes, LUT-ILF-U/V/F, disable filtering on some temporal layers to avoid temporal error accumulation, while LUT-ILF++ enables all layers, a direct comparison may be unfair.)

\subsection{Deep Analyses of Finetuning in LUT-based ILF Solutions}
Based on Section II.A (3), Section VI.B and Section VII.B (5) of main text, LUT finetuning serves as a crucial step for transferring the filtering capacity of deep neural network to the practical LUTs. Specifically, the finetuning with interpolation adaptation is used to bridge these transformations and ensure that the impact on the coding gain remains minimal, ensuring that the LUT-based ILF maintains filtering capacity while significantly reducing computational/time complexity. 

To further analyze the benefits that finetuning brings to the LUT-based ILF solution, we conduct ablation studies on all LUT-based ILF models to demonstrate the critical importance of these prolonged LUT training, transferring, and finetuning processes. Specifically, for the ablation setting of LUT-ILF-U, LUT-ILF-V and LUT-ILF-F\cite{li2024loop}, since they only adopt the uniformly sampled pruning and storage strategies, their filtering LUTs only require a single finetuning stage. Thus, the ablation study for these models focuses on evaluating the effect of this single step. In contrast, as detailed in the Section VI.B and Fig.11 of main text, the training of LUT-ILF++ contains a two-step finetuning process: the first step finetunes the LUTs after uniform clipping, and the second step finetunes them again after non-uniform (diagonal-oriented) sampling. Therefore, we perform ablation studies on both finetuning steps to analyze their contributions to the final performance. 

Based on Table II of the main text, in Table~\ref{tab:Fintuning}\,, here we present the comparison results of different LUT-based ILF models at various stages of the above ablation pipelines under all intra (AI) configuration. In detail, these stages are evaluated: (1) reproduced DNN training stage, representing the baseline performance achieved by the fully trained neural network before any LUT conversion; (2) DNN-to-LUT conversion stage, reflecting the performance degradation caused by directly transferring the DNN into LUT without any finetuning; (3) LUT finetuning stage, the LUTs are finetuned with an interpolation model adaptation to regain lost performance after conversion. As shown in Table~\ref{tab:Fintuning}\,, for LUT-ILF-U, LUT-ILF-V, and LUT-ILF-F, it can be observed that these models achieve notable performance degradation when directly converting from DNNs to LUTs without finetuning, while a single-step finetuning effectively restores their performance to the DNN baseline with minimal additional training cost. For LUT-ILF++, the results further verify that each finetuning step contributes progressively to performance improvement. These findings confirm that the finetuning strategy plays a critical role in maximizing the filtering effectiveness, especially in LUT-ILF++, ensuring that its compacted LUT representation maintains coding performance comparable to that of the reproduced DNN models.

\vspace{-1em}
\subsection{Detailed Calculation Manner of Computational Complexity and Energy Cost}
To clearly show the calculation process of the computational complexity of the proposed framework and facilitate researchers to follow the LUT-based ILF solution, here we detail the specific operation num of the basic framework (LUT-ILF-U/V/F in \cite{li2024loop}) and improved framework (LUT-ILF++) on the pixel (per pixel) and frame level (a image/video frame with 1920 × 1080 spatial resolution), as shown in Table~\ref{tab:cc}\,. For the extension of this solution, computational complexity can be calculated with the reference of the basic architecture of LUT-ILF-U for relative expansion. For the basic architecture, it is constructed by incorporating pattern 1 of Fig.5 of main text for reference indexing mechanism and a two-step cascaded filtering iteration (iter = 2) for progressive indexing, enabling a 5×5 reference range.

As shown in Table~\ref{tab:cc}\,, the detailed construction of operation counts, computational complexity, and energy cost for LUT-ILF-U, LUT-ILF-V, LUT-ILF-F, and LUT-ILF++ is presented. Specifically, we  report the results of LUT-ILF++ before and after applying the LUT compaction scheme (Section VI of main text). It can be observed that the application of the proposed LUT compaction scheme not only greatly reduces the storage cost of the entire framework but also only brings a slight increase in the number of addition operations while keeping the number of multiplication operations unchanged, demonstrating its excellent hardware friendliness.

As shown in Table~\ref{tab:cc2}\,, the detailed construction of operation counts, computational complexity, and energy cost of the 3D channel LUT with trilinear interpolation model and the 4D channel LUT with 4-$simplex$ interpolation model is presented. Compared to the trilinear model for 3D LUT, although the 4-$simplex$ model for 4D channel LUT involves a more complex formulation, the adopted interpolation model only requires partial surrounding storage index interpolation, effectively controlling its overall computational complexity. In addition, the detailed construction of the improved framework (LUT-ILF++) for each chroma component is also presented. Compared to the luma component, we adopt a cross-color-component collaborative scheme for chroma filtering (Section V of main text), which effectively improves performance while consuming fewer computational resources.

\subsection{Deployment Discussion of LUT-based ILF Solution Integrated Into Codec}
Based on our reported LUT storage consumption of the whole LUT-based ILF solutions (LUT-ILF-U/V/F, and LUT-ILF++) mentioned in Table II and III of main text, here we further verify their practical deployability by measuring the peak memory usage on CPU using $Memray$\,\footnote{Memray in the media: \url{https://github.com/bloomberg/memray}}\,, respectively. As shown in Table~\ref{tab:peak}\,, all LUT-based filtering models with only integer precision operations exhibit significantly lower peak memory consumption compared to the consumption of the entire decoder during runtime, occupying only about one-fifth of the VTM decoder's peak memory, demonstrating that LUT-based ILF solutions impose minimal additional runtime memory demands and are thus favorable for practical codec deployment. Note that the proposed LUT compaction and pruning strategies in LUT-ILF++ not only improve its advantage in reducing storage consumption but also maintain lower runtime storage consumption.

\begin{table}
	\renewcommand\arraystretch{1.2}
	\centering
	\fontsize{6.5pt}{9pt}\selectfont
	\caption{The Comparison of Peaking Memory Consumption between \\VTM Decoder and Different LUT-based ILF Models}
	\label{tab:peak}
	\vspace{-0.7em}
	\setlength{\tabcolsep}{0.9mm}
	{
		\begin{tabular}{c|c|c|c|c|c}
			\hline
			\textbf{Schemes} & \textbf{\begin{tabular}[c]{@{}c@{}}Storage \\ Manner\end{tabular}}  & \textbf{\begin{tabular}[c]{@{}c@{}}Peak Operating \\ Memory (MB)\end{tabular}}  & \textbf{\begin{tabular}[c]{@{}c@{}}Computational\\ Complexity\\ (Ops/pixel)\end{tabular}} & \textbf{\begin{tabular}[c]{@{}c@{}}Storge\\ Cost\end{tabular}} & \textbf{\begin{tabular}[c]{@{}c@{}}Energy \\ Cost\\ ($pJ$/pixel)\end{tabular}} \\ \hline
			VTM Decoder      &  --   &        275.4MB                               & --                                                                                        & --                                                                       & --                                                                \\ \hline
			LUT-ILF-V      &  Uniform   &        53.7 MB                               & 0.83K Ops/pixel                                                                                        & 1476 KB                                                                         & 0.75K                                                               \\ \hline
			LUT-ILF-F       &  Uniform  & 59.4 MB                                                          & 1.91K Ops/pixel                                                                                       & 3444 KB                                                                         & 1.69K                                                                 \\ \hline
			\textbf{LUT-ILF++} & Non-uniform  & \textbf{56.3 MB}                                                 & \textbf{5.23K Ops/pixel}                                                                               & \textbf{812 KB}                                                                & \textbf{4.26K}                                                         \\ \hline
		\end{tabular}
	}
	\vspace{-1.7em}
\end{table}

\end{document}